\documentclass[twocolumn,tighten,times]{aastex63}
\hypersetup{linkcolor=magenta,citecolor=blue,filecolor=cyan}

\newcommand{\mytilde}{{\raise.17ex\hbox{$\scriptstyle\mathtt{\sim}$}}}

\usepackage{amsmath}

\usepackage{nicefrac}

\submitjournal{The Astrophysical Journal}

\usepackage{CJKutf8}
\newcommand{\cntextSim}[1]{\begin{CJK*}{UTF8}{gkai}#1\end{CJK*}}
\newcommand{\cntext}[1]{\begin{CJK*}{UTF8}{bkai}#1\end{CJK*}}

\shorttitle{The Impact of the New $^{65\!\!}$As(p,$\gamma$)$^{66\!}$Se Reaction Rate ...}
\shortauthors{Lam et al.}

\begin{document}

\title{The Impact of the New $^{65\!\!}$As(p,$\gamma$)$^{66\!}$Se Reaction Rate on the Two-Proton Sequential Capture of $^{64}$Ge,\\Weak GeAs Cycles, and Type-I X-Ray Bursts such as the Clocked Burster GS~1826$-$24}

\correspondingauthor{Yi Hua Lam, Zi Xin Liu, Alexander Heger}
\email{lamyihua@impcas.ac.cn, \\liuzixin1908@impcas.ac.cn, \\alexander.heger@monash.edu}

\author[0000-0001-6646-0745]{Yi Hua Lam (\cntext{藍乙華})}
\affiliation{Institute of Modern Physics, Chinese Academy of Sciences, Lanzhou 730000, People's Republic of China}
\affiliation{School of Nuclear Science and Technology, University of Chinese Academy of Sciences, Beijing 100049, People's Republic of China}

\author[0000-0001-5652-1516]{Zi Xin Liu (\cntextSim{刘子鑫})}
\affiliation{Institute of Modern Physics, Chinese Academy of Sciences, Lanzhou 730000, People's Republic of China}
\affiliation{School of Nuclear Science and Technology, University of Chinese Academy of Sciences, Beijing 100049, People's Republic of China}
\affiliation{School of Physics Science and Technology, Lanzhou University, Lanzhou 730000, People's Republic of China}

\author[0000-0002-3684-1325]{Alexander Heger}
\affiliation{School of Physics and Astronomy, Monash University, Vic 3800, Australia}
\affiliation{OzGrav-Monash -- Monash Centre for Astrophysics, School of Physics and Astronomy, Monash University, Vic 3800, Australia}
\affiliation{Center of Excellence for Astrophysics in Three Dimensions (ASTRO-3D), School of Physics and Astronomy, Monash University, Vic 3800, Australia}
\affiliation{Joint Institute for Nuclear Astrophysics, Michigan State University, East Lansing, MI 48824, USA}


\author[0000-0002-3445-0451]{Ning Lu (\cntext{盧寧})}
\affiliation{Institute of Modern Physics, Chinese Academy of Sciences, Lanzhou 730000, People's Republic of China}
\affiliation{School of Nuclear Science and Technology, University of Chinese Academy of Sciences, Beijing 100049, People's Republic of China}
\affiliation{School of Nuclear Science and Technology, Lanzhou University, Lanzhou 730000, People's Republic of China}

\author[0000-0002-3580-2420]{Adam Michael Jacobs}
\affiliation{Joint Institute for Nuclear Astrophysics, Michigan State University, East Lansing, MI 48824, USA}
\affiliation{Department of Physics and Astronomy, Michigan State University, East Lansing, MI 48824, USA}


\author[0000-0003-4023-4488]{Zac Johnston}%
 \affiliation{Joint Institute for Nuclear Astrophysics, Michigan State University, East Lansing, MI 48824, USA}%
 \affiliation{Department of Physics and Astronomy, Michigan State University, East Lansing, MI 48824, USA}




\begin{abstract}
%
We re-assess the $^{65\!\!}$As(p,$\gamma$)$^{66\!}$Se reaction rates based on a set of proton thresholds of $^{66\!}$Se, $S_\mathrm{p}$($^{66\!}$Se), estimated from the experimental mirror nuclear masses, theoretical mirror displacement energies, and full $p\!f$-model space shell-model calculation. The self-consistent relativistic Hartree-Bogoliubov theory is employed to obtain the mirror displacement energies with much reduced uncertainty, and thus reducing the proton-threshold uncertainty up to 161~keV compared to the AME2020 evaluation. Using the simulation instantiated by the one-dimensional multi-zone hydrodynamic code, \textsc{Kepler}, that closely reproduces the observed GS~1826$-$24 clocked bursts, the present forward and reverse $^{65\!\!}$As(p,$\gamma$)$^{66\!}$Se reaction rates based on a selected $S_\mathrm{p}$($^{66\!}$Se) $=2.469\mathord\pm0.054$~MeV, and the latest $^{22}$Mg($\alpha$,p)$^{25}\!$Al, $^{56}$Ni(p,$\gamma$)$^{57}$Cu, $^{57}$Cu(p,$\gamma$)$^{58}$Zn, $^{55}$Ni(p,$\gamma$)$^{56}$Cu, and $^{64}$Ge(p,$\gamma$)$^{65\!\!}$As reaction rates, we find that though the GeAs cycles is weakly established in the rapid-proton capture process path, the $^{65\!\!}$As(p,$\gamma$)$^{66\!}$Se reaction still strongly characterizes the burst tail end due to the two-proton sequential capture on $^{64}$Ge, not found by \citet{Cyburt2016} sensitivity study. The $^{65\!\!}$As(p,$\gamma$)$^{66\!}$Se reaction influences the abundances of nuclei $A=64$, $68$, $72$, $76$, and $80$ up to a factor of 1.4. The new $S_\mathrm{p}$($^{66\!}$Se) and the inclusion of the updated $^{22}$Mg($\alpha$,p)$^{25}\!$Al reaction rate increases the production of $^{12}$C up to a factor of $4.5$ that is not observable and could be the main fuel for superburst. The enhancement of $^{12}$C mass fraction alleviates the discrepancy in explaining the origin of superburst. The waiting point status of and two-proton sequential capture on $^{64}$Ge, weak-cycle feature of GeAs at region heavier than $^{64}$Ge, and impact of other possible $S_\mathrm{p}$($^{66\!}$Se) are also discussed.
\end{abstract}



\section{Introduction} \label{sec:intro}

During the accretion of stellar matter from a close companion by the neutron star in a low-mass X-ray binary (LMXB), the accreted stellar matter, mainly comprising H and He fuses to heavier nuclei in steady-state burning \citep{Hansen1975, Schwarzschild1965} and a thermonuclear runaway is likely to occur in the accreted envelope of the neutron star. This causes the observed thermonuclear (Type-I) X-ray bursts (XRBs; \citealt{Woosley1976, Maraschi1977, Joss1977,Lamb1978, Bildsten1998}).
The burst light curve, usually observed in X-rays, is the important observable indicating an episode of XRB, which can be either a series of XRBs or a single XRB, for instance, the XRBs recorded by the \emph{Rossi X-ray Timing Explorer} (RXTE) Proportional Counter Array \citep{Galloway2004,Galloway2008,MINBAR}. For a series of XRBs, we can further deduce the recurrence time between two XRBs, or the averaged recurrence time of a series of XRBs. The burst light curve profile and (averaged) recurrence time can be studied via models best matching with observations to further understand the role of an important reaction, e.g., see \citet{Cyburt2016, Meisel2018a}.


In recent years, the Type-I clocked XRBs from the GS~1826$-$24 X-ray source \citep{Makino1988, Tanaka1989, Ubertini1999} became the primary target of investigation due to its almost constant accretion rate and consistent light curve profile \citep{Heger2007, MESA2015, Meisel2018a, Meisel2019, Johnston2020, Dohi2020, Dohi2021}. The other important feature of GS~1826$-$24 clocked burster is the nuclear reaction flow during its XRB onsets. It can reach up to the SnSbTe cycles according to the models \citep{Woosley2004} used by \citet{Cyburt2016} and \citet{Jacobs2018} to investigate the sensitivity of reactions. The region around SnSbTe cycles was found to be the end point of nucleosynthesis of an XRB \citep{Schatz2001}.

This state-of-the-art GS~1826$-$24 model was even recently advanced by the newly deduced $^{22}$Mg($\alpha$,p)$^{25}\!$Al reaction rate to match with observation \citep{Hu2021}. As the abundances of synthesized isotopes are not observed, we remark here that using a set of high-fidelity XRB models capable to closely reproduce the observed burst light curves is important for us to diagnose the nuclear reaction flow and abundances of synthesized isotopes in the accreted envelope. 

\begin{figure}[b!]
\begin{center}
\includegraphics[width=5.7cm, angle=0]{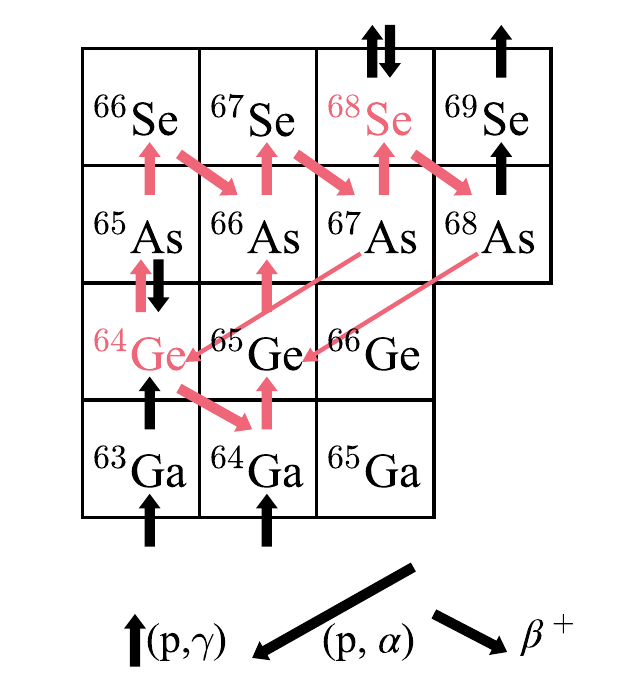}
\singlespace
\caption{\label{fig:Cycles_GeAs}{\footnotesize The rapid-proton capture (rp-) process path passing through the weak GeAs cycles. Waiting points are shown~in red texts. The GeAs cycles are displayed as red arrows. The GeAs~I cycle consists~of $^{64}$Ge(p,$\gamma$)$^{65\!\!}$As(p,$\gamma$)$^{66\!}$Se($\beta^+\nu$)$^{66}\!$As(p,$\gamma$)$^{67}\!$Se($\beta^+\nu$)$^{67}\!$As(p,$\alpha$)$^{64}$Ge reactions \citep{Wormer1994}, and the GeAs~II cycle involves a series~of $^{65}\!$Ge(p,$\gamma$)$^{66}\!$As(p,$\gamma$)$^{67}\!$Se($\beta^+\nu$)$^{67}\!$As(p,$\gamma$)$^{68}$Se($\beta^+\nu$)$^{68}\!$As(p,$\alpha$)$^{65}$Ge reactions.~The~other~sub-GeAs~II cycle, $^{64}$Ge($\beta^+\nu$)$^{64}$Ga(p,$\gamma$)$^{65}$Ge (p,$\gamma$)$^{66}$As(p,$\gamma$)$^{67}$Se($\beta^+\nu$)$^{67}$As(p,$\alpha$)$^{64}$Ge, may also be established. The weak (p,$\alpha$) reactions are represented by thinner arrows.}}
\end{center}
\vspace{-5mm}
\end{figure}

According to this GS~1826$-$24 model \citep{Heger2007, Cyburt2016, Johnston2020, Jacobs2018, Hu2021, Lam2021a}, throughout the course of the XRBs of GS~1826$-$24 clocked burster, the nuclear reaction flow has to break out from the ZnGa cycles to reach the GeAsSe region. The reaction flow inevitably goes through the $^{64}$Ge waiting point and also the GeAs cycles that might exist \citep{Wormer1994} before surging through the region heavier than the Ge and Se isotopes where hydrogen is intensively burned. The GeAs cycles are supposedly weaker than the NiCu and ZnGa cycles due to the competition between the weak (p,$\alpha$) and the strong (p,$\gamma$) reactions at As isotopes, see the weak GeAs I, II, and sub-II cycles presented in Fig.~\ref{fig:Cycles_GeAs}. Thus, the establishment of the two-proton- (2p) sequential capture (hereafter 2p-capture) on $^{64}$Ge could be crucial to draw synthesized materials to follow the $^{64}$Ge(p,$\gamma$)$^{65\!\!}$As(p,$\gamma$)$^{66\!}$Se($\beta^+\nu$)$^{66}$As(p,$\gamma$)$^{67}$Se($\beta^+\nu$)$^{67}$As (p,$\gamma$)$^{68}$Se($\beta^+\nu$)$^{68}$As(p,$\gamma$)$^{69}$Se path for the strong rp-process of hydrogen (H)-burning that happens in nuclei heavier than Se isotopes.


Recently, the role of $^{64}$Ge as an important waiting point was questioned when the proton threshold of $^{65\!\!}$As was experimentally deduced, opening a 2p-capture channel \citep{Tu2011}; its significance of an important waiting point also becomes uncertain in a study using an one-zone XRB model with the $^{64}$Ge(p,$\gamma$)$^{65\!\!}$As and $^{65\!\!}$As(p,$\gamma$)$^{66\!}$Se reaction rates deduced from the evaluated $^{65\!\!}$As and $^{66\!}$Se proton thresholds \citep{Tu2011, AME2016} and the full $p\!f$-model space shell-model calculation \citep{Lam2016}. Moreover, \citet{Schatz2006a} and \citet{Schatz2017} found that the $^{65\!\!}$As and $^{66\!}$Se masses contributed to the respective proton thresholds affect the (p,$\gamma$) reverse rate, and hence influence the modeled XRB light curves. We remark that the investigations of the role of $^{64}$Ge as an important waiting point and the impact of $^{65\!\!}$As and $^{66\!}$Se masses were, however, based on the computationally effective zero-dimensional one-zone XRB model  \citep{Schatz2017} with extreme parameters, i.e., very high accretion rate and very low crustal heating, of which the abundance of synthesized nuclei is only provided for a single mass zone but not along the mass coordinate in the accreted envelope. A finer parameter range can be chosen to improve the capability of one-zone XRB models in estimating the influence of a particular reaction on XRB with agreement close to one-dimensional multi-zone hydrodynamic XRB model \citep{Schatz2021}. Notably, the hydrodynamic data generated from a more constrained one-dimensional multi-zone XRB model that is capable of reconciling theoretical light curves with observations could be beneficial for post-processing and one-zone models to obtain the nuclear energy generation (XRB flux) consistent with the input hydrodynamic snapshots and to assess inventory of abundances and the reaction flow during the burst.


With the advantage of new and high intensity exotic proton-rich isotopes production and advances in experimental techniques of isochronous mass spectrometry in storage rings \citep{Stadlmann2004, Zhang2018}  and multi-reflection time-of-flight (MRTOF) mass spectrometers \citep{Wolf2013, Dickel2015, Jesch2015, Rosenbusch2020}, the $^{66\!}$Se proton threshold, $S_\mathrm{p}$($^{66\!}$Se), could be more precisely determined to replace the $S_\mathrm{p}$($^{66\!}$Se)~$=2.010\pm220$~MeV predicted by AME2020 extrapolation (\citealt{AME2020}; AME2020). The extrapolation is based on the Trend from the Mass Surface, which is weak in predicting proton-rich nuclear masses due to the shell effect originated from the respective shell structure or the configuration of the ground state \citep{Huang2021}. Therefore, a set of forefront investigations on astrophysical impacts due to the forward and reverse $^{65\!\!}$As(p,$\gamma$)$^{66\!}$Se reactions based on a set of $S_\mathrm{p}$($^{66\!}$Se) with lower uncertainty than the one predicted by AME2020 and XRB models best matching with observation is highly desired. 

In this work, in Section~\ref{sec:rate_formalism}, we present the formalism obtaining the forward and reverse $^{65\!\!}$As(p,$\gamma$)$^{66\!}$Se reaction rates, and discuss these newly deduced forward and reverse reaction rates. We then employ the one-dimensional multi-zone hydrodynamic \textsc{Kepler} code \citep{Weaver1978,Woosley2004,Heger2007} to instantiate XRB simulations that produce a set of XRB episodes matched with the GS~1826$-$24 clocked burster with the newly deduced $^{65\!\!}$As(p,$\gamma$)$^{66\!}$Se forward and reverse reaction rates. Other recently updated reaction rates, i.e., $^{55}$Ni(p,$\gamma$)$^{56}$Cu \citep{Valverde2019}, $^{56}$Ni(p,$\gamma$)$^{57}$Cu \citep{Kahl2019}, $^{57}$Cu(p,$\gamma$)$^{58}$Zn \citep{Lam2021a}, and $^{64}$Ge(p,$\gamma$)$^{65\!\!}$As \citep{Lam2016} around the historic $^{56}$Ni and $^{64}$Ge waiting points, and $^{22}$Mg($\alpha$,p)$^{25}\!$Al \citep{Hu2021} at the important $^{22}$Mg branch point are also taken into account. 

In Section~\ref{sec:Astro}, we study the influence of these $^{65\!\!}$As(p,$\gamma$)$^{66\!}$Se forward and reverse rates, and also investigate the influence of the 2p-capture on $^{64}$Ge and GeAs cycles on XRB light curves, on the rp-process path during the thermonuclear runaway of the GS~1826$-$24 clocked XRBs, and on the nucleosyntheses in and evolution of the accreted envelope along the mass coordinate for nuclei of mass $A=59$, $\dots$, $68$. The conclusions of this work are given in Section~\ref{sec:summary}.

\section{Reaction rate calculations}
\label{sec:rate_formalism}

We deduce the $S_\mathrm{p}$($^{66\!}$Se) value based on the experimental $^{66}$Ge mirror mass and theoretical Coulomb displacement energy (and the term should be replaced as ``mirror displacement energy'' due to the important role of isospin non-conserving forces from nuclear origin \citep{Zuker2002}). The mirror displacement energy (MDE) is obtained from the self-consistent relativistic Hartree-Bogoliubov (RHB) theory \citep{Kucharek1991, Ring1996, Meng1996, Poschl1997}.
The MDE for a given pair of mirror nuclei is expressed as \citep{Brown2000, Brown2002,Zuker2002}
\begin{equation}
\label{eq:MDE}
\mathrm{MDE}(A,I)=\mathrm{BE}(A,I^{<}_{z}) - \mathrm{BE}(A,I^{>}_{z}) \, ,
\end{equation}
where $A$ is the nuclear mass number and $I$ is the isospin, BE$(A,I^{<}_{z})$ and BE$(A,I^{>}_{z})$ are the binding energies of the proton and neutron rich nuclei, respectively. We implement the explicit density-dependent meson-nucleon couplings (DD-ME2) effective interaction \citep{Lalazissis2005} in RHB calculations to obtain the MDE$(A,I)$ of $I=1/2$, $1$, $3/2$, and $2$, $A=41$~--~$75$ mirror nuclei. In order to constrict the complicated problem of a cut-off at large energies inherent in the zero range pairing forces, a separable form of the finite range Gogny pairing interaction is adopted \citep{Tian2009}. To describe the nuclear structure in odd-$N$ and/or odd-$Z$ nuclei, we take into account of the blocking effects of the unpaired nucleon(s). The ground state of a nucleus with an odd neutron and/or proton numbers is a one-quasiparticle state, $|\Phi_{1}\rangle = \beta_{i_\mathrm{b}}^{\dag}|\Phi_{0}\rangle$, which is constructed based on the ground state of an even-even nucleus $|\Phi_{0}\rangle$, where $\beta^{\dag}$ is the single-nucleon creation operator and $i_\mathrm{b}$ denotes the blocked quasiparticle state occupied by the unpaired nucleon(s). The detail description of implementing the blocking effect is explained by \citet{Ring1980}.

\begin{table*}[t]
\footnotesize
\caption{\label{tab:MDE} Mirror displacement energies and binding energies of $A=65$ and $66$, and $S_\mathrm{p}$($^{66\!}$Se).}
\begin{tabular*}{\linewidth}{@{\hspace{1mm}\extracolsep{\fill}}llllllll@{\hspace{1mm}}}
\hline
\hline
     & MDE$_{A=66}$ & BE($^{66}$Ge) & BE($^{66\!}$Se)$^a$ & MDE$_{A=65}$ & BE($^{65}$Ge) & BE($^{65\!\!}$As)$^a$ & $S_\mathrm{p}$($^{66\!}$Se) \\
\hline
SHF (Sk$X_\mathrm{csb}$)$^b$ &&&&&&&\\
\citet{Brown2002}                  & 21.340 (100)$^b$ & $-$569.293 (30)$^c$  & $-$547.953 (104)      & 10.491 (100)$^b$ & $-$556.011 (100)$^c$ & $-$545.520 (100)     & 2.433 (144) \\
SHF$^\mathrm{i}$  (AME2020)        & 21.340 (100)$^b$ & $-$569.279 (2)$^d$   & $-$547.939 (100)      & 10.326 (80)$^d$  & $-$556.079 (2)$^d$   & $-$545.753 (80)$^d$  & 2.186 (128) \\
SHF$^\mathrm{ii}$ (AME2020)        & 21.340 (100)$^b$ & $-$569.279 (2)$^d$   & $-$547.939 (100)      & 10.491 (100)$^b$ & $-$556.079 (2)$^d$   & $-$545.588 (100)     & 2.351 (144) \\
\hline
RHB (DD-ME2)$^e$             &&&&&&&\\
Spherical$^\mathrm{i}$ (AME2020)   & 21.083 (62)$^{e_{\mathrm{(i)}}}$  & $-$569.279 (2)$^d$   & $-$548.196 (62)  & 10.326 (80)$^d$                   & $-$556.079 (2)$^d$   & $-$545.753 (80)$^d$ & 2.443 (101) \\
Spherical$^\mathrm{ii}$ (AME2020)  & 21.083 (62)$^{e_{\mathrm{(i)}}}$  & $-$569.279 (2)$^d$   & $-$548.196 (62)  & 10.389 (33)$^{e_{\mathrm{( i)}}}$ & $-$556.079 (2)$^d$   & $-$545.690 (33)     & 2.507 (70)  \\
&&&&&&&\\
Axial$^\mathrm{i}$    (AME2020)    & 21.242 (46)$^{e_{\mathrm{(ii)}}}$ & $-$569.279 (2)$^d$   & $-$548.037 (46)  & 10.326 (80)$^d$                   & $-$556.079 (2)$^d$   & $-$545.753 (80)$^d$ & 2.284 (92)  \\
Axial$^\mathrm{ii}$   (AME2020)    & 21.242 (46)$^{e_{\mathrm{(ii)}}}$ & $-$569.279 (2)$^d$   & $-$548.037 (46)  & 10.510 (29)$^{e_{\mathrm{(ii)}}}$ & $-$556.079 (2)$^d$   & $-$545.568 (29)     & 2.469 (54)  \\
\hline
\end{tabular*}
\begin{minipage}{\linewidth}
\footnotesize
\vskip5pt
{\sc Note}---\\
$^a$ deduced from $\mathrm{BE}(A,I^{<}_{z}) = \mathrm{MDE}(A,I) + \mathrm{BE}(A,I^{>}_{z})$, c.f., Eq.~(\ref{eq:MDE}), except otherwise quoted from experiment. \\
$^b$ Skyrme Hartree-Fock calculations with Sk$X_\mathrm{csb}$ parameters \citep{Brown2021, Brown2002, Brown2000, Brown1998}. \\
$^c$ quoted from AME2000, which is a preliminary data set of AME2003 \citep{AME2003}. \\
$^d$ quoted from AME2020 \citep{AME2020}. \\
$^e$ relativistic Hartree-Bogoliubov \citep{Kucharek1991, Ring1996, Meng1996, Poschl1997} calculations with DD-ME2 effective interaction \citep{Lalazissis2005}, using (I) spherical harmonic oscillator basis, (II) axially symmetric quadrupole deformation. The 62-keV and 46-keV uncertainties of MDE$_{A=66}$ are from the root-mean-square (rms) deviation value of comparing the theoretical and experimental MDE$(A,I)$ values for $I=1$, $A=42$~--~58 mirror nuclei. We apply the same procedure to quantify the 33-keV and 29-keV uncertainties for the MDE$_{A=65}$ considering $I=1/2$, $A=41$~--~75 mirror nuclei. The uncertainties of BE($^{66\!}$Se), BE($^{65\!\!}$As), $S_\mathrm{p}$($^{66\!}$Se) are from the combination of rms deviation and the respective experimental uncertainty.
\end{minipage}
\end{table*}

\begin{figure}[t]
\begin{center}
\includegraphics[width=8.0cm, angle=0]{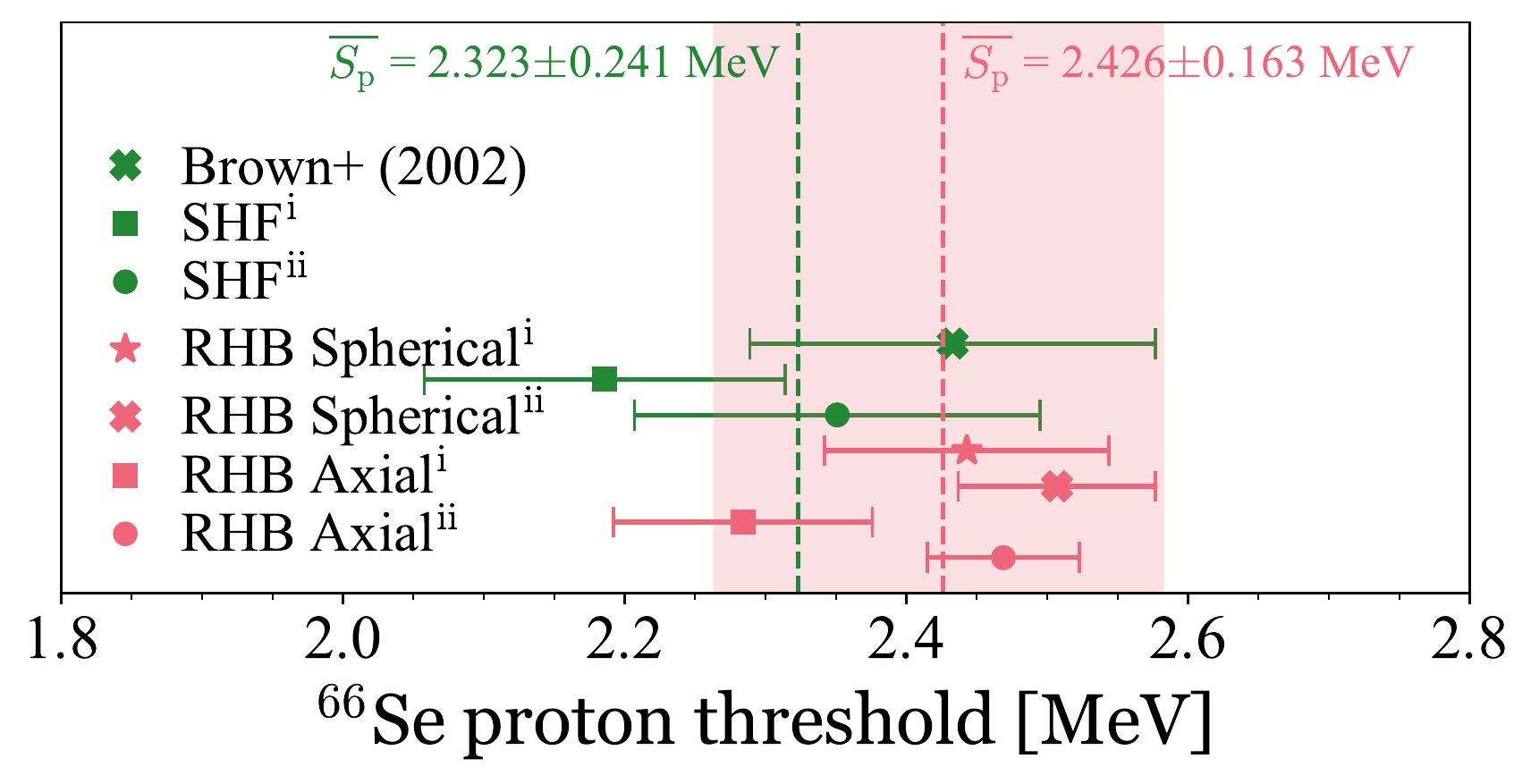}
\singlespace
\caption{\label{fig:Sp}{\footnotesize The estimated $S_\mathrm{p}$($^{66\!}$Se) from the SHF with Sk$X_\mathrm{csb}$ interaction and from the RHB with DD-ME2 interaction using spherical harmonic oscillator basis or axially symmetric quadrupole deformation. The averaged $S_\mathrm{p}$($^{66\!}$Se) from the SHF and from the RHB are $\bar{S_\mathrm{p}}$($^{66\!}$Se)~$=2.323\mathord\pm0.241$~MeV (green dashed line) and $\bar{S_\mathrm{p}}$($^{66\!}$Se)~$=2.426\mathord\pm0.163$~MeV (red dashed line with red uncertainty zone), respectively. See text and Table~\ref{tab:MDE}.}}
\end{center}
\vspace{-5mm}
\end{figure}

Table~\ref{tab:MDE} and Figure~\ref{fig:Sp} present the $S_\mathrm{p}$($^{66\!}$Se) values estimated from two mean field approaches, i.e., Skyrme Hartree-Fock (SHF) and RHB. For the SHF framework with the Sk$X_\mathrm{csb}$ parameters \citep{Brown2021, Brown2002, Brown2000, Brown1998}, the previously estimated $S_\mathrm{p}$($^{66\!}$Se) with the AME2000 $^{66}$Ge and $^{65}$Ge binding energies is listed in the first row of Table~\ref{tab:MDE}, whereas the presently recalculated $S_\mathrm{p}$($^{66\!}$Se) values with the AME2020 $^{66}$Ge and $^{65}$Ge binding energies \citep{AME2020} (SHF$^\mathrm{i}$), or with the AME2020 $^{66}$Ge, $^{65}$Ge, and SHF MDE$_{A=65}$ (SHF$^\mathrm{ii}$), are arranged in the second and third rows of Table~\ref{tab:MDE}, respectively. For the RHB framework with DD-ME2 effective interaction, we deduce the $S_\mathrm{p}$($^{66\!}$Se) using (I) the spherical harmonic oscillator basis and the AME2020 $^{66}$Ge binding energy and theoretical or experimental MDE$_{A=65}$ (RHB Spherical$^\mathrm{i,ii}$; the fourth and fifth rows in Table~\ref{tab:MDE}), (II) the axially symmetric quadrupole deformation and the AME2020 $^{66}$Ge binding energy and theoretical or experimental MDE$_{A=65}$ (RHB Axial$^\mathrm{i,ii}$; the sixth and seventh rows in Table~\ref{tab:MDE}).

The region around nuclei $A=65$ and $A=66$ is subject to deformation as found by \citet{Hilaire2007}, however, the SHF calculations do not take deformation into account. This somehow causes the SHF MDEs having large rms deviation of around $100$~keV. The $S_\mathrm{p}$($^{66\!}$Se) uncertainties of the SHF calculations are larger than the RHB calculations, whereas the RHB Axial$^\mathrm{i,ii}$ uncertainties are lower than the ones from RHB Spherical$^\mathrm{i,ii}$. The uncertainty (rms deviation) of MDEs from RHB Spherical$^\mathrm{i,ii}$ suggest that refitting the Sk$X_\mathrm{csb}$ parameters and revising the blocking effect for calculating even-odd and odd-odd nuclei are likely to reduce the SHF uncertainty to around $60$~keV. Nevertheless, the deformation of nuclei is not considered in both SHF and RHB Spherical$^\mathrm{i,ii}$ calculations. The RHB Axial$^\mathrm{i,ii}$ that take into account of deformation further alleviate the discrepancy between theoretical and experimental MDEs, and thus reducing the uncertainty of the deduced $S_\mathrm{p}$($^{66\!}$Se) values. This is exhibited by comparing the RHB Axial$^\mathrm{i,ii}$ MDE$_{A=66}$ $46$-keV and MDE$_{A=65}$ $29$-keV uncertainties with the $62$-keV and $33$-keV (RHB Spherical$^\mathrm{i,ii}$) and with the $100$-keV (SHF) uncertainties, see Table~\ref{tab:MDE} and Fig.~\ref{fig:Sp}. In additions, the RHB with DD-ME2 calculations having an advantage that the meson-nucleon effective interaction is explicitly constructed without additional extra terms as included in the SHF calculation via the Sk$X_\mathrm{csb}$ terms which have to be continually refitted with updated experiments.

For the RHB calculations, we find that the RHB MDEs coupled with highly precisely measured $^{66}$Ge and $^{65}$Ge binding energies yield a set of $S_\mathrm{p}$($^{66\!}$Se) values with low uncertainties, i.e., $2.507\mathord\pm0.070$~MeV and $2.469\mathord\pm0.054$~MeV (fifth and seventh rows in Table~\ref{tab:MDE}). This indicates that the $80$-keV uncertainty of $^{65\!\!}$As binding energy is somehow large enough to influence the deduced $S_\mathrm{p}$($^{66\!}$Se), see fourth and sixth rows in Table~\ref{tab:MDE} and Fig.~\ref{fig:Sp}, suggesting a highly precisely measured $^{65\!\!}$As mass is demanded. Furthermore, we also find that consistently using the RHB MDEs and precisely measured $^{66}$Ge and $^{65}$Ge binding energies maintains the global description feature provided by the RHB framework to describe MDEs along the nuclear region with symmetric neutron-proton number.

The $S_\mathrm{p}$($^{66\!}$Se) of RHB listed in Table~\ref{tab:MDE} is averaged to be $2.426\mathord\pm0.163$~MeV of which its uncertainty covers all RHB central $S_\mathrm{p}$($^{66\!}$Se) values (red zone in Fig.~\ref{fig:Sp}). We caution that this is an averaged $S_\mathrm{p}$($^{66\!}$Se) from two sets of independent RHB frameworks. The $S_\mathrm{p}$($^{66\!}$Se) of SHF (Table~\ref{tab:MDE}) is averaged to be $2.323\mathord\pm0.241$~MeV, or to be $2.269\mathord\pm0.193$~MeV if we only consider the $S_\mathrm{p}$($^{66\!}$Se) of SHF$^\mathrm{i}$ and SHF$^\mathrm{ii}$. Nevertheless, the uncertainties of both averaged $S_\mathrm{p}$($^{66\!}$Se) of SHF framework are still larger than the one from RHB. The $S_\mathrm{p}$($^{66\!}$Se) with the lowest uncertainty among all estimations is $2.469\mathord\pm0.054$~MeV, estimated from the RHB Axial$^\mathrm{ii}$, and is up to 90~keV lower than the ones proposed by the SHF. With the advantage of the consideration of axial deformation and low uncertainty, hereafter, we select $S_\mathrm{p}$($^{66\!}$Se)~$=2.469\mathord\pm54$~keV as our reference. This selection also maintains the global description of MDEs provided by RHB and refrains the influence of high uncertainty $^{65\!\!}$As mass. We present the influence of the selected $S_\mathrm{p}$($^{66\!}$Se) on the $^{65\!\!}$As(p,$\gamma$)$^{66\!}$Se forward and reverse reaction rates and on the GS~1826$-$24 clocked XRBs. We qualitatively discuss the estimated influence from other $S_\mathrm{p}$($^{66\!}$Se) listed in Table~\ref{tab:MDE} as well in the following discussion and Section~\ref{sec:Astro}.

With using the selected $S_\mathrm{p}$($^{66\!}$Se), the resonant energies correspond to the new Gamow window are shifted up to $E_\mathrm{res}=2.40$~--~$4.70$~MeV, and the dominant resonance states are shifted to the excited state region of $E_\mathrm{x}\mytilde3.50$~MeV accordingly. As an assignment of a 2.1~\!\% uncertainty does not produce a large impact on the final uncertainty, for the present work, we extend the present uncertainty of $S_\mathrm{p}$($^{66\!}$Se) to 100~keV as proposed by \citet{Brown2002} and by the $S_\mathrm{p}$($^{66\!}$Se) uncertainty based on the RHB with DD-ME2 using spherical harmonic oscillator basis (fourth row in Table~\ref{tab:MDE}). Such extended uncertainty could be more reasonable and more conservative for us to estimate the uncertainty of the $^{65\!\!}$As(p,$\gamma$)$^{66\!}$Se forward and reverse reaction rates, and also to cover other reaction rates due to other $S_\mathrm{p}$($^{66\!}$Se) listed in Table~\ref{tab:MDE}, i.e., $S_\mathrm{p}$($^{66\!}$Se)~$=2.433$~MeV, $2.443$~MeV, and $2.507$~MeV, whereas the reaction rates implement $S_\mathrm{p}$($^{66\!}$Se)~$=2.186$~MeV, or $2.284$~MeV, or $2.351$~MeV are separately calculated, presented, and discussed in the following part of this Section. The averaged $S_\mathrm{p}$($^{66\!}$Se) in Table~\ref{tab:MDE} is $2.381\pm0.041$~MeV. The influence of $S_\mathrm{p}$($^{66\!}$Se)~$=2.381\pm0.041$~MeV on the $^{65\!\!}$As(p,$\gamma$)$^{66\!}$Se forward and reverse reaction rates and on XRB can be analogous to the influence of statistical model (\emph{ths}8 or NON-SMOKER) $^{65\!\!}$As(p,$\gamma$)$^{66\!}$Se rate based on $S_\mathrm{p}$($^{66\!}$Se)~$=2.349$~MeV.




Instead of scaling the rate as done by \citet{Valverde2018}, which may introduce unknown uncertainty to the reaction rate, we follow the procedure implemented by \citet{Lam2016} to obtain the new $^{65\!\!}$As(p,$\gamma$)$^{66\!}$Se reaction rate that is expressed as the sum of resonant- (res) and direct (DC) proton capture on the ground state and thermally excited states of the $^{65\!\!}$As target nucleus \citep{Fowler1964,Rolfs1988}, 
\begin{eqnarray}
\label{eq:total}
N_\mathrm{A}\langle \sigma v \rangle = && \sum_i(N_\mathrm{A}\langle \sigma v \rangle_\mathrm{res}^i+N_\mathrm{A}\langle \sigma v \rangle_\mathrm{DC}^i)\nonumber\\
&&\times\frac{(2J_i+1)e^{-E_i/kT}}{\sum_n(2J_n+1)e^{-E_n/kT}} \, .
\end{eqnarray}
Each proton capture is weighted with its partition functions of initial and final nuclei (see \citet{Lam2016} for the detailed notation and formalism). The direct-capture rate of the $^{65\!\!}$As(p,$\gamma$)$^{66\!}$Se reaction can be neglected as its contribution to the total rate is exponentially lower than the contribution of resonant rate. The resonant rate for proton capture on a $^{65\!\!}$As nucleus in its initial state $i$, in units of $\mathrm{cm^3s^{-1}mol^{-1}}$, is expressed as~\citep{Fowler1967,Rolfs1988,Schatz2005,Iliadis2007}, 
\begin{eqnarray}
\label{eq:res}
N_\mathrm{A}\langle \sigma v \rangle_\mathrm{res}^i = && 1.54 \times 10^{11} (\mu T_9)^{-3/2} \nonumber\\
                                && \times \sum_j \omega\gamma_{ij} \mathrm{exp} \left (-\frac{11.605E^{ij}_\mathrm{res}}{T_9} \right)\, ,
\end{eqnarray}
where the resonance energy in the center-of-mass system is $E^{ij}_\mathrm{res}$=$E^j_\mathrm{x} - S_\mathrm{p}(^{66}\mathrm{Se}) -E_i$; $E^j_\mathrm{x}$ is the excited state energy of a state $j$ for the $^{65\!\!}$As$+$p compound nucleus system, 
$E_i$ is the initial state energy, $E_i=0$ for the capture on the ground state (g.s.) of $^{65\!\!}$As, the 11.605 constant is in units of $\nicefrac{10^{9}\mathrm{K}}{k_\mathrm{B}}$, $\mu$ is the reduced mass of the entrance channel, $T_9 = T 10^{-9}\,\mathrm{K}^{-1}$, and $\omega\gamma$ is the resonance strength. We consider only up to the proton capture on the first excited state of $^{65\!\!}$As, $5/2^-_1$, whereas the contribution from the proton capture on higher excited states of $^{65\!\!}$As is negligible. 

\begin{figure}[t!]
\begin{center}
\includegraphics[width=8.7cm, angle=0]{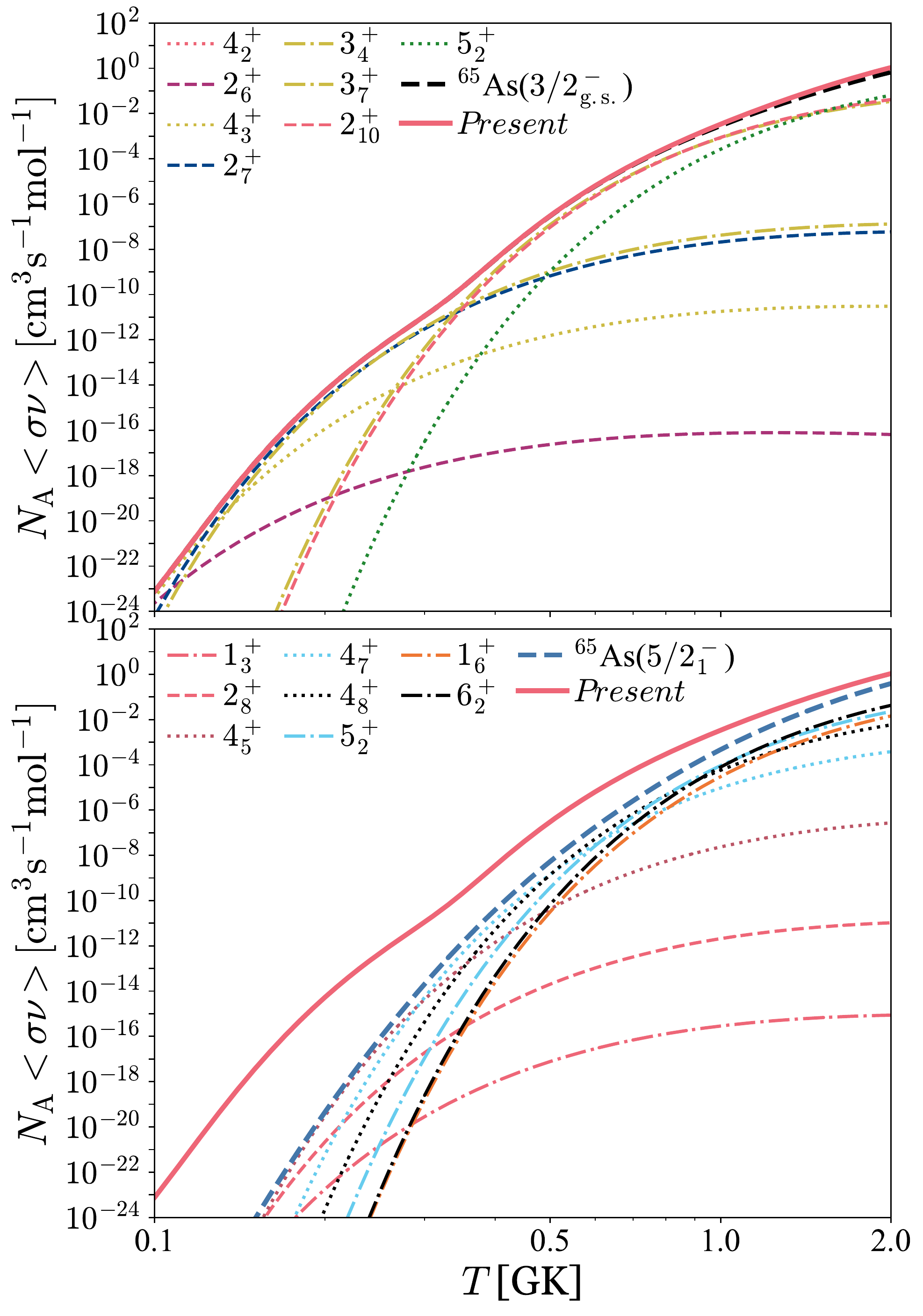}
\caption{\label{fig:65As_66Se_contri}{\footnotesize The dominant resonances contributing to the $^{65\!\!}$As(p,$\gamma$)$^{66\!}$Se reaction rates. \textsl{Top Panel:} The main contributing resonances of proton captures on the $3/2^-_\mathrm{g.s.}$ state of $^{65\!\!}$As. \textsl{Bottom Panel:} The main contributing resonances of proton captures on the $5/2^-_{1}$ state of $^{65\!\!}$As. The \emph{Present} rate (red line) is plotted together with the rates contributed from the proton captures on the $3/2^-_\mathrm{g.s.}$ (black dashed line) and $5/2^-_{1}$ (blue dashed line) states of $^{65\!\!}$As in each panel.}}
\end{center}
\end{figure}

\begin{figure}[t]
\begin{center}
\includegraphics[width=8.7cm, angle=0]{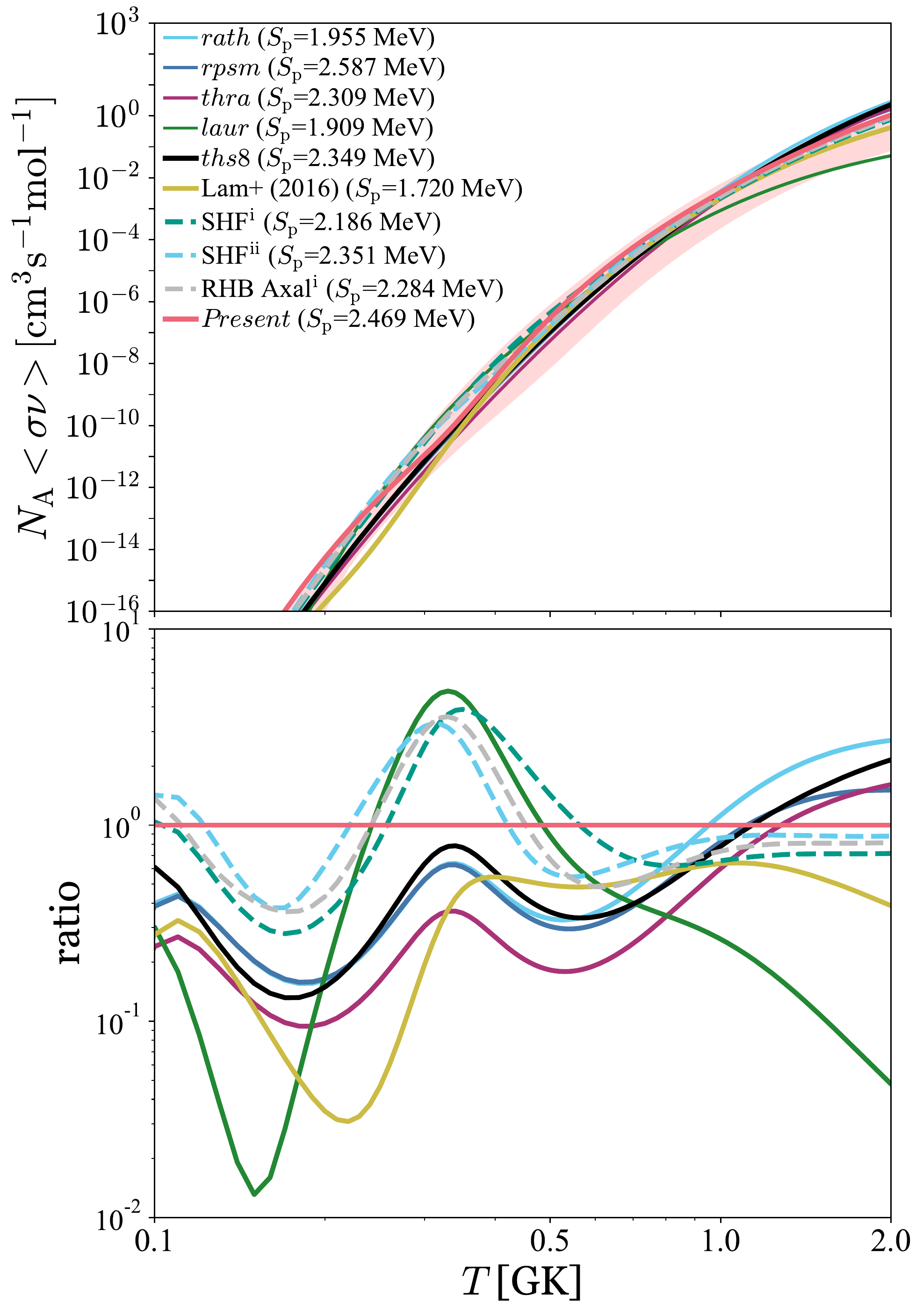}
\caption{\label{fig:rp_65As_66Se}{\footnotesize The $^{65\!\!}$As(p,$\gamma$)$^{66\!}$Se thermonuclear reaction rates in the temperature region of XRB interest. \textsl{Top Panel:} the \emph{rath}, \emph{rpsm}, \emph{thra}, \emph{laur}, and \emph{ths}8 are the available rates compiled by \citet{Cyburt2010} and \emph{ths}8 is the recommended rate published in part of the JINA REACLIB~v2.2 release \citep{Cyburt2010}. \citet{Lam2016} rate is based on $S_\mathrm{p}$($^{66\!}$Se) $=1.720$~MeV (\citealt{AME2012}; AME2012). The uncertainties of the \emph{Present} rate is indicated as light red zone. \textsl{Bottom Panel:} the comparison of the \emph{Present} rate with all reaction rates presented in \textsl{Top Panel}.}}
\vspace{-3mm}
\end{center}
\end{figure}

The nuclear structure information for the proton widths, $\Gamma_{\!\!\mathrm{p}}$, and gamma widths $\Gamma_{\!\!\mathrm{\gamma}}$ at the Gamow window corresponding to the XRB temperature range is deduced based on the full $p\!f$-model space shell-model calculation using the \textsc{KShell} code~\citep{Shimizu2019} and \textsc{NuShellX@MSU} code \citep{NuShellX} with the GXPF1a Hamiltonian \citep{Honma2004,Honma2005}. Hamiltonian matrices of dimensions up to $6.56\times10^8$ for nuclear structure properties of $A=65$ and $66$ have been diagonalized using the thick-restart block Lanczos method. These $\Gamma_{\!\!\mathrm{p}}$ are mainly estimated from proton scattering cross sections in adjusted Woods-Saxon potentials that reproduce known proton energies \citep{WSPOT}. Alternatively, we also employ the potential barrier penetrability calculation \citep{Wormer1994, Herndl1995} to estimate these $\Gamma_{\!\!\mathrm{p}}$. The $\Gamma_{\!\!\mathrm{p}}$ estimated from both methods vary only up to a factor of 1.6. We only take into account the $\gamma$-decay widths from the M1 and E2 electromagnetic transitions for the resonance states as their contribution are exponentially higher than the M3 and E4 transitions. 

\renewcommand{\arraystretch}{0.8}
\begin{table}
\footnotesize
\caption{\label{tab:rates} Thermonuclear reaction rates of $^{65\!\!}$As(p,$\gamma$)$^{66\!}$Se.}
\begin{tabular*}{\linewidth}{@{\hspace{2mm}\extracolsep{\fill}}clll@{\hspace{2mm}}}
\hline
\hline
$T_{9}$ & centroid & lower limit & upper limit \\
&[cm$^{3}$s$^{-1}$mol$^{-1}$]&[cm$^{3}$s$^{-1}$mol$^{-1}$]&[cm$^{3}$s$^{-1}$mol$^{-1}$]\\
\hline
0.1  &   $7.56\times10^{-24}$ &   $3.53\times10^{-25}$ & $2.09\times10^{-23}$ \\     
0.2  &   $5.34\times10^{-15}$ &   $9.90\times10^{-17}$ & $7.12\times10^{-15}$ \\     
0.3  &   $1.13\times10^{-11}$ &   $1.59\times10^{-12}$ & $6.21\times10^{-11}$ \\     
0.4  &   $3.61\times10^{-09}$ &   $1.96\times10^{-10}$ & $2.04\times10^{-08}$ \\     
0.5  &   $3.04\times10^{-07}$ &   $7.03\times10^{-09}$ & $8.20\times10^{-07}$ \\     
0.6  &   $6.33\times10^{-06}$ &   $1.56\times10^{-07}$ & $1.19\times10^{-05}$ \\     
0.7  &   $5.63\times10^{-05}$ &   $1.94\times10^{-06}$ & $1.01\times10^{-04}$ \\     
0.8  &   $2.97\times10^{-04}$ &   $1.37\times10^{-05}$ & $5.45\times10^{-04}$ \\     
0.9  &   $1.12\times10^{-03}$ &   $6.50\times10^{-05}$ & $2.12\times10^{-03}$ \\     
1.0  &   $3.38\times10^{-03}$ &   $2.32\times10^{-04}$ & $6.46\times10^{-03}$ \\     
1.1  &   $8.65\times10^{-03}$ &   $6.66\times10^{-04}$ & $1.63\times10^{-02}$ \\     
1.2  &   $1.96\times10^{-02}$ &   $1.63\times10^{-03}$ & $3.55\times10^{-02}$ \\     
1.3  &   $4.01\times10^{-02}$ &   $3.50\times10^{-03}$ & $6.90\times10^{-02}$ \\     
1.4  &   $7.57\times10^{-02}$ &   $6.66\times10^{-03}$ & $1.23\times10^{-01}$ \\     
1.5  &   $1.34\times10^{-01}$ &   $1.17\times10^{-02}$ & $2.02\times10^{-01}$ \\     
1.6  &   $2.22\times10^{-01}$ &   $1.90\times10^{-02}$ & $3.13\times10^{-01}$ \\     
1.7  &   $3.50\times10^{-01}$ &   $2.86\times10^{-02}$ & $4.59\times10^{-01}$ \\     
1.8  &   $5.29\times10^{-01}$ &   $4.05\times10^{-02}$ & $6.46\times10^{-01}$ \\     
1.9  &   $7.68\times10^{-01}$ &   $5.53\times10^{-02}$ & $8.75\times10^{-01}$ \\     
2.0  &   $1.08              $ &   $7.33\times10^{-02}$ & $1.15              $ \\  
\hline
\end{tabular*}
\end{table}
\renewcommand{\arraystretch}{1.0}

\begin{table*}[t]
\footnotesize
\caption{\label{tab:parameter} Parameters$^a$ of $^{65\!\!}$As(p,$\gamma$)$^{66\!}$Se centroid reaction rate.}
\begin{tabular*}{\linewidth}{@{\hspace{2mm}\extracolsep{\fill}}cccccccc@{\hspace{2mm}}}
\hline
\hline
$i$ & $a_0$ & $a_1$ & $a_2$ & $a_3$ & $a_4$ & $a_5$ & $a_6$ \\
\hline
%
%
%
%
1 & $-2.98258\times10^{+1}$ & $-2.35942\times10^{+0}$ & $+3.94649\times10^{+0}$ & $-9.86873\times10^{+0}$ & $1.45916\times10^{-1}$ & $+1.01711\times10^{-1}$ & $2.46893\times10^{+0}$ \\
2 & $-1.49044\times10^{+1}$ & $-3.68082\times10^{+0}$ & $+3.93081\times10^{+0}$ & $-9.86417\times10^{+0}$ & $1.55748\times10^{-1}$ & $+1.02469\times10^{-1}$ & $2.45374\times10^{+0}$ \\
3 & $+1.29861\times10^{-1}$ & $-4.94352\times10^{+0}$ & $+7.67143\times10^{+0}$ & $-2.00710\times10^{+1}$ & $1.29501\times10^{+0}$ & $-2.73522\times10^{-2}$ & $6.16001\times10^{+0}$ \\
4 & $+2.60957\times10^{+1}$ & $-1.04676\times10^{+1}$ & $+1.64542\times10^{+1}$ & $-4.16331\times10^{+1}$ & $3.26017\times10^{+0}$ & $-3.38444\times10^{-1}$ & $1.34811\times10^{+1}$ \\
5 & $+3.41675\times10^{+1}$ & $-1.41042\times10^{+1}$ & $+2.12287\times10^{+1}$ & $-5.17133\times10^{+1}$ & $4.54240\times10^{+0}$ & $-3.08117\times10^{-1}$ & $1.87553\times10^{+1}$ \\
\hline
\end{tabular*}
\begin{minipage}{\linewidth}
\footnotesize
\vskip5pt
{\sc Note}---\\
$^a$ The $a_0, \dots, a_6$ parameters are substituted in
$N_\mathrm{A}\langle\sigma v\rangle = \sum_i \mathrm{exp}(a^i_0 + a^i_1/T_9 + a^i_2/T_9^{1/3} + a^i_3 T_9^{1/3} + a^i_4 T_9 + a^i_5 T_9^{5/3} + a^i_6 \ln{T_9})$ to reproduce the forward reaction rate with an accuracy quantity, $\zeta=\frac{1}{n}\sum_{m=1}^n ( \frac{r_{m} - f_{m}}{f_{m}} )^2$, where $n$ is the number of data points, $r_{m}$ are the original \emph{Present} rate calculated for each respective temperature, and $f_{m}$ are the fitted rate at that temperature \citep{Rauscher2000}, see text. \\
\end{minipage}
\end{table*}
%
%
%
%
%

\begin{figure}[t]
\begin{center}
\includegraphics[width=8.7cm, angle=0]{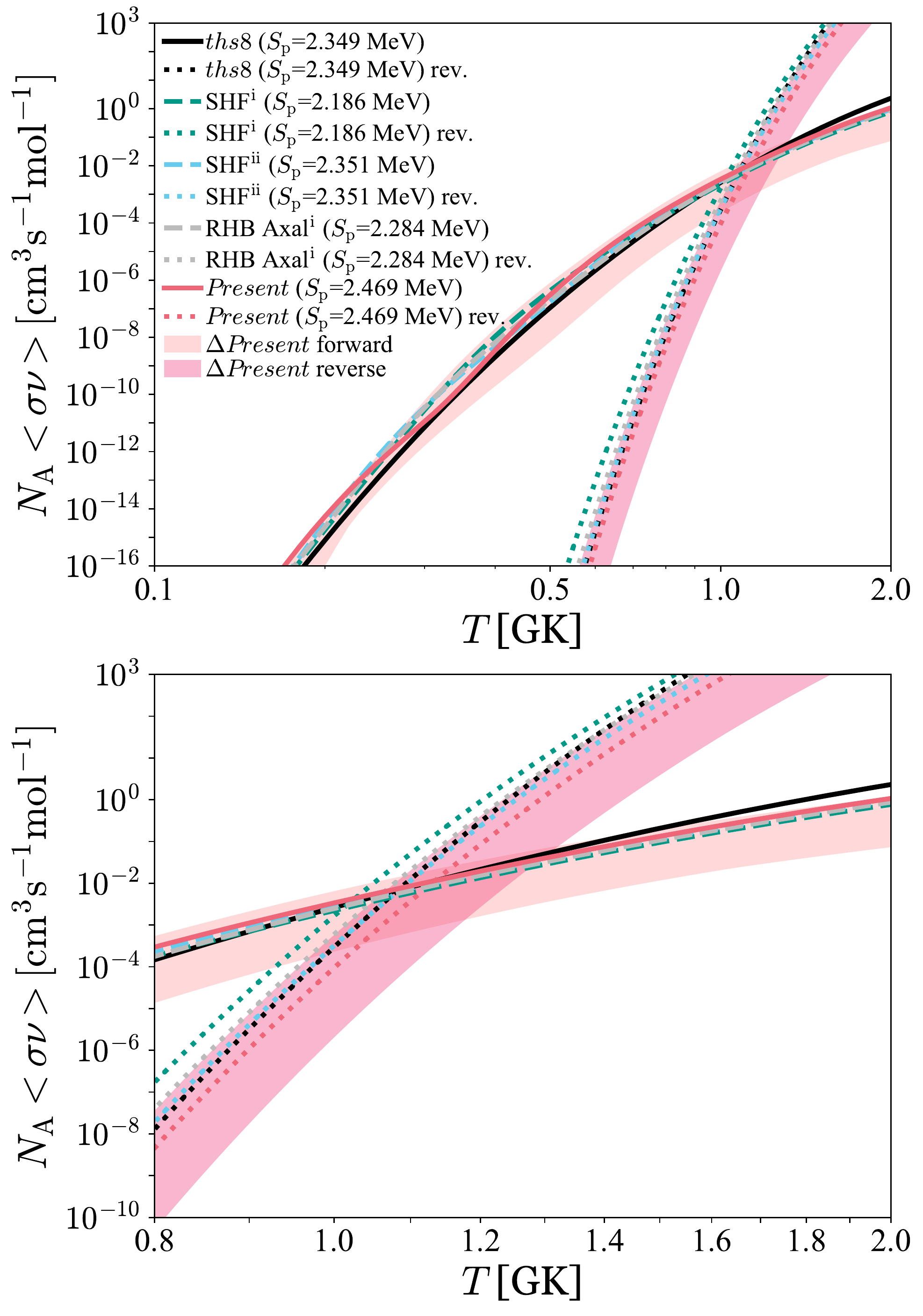}
\caption{\label{fig:rp_65As_66Se_rev}{\footnotesize The $^{65\!\!}$As(p,$\gamma$)$^{66\!}$Se forward and reverse thermonuclear reaction rates in the temperature region of XRB interest. \textsl{Top Panel:} the \emph{ths}8 forward and reverse rates are the recommended rates in the JINA REACLIB~v2.2 release \citep{Cyburt2010}. The uncertainties of the \emph{Present} ($S_\mathrm{p}=2.469\mathord\pm0.100$~MeV) forward and reverse rates are indicated as light red and light pink zones, respectively. \textsl{Bottom Panel:} the magnified portion in \textsl{Top Panel} for $T=0.8$~--~2~GK.}}
\vspace{3mm}
\end{center}
\end{figure} 

The dominant resonances of proton captures on the ground and first excited states of $^{65\!\!}$As are plotted in Fig.~\ref{fig:65As_66Se_contri}, whereas Fig.~\ref{fig:rp_65As_66Se} displays the comparison of the present (\emph{Present}, hereafter) reaction rate with other available rates compiled by \citet{Cyburt2010} for JINA REACLIB v2.2, i.e., \emph{rath}, \emph{rpsm}, \emph{thra}, \emph{laur}, \emph{ths}8, the previous rate \citep{Lam2016}. The $^{65\!\!}$As(p,$\gamma$)$^{66\!}$Se forward rates generated from the $S_\mathrm{p}$($^{66\!}$Se)~$=2.351\mathord\pm0.144$~MeV, $2.284\mathord\pm0.092$~MeV, and $2.186\mathord\pm0.128$~MeV are about a factor of $0.5$~--~$0.8$ below the \emph{Present} forward rate at temperature $T=0.5$~--~$2$~GK, and are within the uncertainty zone of the \emph{Present} rate, see Fig.~\ref{fig:rp_65As_66Se}. The \emph{Present} rate is up to a factor of 7 higher than the \emph{ths}8 (or NON-SMOKER) rate recommended in JINA REACLIB v2.2 release. The \emph{Present} uncertainty region deduced from folding the $S_\mathrm{p}$($^{66\!}$Se) 100-keV uncertainty with the 200-keV uncertainty from shell-model estimated $E^j_\mathrm{x}$ (light red zone in top panel of Fig.~\ref{fig:rp_65As_66Se}) is reduced up to $\mytilde$1.5 order of magnitude compared to the uncertainty (green zone) in Fig.~2 of \citet{Lam2016}. This is due to the presently folded uncertainty from $S_\mathrm{p}$($^{66\!}$Se) and shell-model calculation which is 129~keV lower than the folded uncertainty suggested by \citet{Lam2016} that combines the AME2012 \citep{AME2012} $S_\mathrm{p}$($^{66\!}$Se) and the uncertainties of energy levels from shell-model calculation.


The \emph{Present} reaction rate are presented in Table~\ref{tab:rates} and the parameters listed in Table~\ref{tab:parameter} can be used to reproduce the \emph{Present} centroid reaction rate with $n=191$, a fitting error of $4.5\,\%$, and an accuracy quantity of $\zeta=0.003$ for the temperature range from $0.1$~--~$2$~GK according to the format and evaluation procedure proposed by \citet{Rauscher2000}. The parameterized rate is obtained using the Computational Infrastructure for Nuclear Astrophysics (CINA;~\citealt{CINA}). For the rate above $2$~GK, we refer to statistical-model calculations to match with the \emph{Present} rate, see NACRE~\citep{Angulo1999}.

The new reverse $^{65\!\!}$As(p,$\gamma$)$^{66\!}$Se reaction rate based on the $S_\mathrm{p}$($^{66\!}$Se$)= 2.469$~MeV is related to the respective forward reaction rate, Eq.~(\ref{eq:total}), via the expression \citep{Rauscher2000, Schatz2017}, 
\begin{eqnarray}
\label{eq:rev}
\lambda_{(\gamma,\mathrm{p})} = && \frac{2 G_f}{G_i} \left( \frac{\mu kT}{2 \pi \hbar^2} \right)^{3/2} \mathrm{exp}\left( - \frac{ S_\mathrm{p} }{kT} \right) N_\mathrm{A}\langle \sigma v \rangle \, ,
\end{eqnarray}
where $G_i$ and $G_f$ are the partition functions of initial and final nuclei. The new reverse rates are presented in Fig.~\ref{fig:rp_65As_66Se_rev}, and compared with the \emph{ths}8 statistical-model reverse rate. The respective uncertainty (lower and upper limits) of the \emph{Present} reverse rate corresponds to the uncertainty (lower and upper limits) of the \emph{Present} forward rate with the consideration of $100$-keV uncertainty imposed from the present $S_\mathrm{p}$($^{66\!}$Se). Although this $100$-keV uncertainty is rather extreme, its range is capable to cover possible reverse rates due to other estimated $S_\mathrm{p}$($^{66\!}$Se), i.e., $S_\mathrm{p}$($^{66\!}$Se)~$=2.433\mathord\pm0.144$~MeV, $2.443\mathord\pm0.101$~MeV, and $2.507\mathord\pm0.070$~MeV, see Table~\ref{tab:MDE} and Fig.~\ref{fig:Sp}. The corresponding reverse rates of $S_\mathrm{p}$($^{66\!}$Se)~$=2.351\mathord\pm0.144$~MeV and $2.284\mathord\pm0.092$~MeV (cyan and gray dotted lines in Fig.~\ref{fig:rp_65As_66Se_rev}) are at the upper limit of the \emph{Present} reverse rate, whereas the reverse rate of $S_\mathrm{p}$($^{66\!}$Se)~$=2.186\mathord\pm0.128$~MeV (green dotted lines in Fig.~\ref{fig:rp_65As_66Se_rev}) is about one order of magnitude higher than the \emph{Present} reverse rate.
\newline
\newline



\section{Implication on one-dimensional multi-zone GS~1826$-$24 clocked burst models}
\label{sec:Astro}

We use \textsc{Kepler} code~\citep{Weaver1978,Woosley2004,Heger2007} to construct the theoretical XRB models matched with the periodic XRBs, \emph{Epoch Jun 1998}, of GS~1826$-$24 X-ray source compiled by \citet{Galloway2017}. These XRB models are fully self-consistent. The evolution of chemical inertia and hydrodynamics that power the nucleosynthesis along the rp-process path are correlated with the mutual feedback between the nuclear energy generation in the accreted envelope and the rapidly evolving astrophysical conditions. Meanwhile, \textsc{Kepler} code uses an adaptive reaction network of which the relevant reactions out of the more than 6000 isotopes from JINA REACLIB v2.2~\citep{Cyburt2010} are automatically included or discarded throughout the evolution of thermonuclear runaway in the accreted envelope. 

The XRB models keep tracking the evolution of a grid of Lagrangian zones, of which each zone has its own isotopic composition and thermal properties. We use the time-dependent mixing length theory~\citep{Heger2000} to describe the convection that transfers synthesized and accreted nuclei and heat between these Lagrangian zones. We remark that this important feature is not considered in zero-dimensional one-zone and post-processing XRB models. 

We adopt the astrophysical settings of GS~1826$-$24 model from~\citet{Jacobs2018} to match with the observed burst light curve properties of \emph{Epoch Jun 1998}. To obtain the best-match modeled light curve profile with the observed profile and averaged recurrence time, $\Delta t_\mathrm{rec}=5.14\mathord\pm0.7$~h, we assign the accreted $^1$H, $^4$He, and CNO metallicity fractions to 0.71, 0.2825, and 0.0075, respectively, whereas the accretion rate is adjusted to a factor of 0.114 of the Eddington-limited accretion rate, $\dot{M}_\mathrm{Edd}$. This refined GS~1826$-$24 XRB model with the JINA REACLIB v2.2 defines the \emph{baseline} model in this work. Other models that adopt the same astrophysical settings but implement 
the \emph{Present} $^{65\!\!}$As(p,$\gamma$)$^{66\!}$Se forward and reverse rate (solid and dotted red lines in the top panel of Fig.~\ref{fig:rp_65As_66Se_rev}), 
or the lower limit of \emph{Present} $^{65\!\!}$As(p,$\gamma$)$^{66\!}$Se forward and reverse rates (lower borders of red and pink zones in the top panel of Fig.~\ref{fig:rp_65As_66Se_rev}), 
or the \emph{Present} $^{65\!\!}$As(p,$\gamma$)$^{66\!}$Se forward and reverse rate (solid and dotted red lines in the top panel of Fig.~\ref{fig:rp_65As_66Se_rev}) and $^{22}$Mg($\alpha$,p)$^{25}\!$Al \citep{Hu2021} rates
are labeled as \emph{Present}$^\dag$, \emph{Present}$^\ddag$, and \emph{Present}$^\S$ models, respectively. These \emph{Present}$^{\dag,\ddag,\S}$ models take into account of the recently updated reaction rates around historical $^{56}$Ni and $^{64}$Ge waiting points, namely $^{55}$Ni(p,$\gamma$)$^{56}$Cu \citep{Valverde2019}, $^{57}$Cu(p,$\gamma$)$^{58}$Zn \citep{Lam2021a}, $^{56}$Ni(p,$\gamma$)$^{57}$Cu \citep{Kahl2019}, and $^{64}$Ge(p,$\gamma$)$^{65\!\!}$As \citep{Lam2016} reactions.

We select the latest $^{22}$Mg($\alpha$,p)$^{25}\!$Al reaction rate which was experimentally deduced by \citet{Hu2021} with the important nuclear structure properties corresponding to the XRB Gamow window instead of the $^{22}$Mg($\alpha$,p)$^{25}\!$Al reaction rate deduced by \citet{Randhawa2020} via direct measurement. It is because \citet{Randhawa2020} rate was extrapolated from non-XRB energy region and may have an additional and large uncertainty that was not shown \citep{Hu2021}. The recent direct measurement performed by \citet{Jayatissa2021} could be helpful to further constrain the $^{22}$Mg($\alpha$,p)$^{25}\!$Al reaction rate as long as the measurement directly corresponds to the XRB Gamow window.

We follow the simulation procedure implemented by \citet{Lam2021a} and \citet{Hu2021}, of which we run a series of 40 consecutive XRBs for \emph{baseline}, \emph{Present}$^\S$, \emph{Present}$^\ddag$, \emph{Present}$^\dag$ models. The first 10 bursts of each model are discarded because these bursts transit from a chemically fresh envelope to an enriched envelope with burned-in burst ashes and stable burning. The burst ashes are recycled in the following burst heatings and stabilize the succeeding bursts. Therefore, we only sum up the last 30 bursts and then average them to produce a modeled burst light-curve profile. This averaging procedure was applied by \citet{Galloway2017} to yield an averaged-observed light-curve profile of \emph{Epoch Jun 1998}, which was recorded by the \emph{Rossi X-ray Timing Explorer} (RXTE) Proportional Counter Array~\citep{Galloway2004,Galloway2008,MINBAR} and was compiled into the Multi-Instrument Burst Archive\footnote{\href{https://burst.sci.monash.edu/minbar/}{https://burst.sci.monash.edu/minbar/}} by \citet{MINBAR}. Our averaging procedure was also implemented in the works of \citet{Lam2021a} and \citet{Hu2021}.

\begin{figure}[t]
\begin{center}
\includegraphics[width=8.7cm,angle=0]{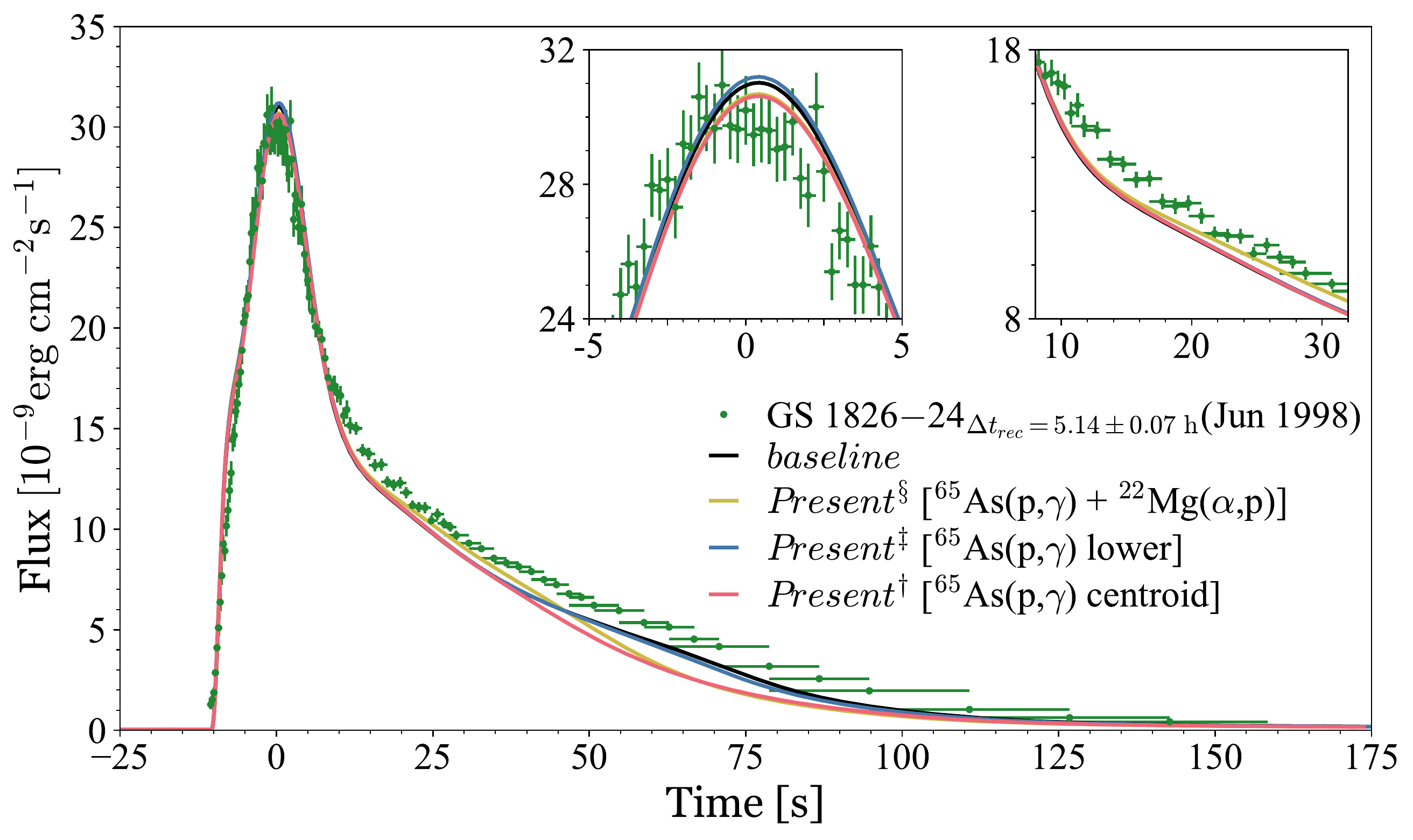}
\caption{\label{fig:Flux_GS1826}{\footnotesize The averaged light curves of GS~1826$-$24 clocked burster as a function of time. \textsl{Top Panel:} the best-fit \emph{baseline}, \emph{Present}$^\dag$, \emph{Present}$^\ddag$, and \emph{Present}$^\S$ modeled light curves to the observed light curve and recurrence time of \emph{Epoch Jun 1998}. Both insets in the \textsl{Top Panel} magnify the light curve portions at $t=-5$ to $5$~s (left inset) and $t=8$ to $32$~s (right inset).}}
\vspace{-3mm}
\end{center}
\end{figure}

\subsection{Clocked bursts of GS~1826$-$24 X-ray source}
\label{sec:XRB}

The observed flux, $F_\mathrm{x}$, is expressed as \citep{Johnston2020}
$F_\mathrm{x} = L_\mathrm{x}/\left\{{4\pi d^2\xi_\mathrm{b}(1+z)^2}\right\}$, 
where $L_\mathrm{x}$ is the burst luminosity generated from each model; $d$ is the distance; $\xi_\mathrm{b}$ takes into account of the possible deviation of the observed flux from an isotropic burster luminosity \citep{Fujimoto1988,He2016}; and the redshift, $z$, adjusts the light curve when transforming into an observer's frame. Assuming that the anisotropy factors of burst and persistent emissions are degenerate with distance, the $d$ and $\xi_\mathrm{b}$ can be combined to form the modified distance $d\sqrt{\xi_\mathrm{b}}$. Instead of specifically selecting data close to the burst peak at $t=-10$~s to $40$~s as done by \citet{Meisel2018a} and \citet{Randhawa2020}, we impartially fit the modeled burst light curves generated from each model to the entire burst timespan of the averaged-observed light curve to avoid artifactually expanding the modeled burst light curve and shift the modeled burst peak, imposing unknown uncertainty. The best-fit $d\sqrt{\xi_\mathrm{b}}$ and $(1+z)$ factors of the \emph{baseline}, \emph{Present}$^\S$, \emph{Present}$^\ddag$, and \emph{Present}$^\dag$ modeled light curves to the averaged-observed light curve and recurrence time of \emph{Epoch Jun 1998} are 
$7.38$~kpc and $1.27$, 
$7.46$~kpc and $1.26$, 
$7.34$~kpc and $1.28$, 
$7.50$~kpc and $1.27$, 
respectively. 
The averaged-modeled recurrence times of \emph{baseline}, \emph{Present}$^\S$, \emph{Present}$^\ddag$, and \emph{Present}$^\dag$ are 
$5.11$~h, 
$4.97$~h, 
$5.13$~h, and 
$5.03$~h, 
respectively. The modeled recurrence times of the \emph{baseline} and \emph{Present}$^\ddag$ scenarios are in good agreement with the observed recurrence time, whereas the modeled recurrence times of the \emph{Present}$^\S$ and \emph{Present}$^\dag$ scenarios are lower than the observed recurrence time by $0.17$~h and $0.11$~h, respectively, suggesting a $1$~--~$2\,\%$ decrement can be applied for the accretion rate of the \emph{Present}$^\S$ and \emph{Present}$^\dag$ models. Such decrement also indicates the new reaction rates used in the \emph{Present}$^\S$ and \emph{Present}$^\dag$ models shorten up to $3\,\%$ of the recurrence time.
\begin{figure*}
\gridline{\fig{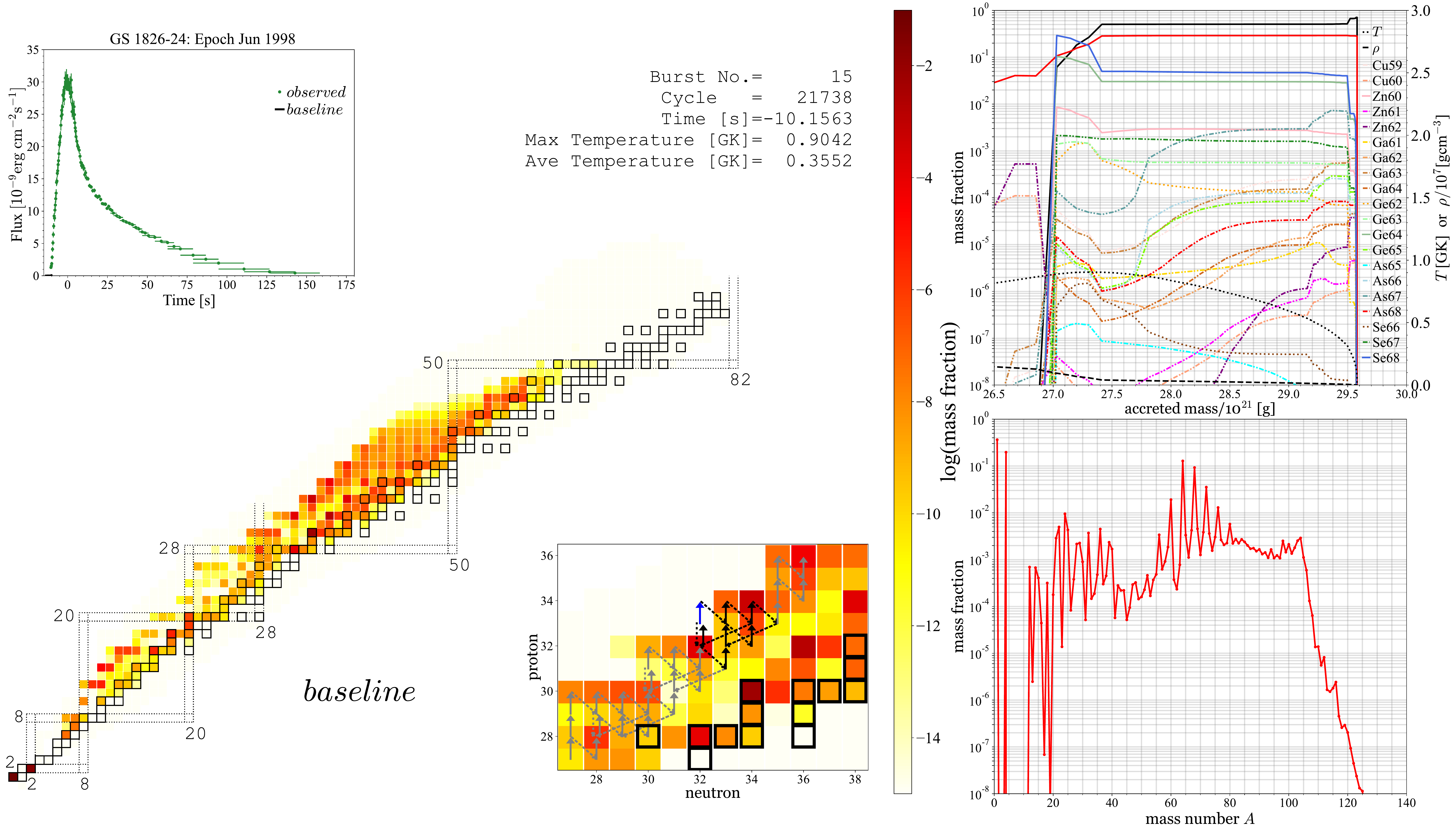}{18cm}{}}
\vspace{-10mm}
\gridline{\fig{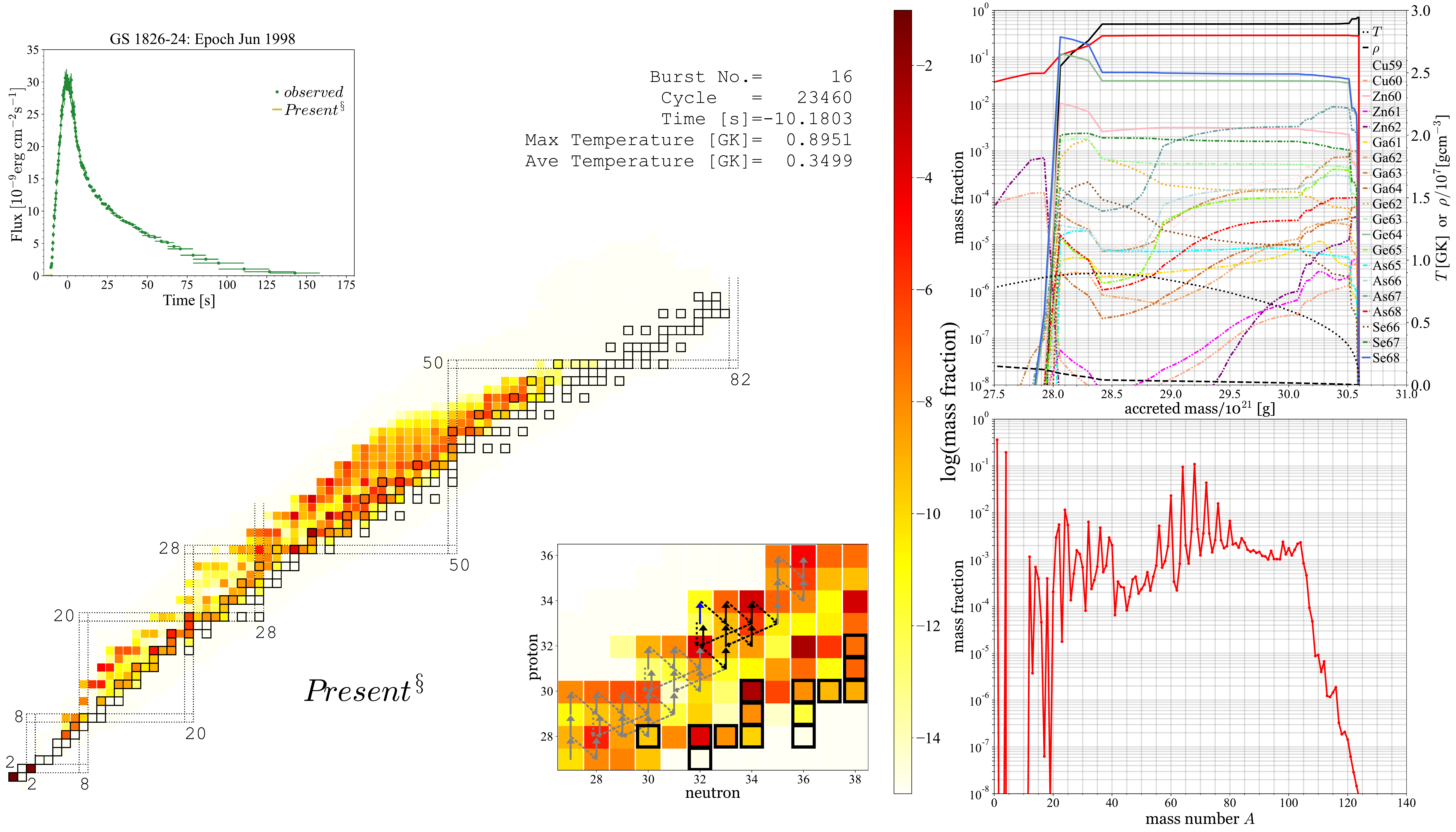}{18cm}{}}
\vspace{-10mm}
\caption{\label{fig:Onset1}{\footnotesize The nucleosynthesis and evolution of envelope corresponds to the moment just before the onset of the 15$^\mathrm{th}$ burst for \emph{baseline} (\textsl{Top Panel}) and of the 16$^\mathrm{th}$ burst for \emph{Present}$^\S$ (\textsl{Bottom Panel}) scenarios. The cycle is the iteration step of the XRB simulation and the time is relative to the burst peak $t = 0\,$s. The averaged abundances of synthesized nuclei are represented by color tones referring to the right color scale in the nuclear chart of each panel. The black squares are stable nuclei. \textsl{Bottom Right Insets} of nuclear chart in each panel: the magnified regions of the NiCu, ZnGa, and GeAs cycles. The arrows are merely used to guide the eyes. \textsl{Top Left Insets} of nuclear chart in each panel: the time snapshot of modeled burst light curve. \textsl{Top Right Insets} in each panel: the corresponding temperature (black dotted line) and density (black dashed line) of each mass zone, referring to the right $y$-axis, and the abundances of synthesized nuclei, referring to the left $y$-axis. The abundances of H and He are represented by black and red solid lines, respectively. \textsl{Bottom Right Insets} in each panel: the averaged mass fractions for each nuclear mass, $A$. The comparisons of averaged mass fractions between \emph{baseline} and \emph{Present}$^{\ddag,\S,\dag}$ scenarios are presented in Fig.~\ref{fig:abundances}.}
}
\end{figure*}

\begin{figure*}
\gridline{\fig{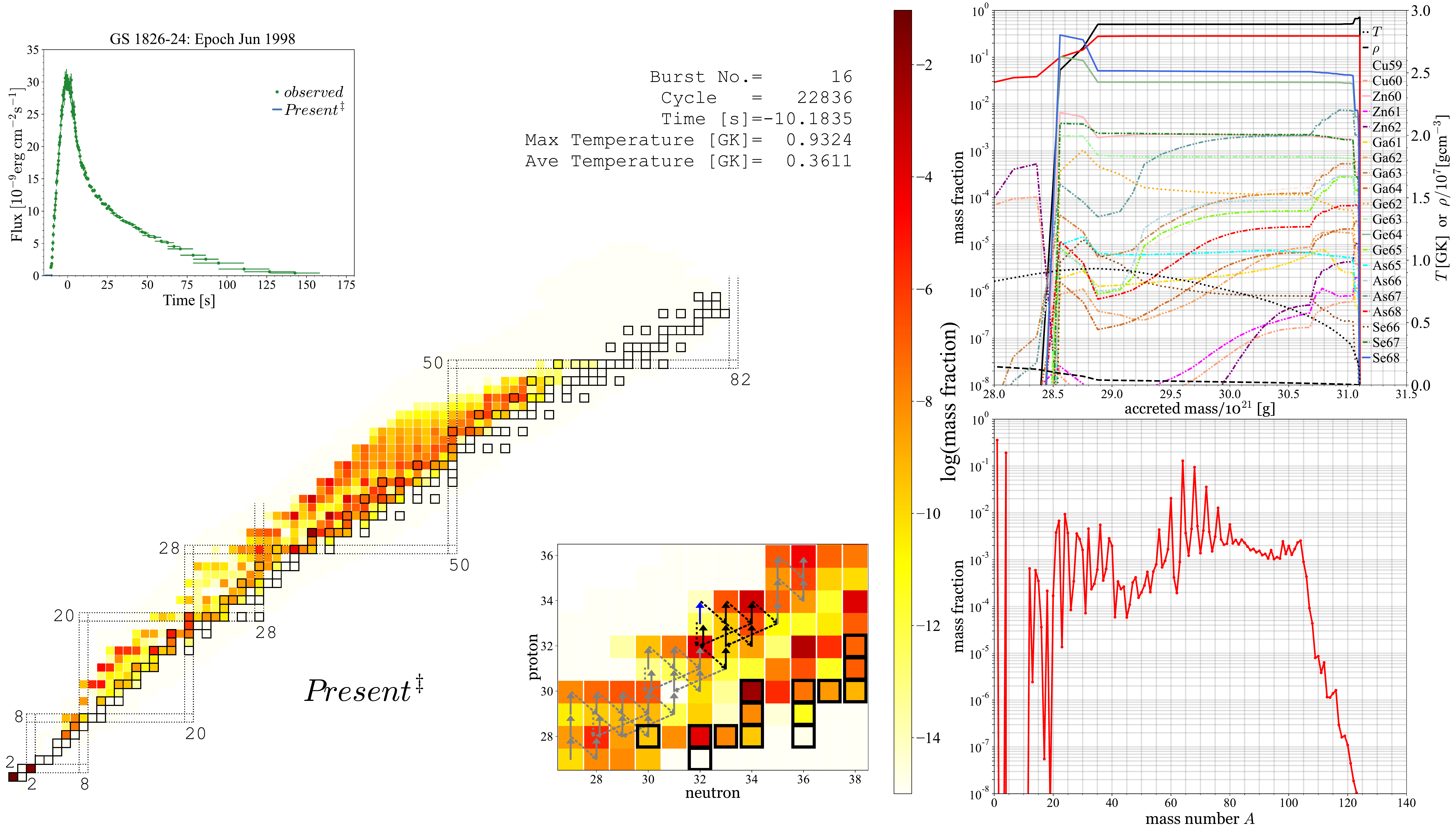}{18cm}{}}
\vspace{-10mm}
\gridline{\fig{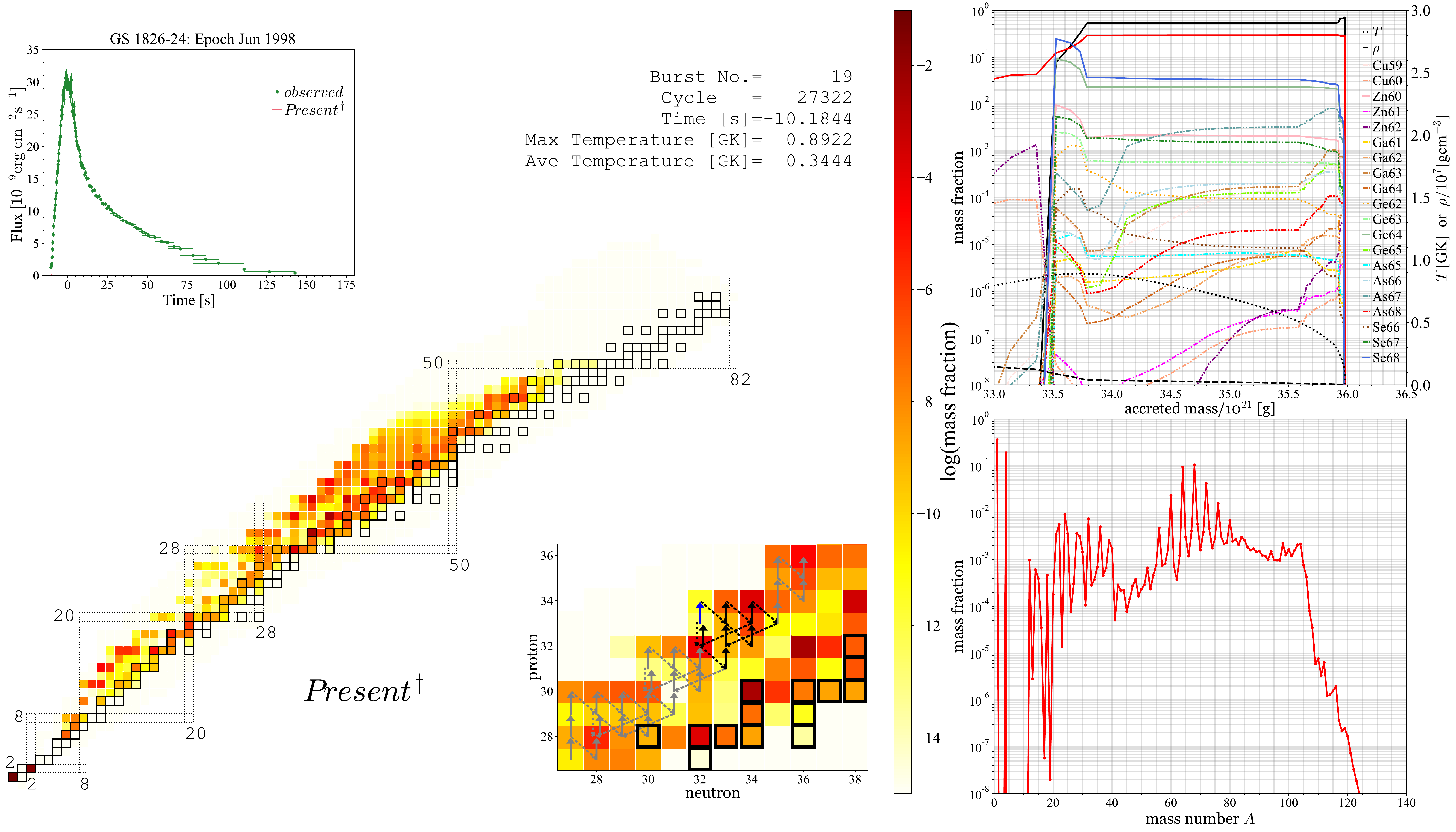}{18cm}{}}
\vspace{-10mm}
\caption{\label{fig:Onset2}{\footnotesize The nucleosynthesis and evolution of envelope corresponds to the moment just before the onset of the 16$^\mathrm{th}$ burst for \emph{Present}$^\ddag$ (\textsl{Top Panel}) and of the 19$^\mathrm{th}$ burst for \emph{Present}$^\dag$ (\textsl{Bottom Panel}) scenarios. See Fig.~\ref{fig:Onset1} for further description.}}
\end{figure*}

\begin{figure*}
\gridline{\fig{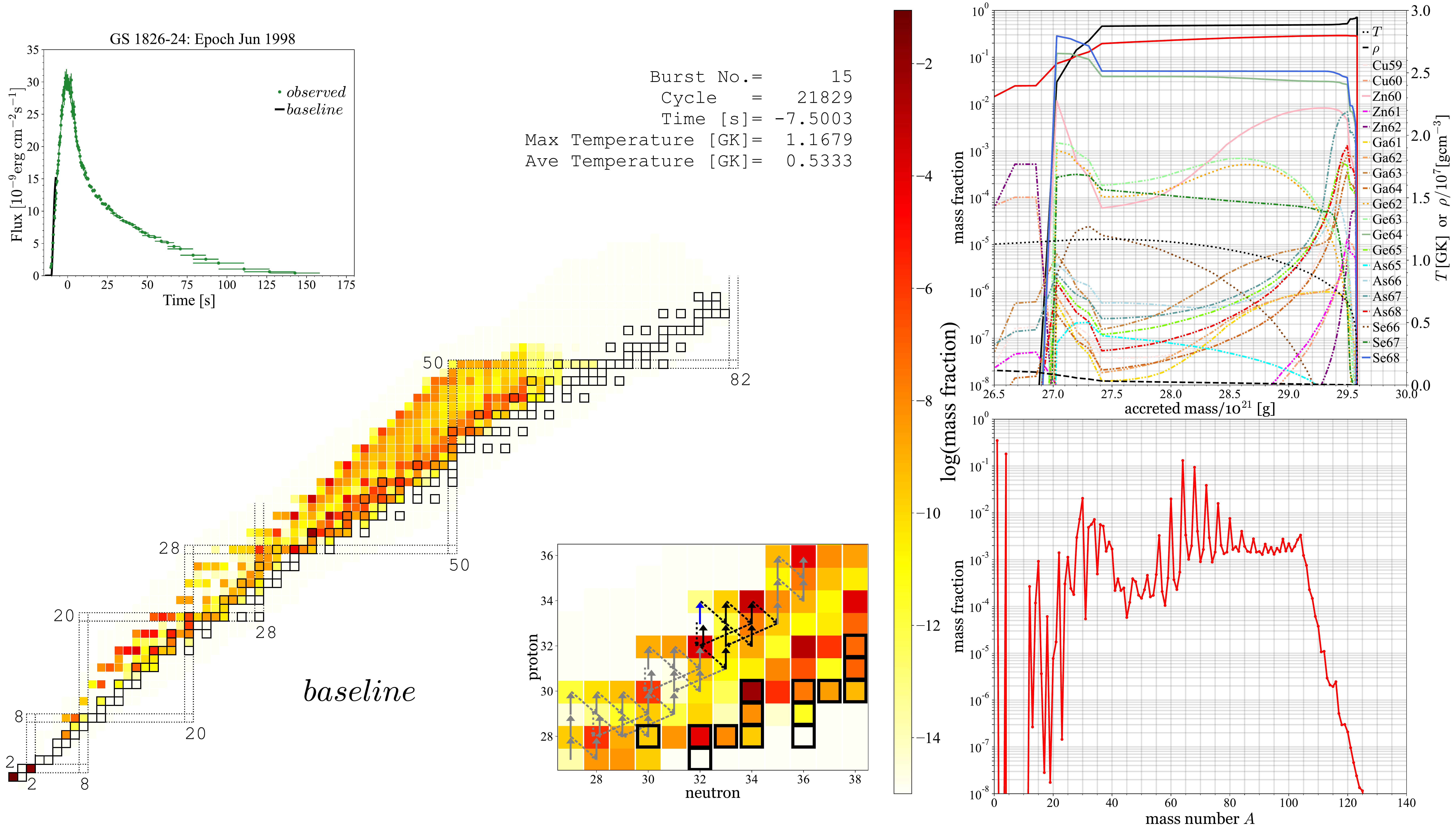}{18cm}{}}
\vspace{-10mm}
\gridline{\fig{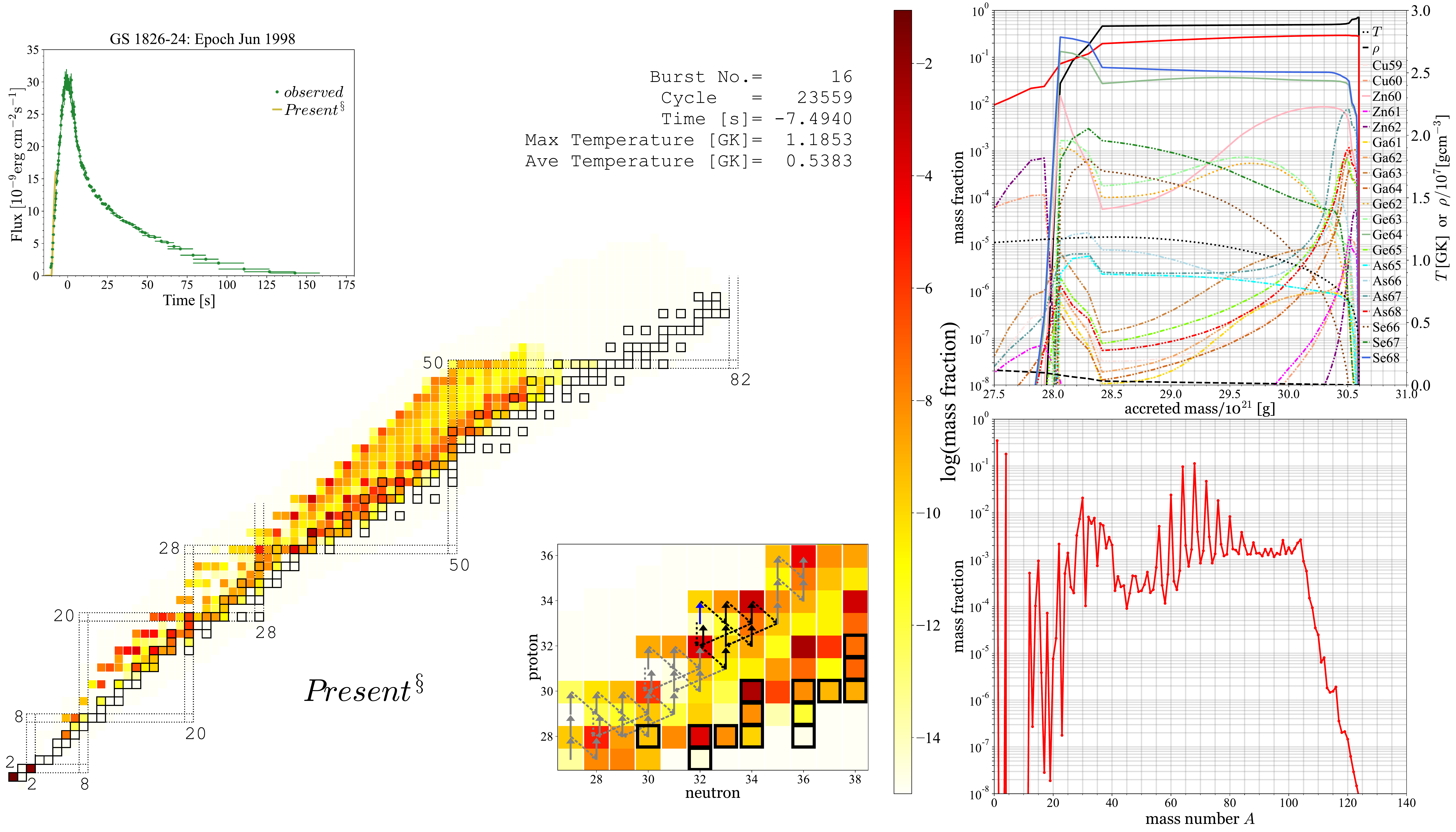}{18cm}{}}
\vspace{-10mm}
\caption{\label{fig:PrePeak1}{\footnotesize The nucleosynthesis and evolution of envelope corresponds to the moment $t=-7.5$~s for \emph{baseline} (\textsl{Top Panel}) and \emph{Present}$^\S$ (\textsl{Bottom Panel}) scenarios. See Fig.~\ref{fig:Onset1} for further description.}}
\end{figure*}

\begin{figure*}
\gridline{\fig{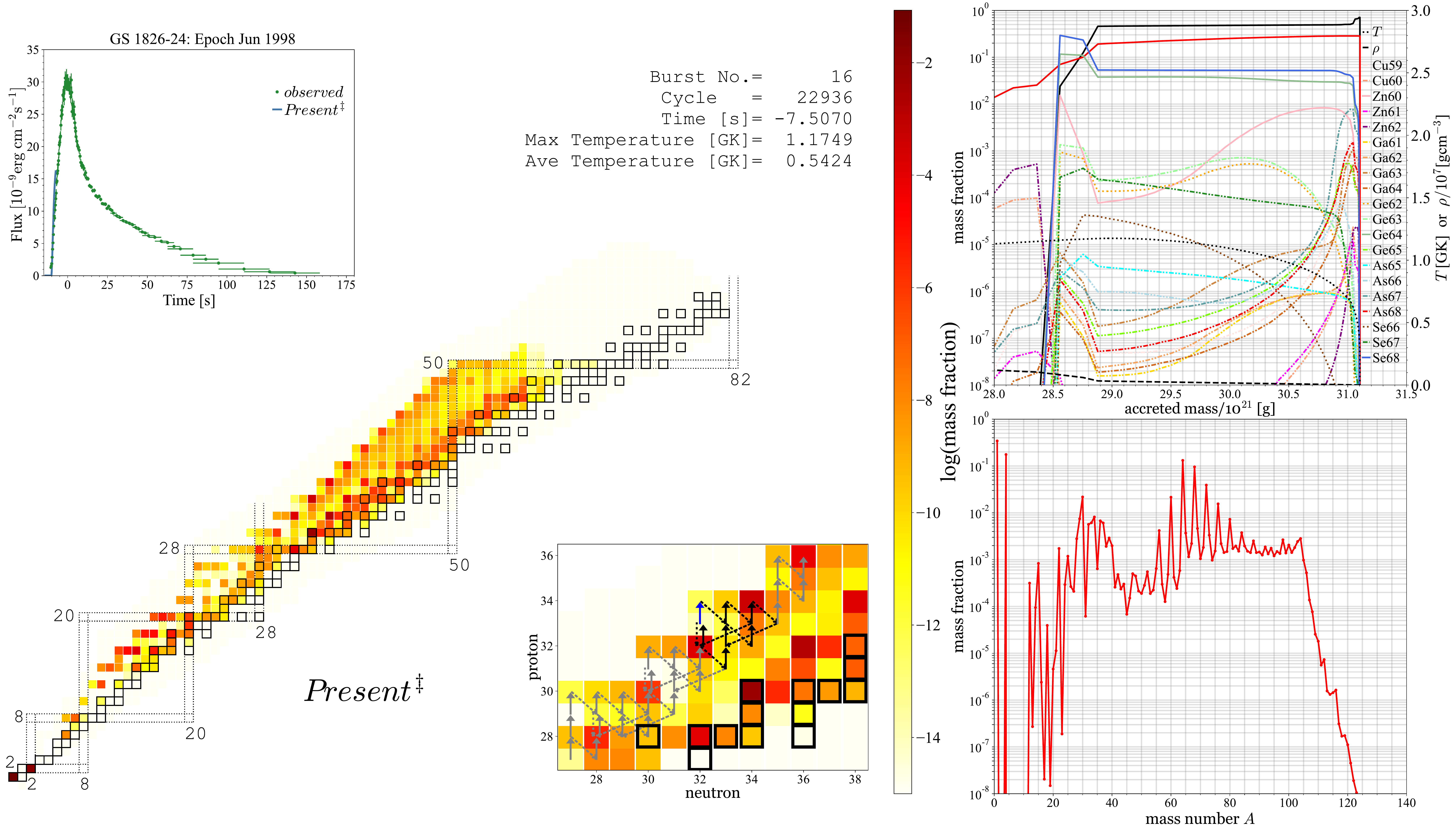}{18cm}{}}
\vspace{-10mm}
\gridline{\fig{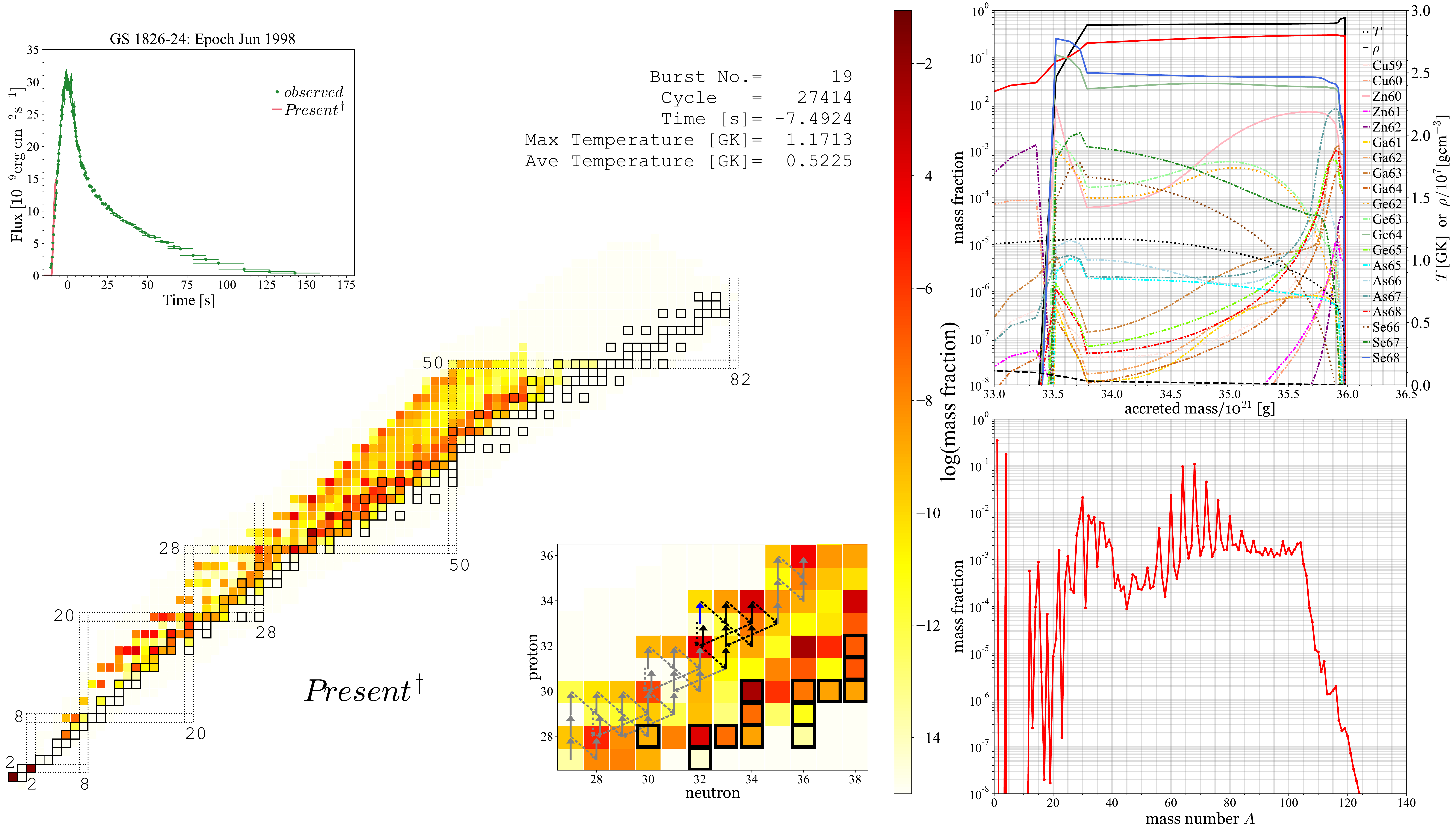}{18cm}{}}
\vspace{-10mm}
\caption{\label{fig:PrePeak2}{\footnotesize The nucleosynthesis and evolution of envelope corresponds to the moment $t=-7.5$~s for \emph{Present}$^\ddag$ (\textsl{Top Panel}) and \emph{Present}$^\dag$ (\textsl{Bottom Panel}) scenarios. See Fig.~\ref{fig:Onset1} for further description.}}
\end{figure*}

\begin{figure*}
\gridline{\fig{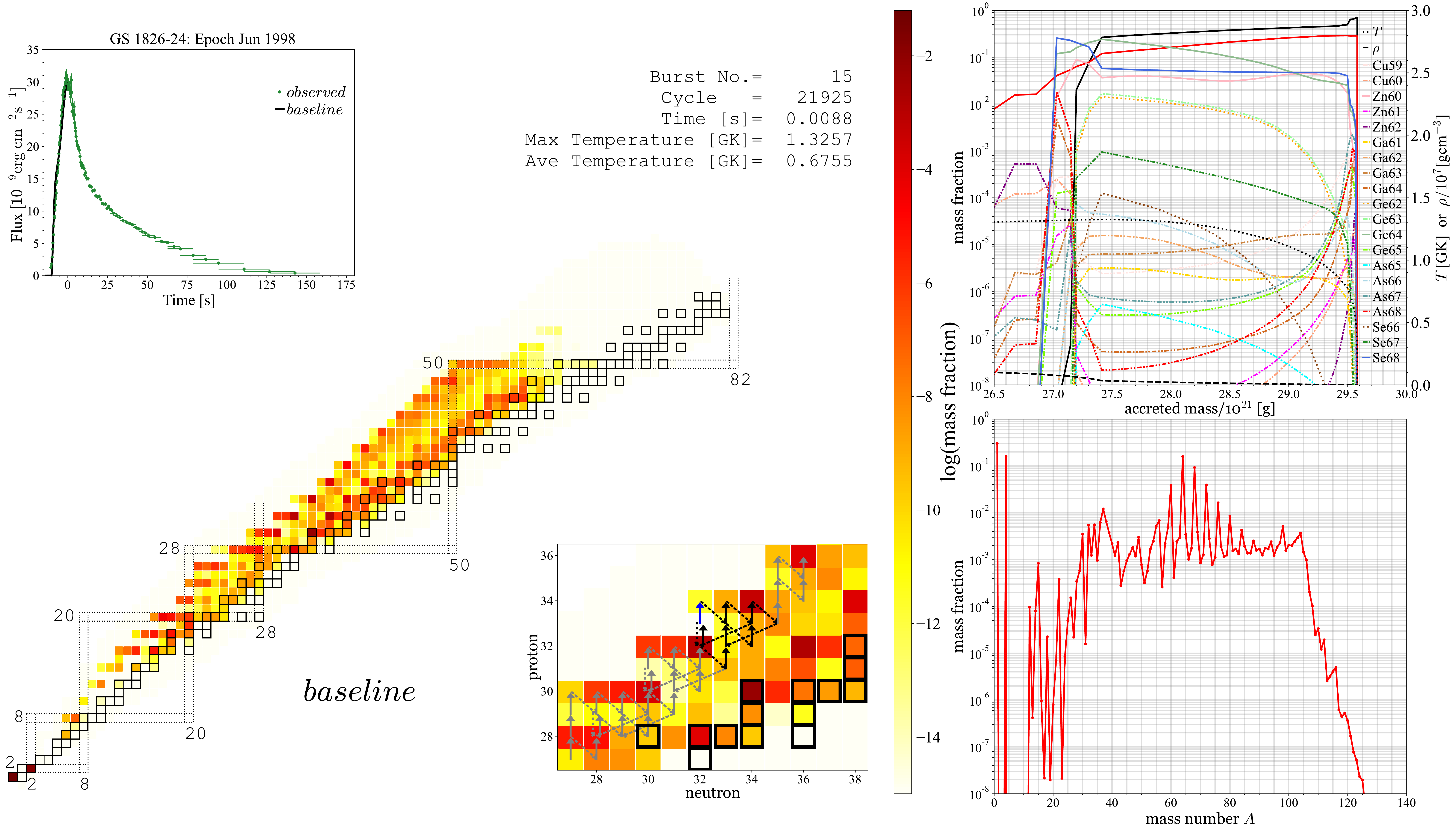}{18cm}{}}
\vspace{-10mm}
\gridline{\fig{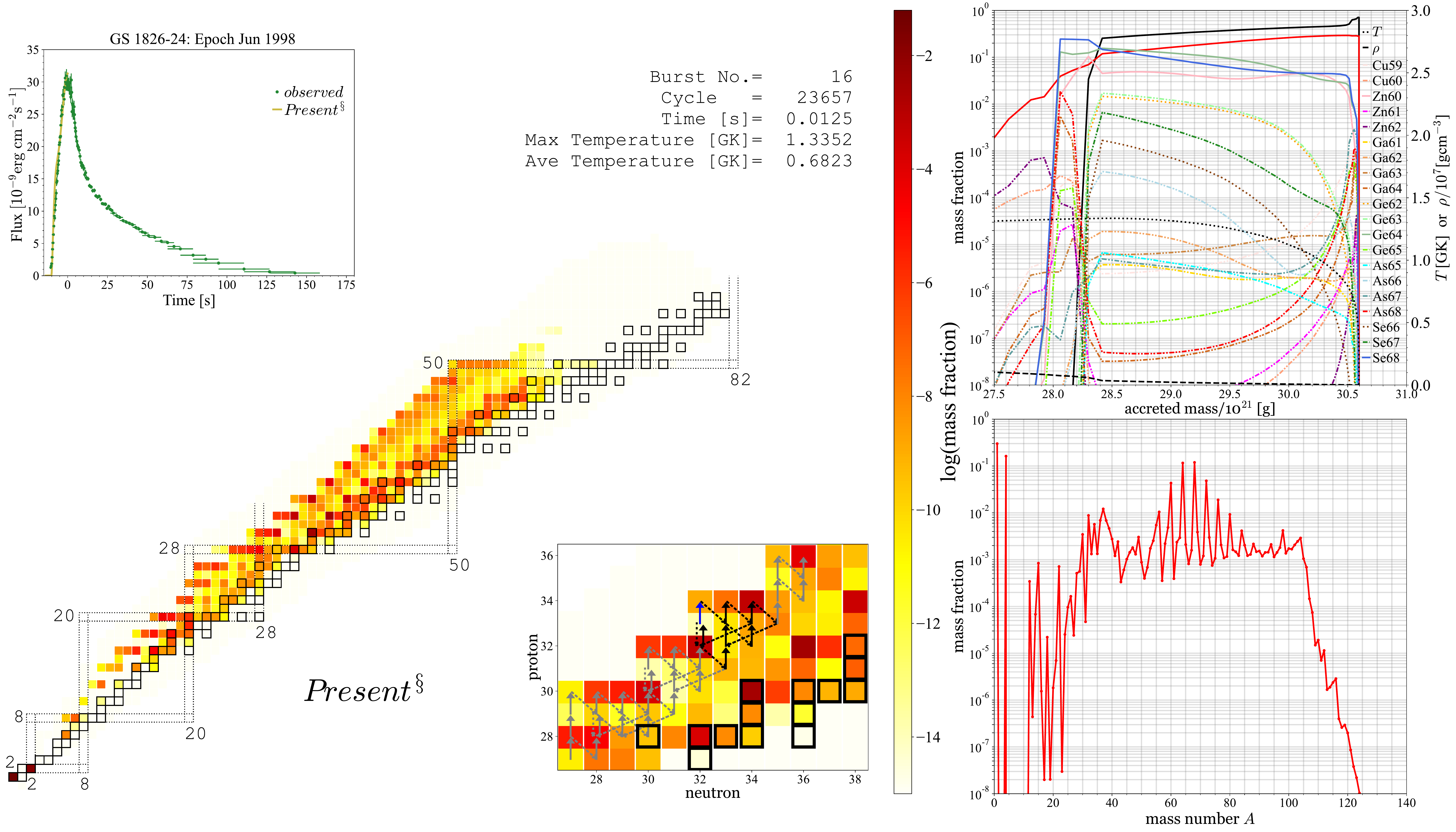}{18cm}{}}
\vspace{-10mm}
\caption{\label{fig:Peak1}{\footnotesize The nucleosynthesis and evolution of envelope corresponds to the moment at the immediate vicinity of the burst peak for \emph{baseline} (\textsl{Top Panel}) and \emph{Present}$^\S$ (\textsl{Bottom Panel}) scenarios. See Fig.~\ref{fig:Onset1} for further description.}}
\end{figure*}

\begin{figure*}
\gridline{\fig{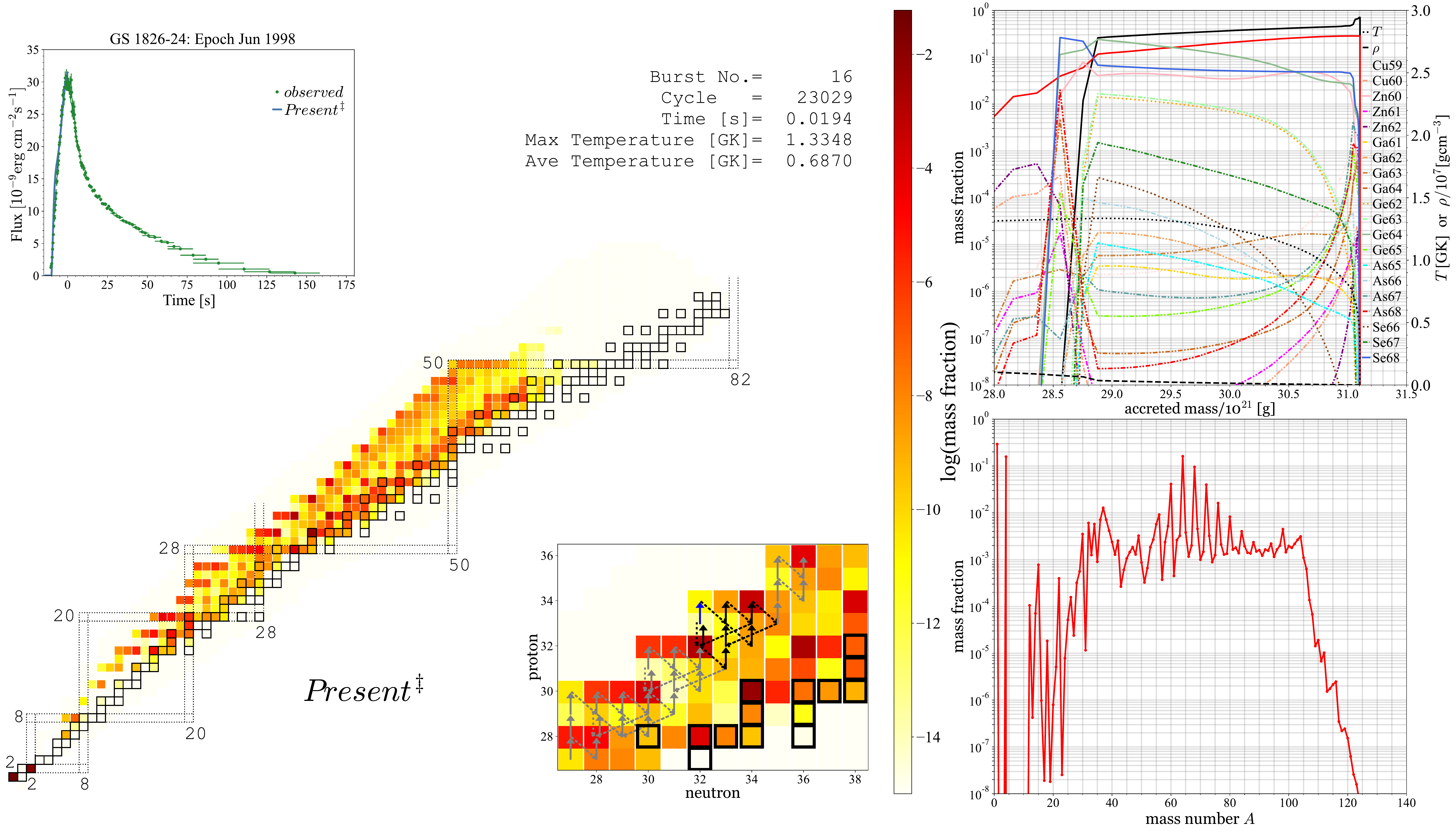}{18cm}{}}
\vspace{-10mm}
\gridline{\fig{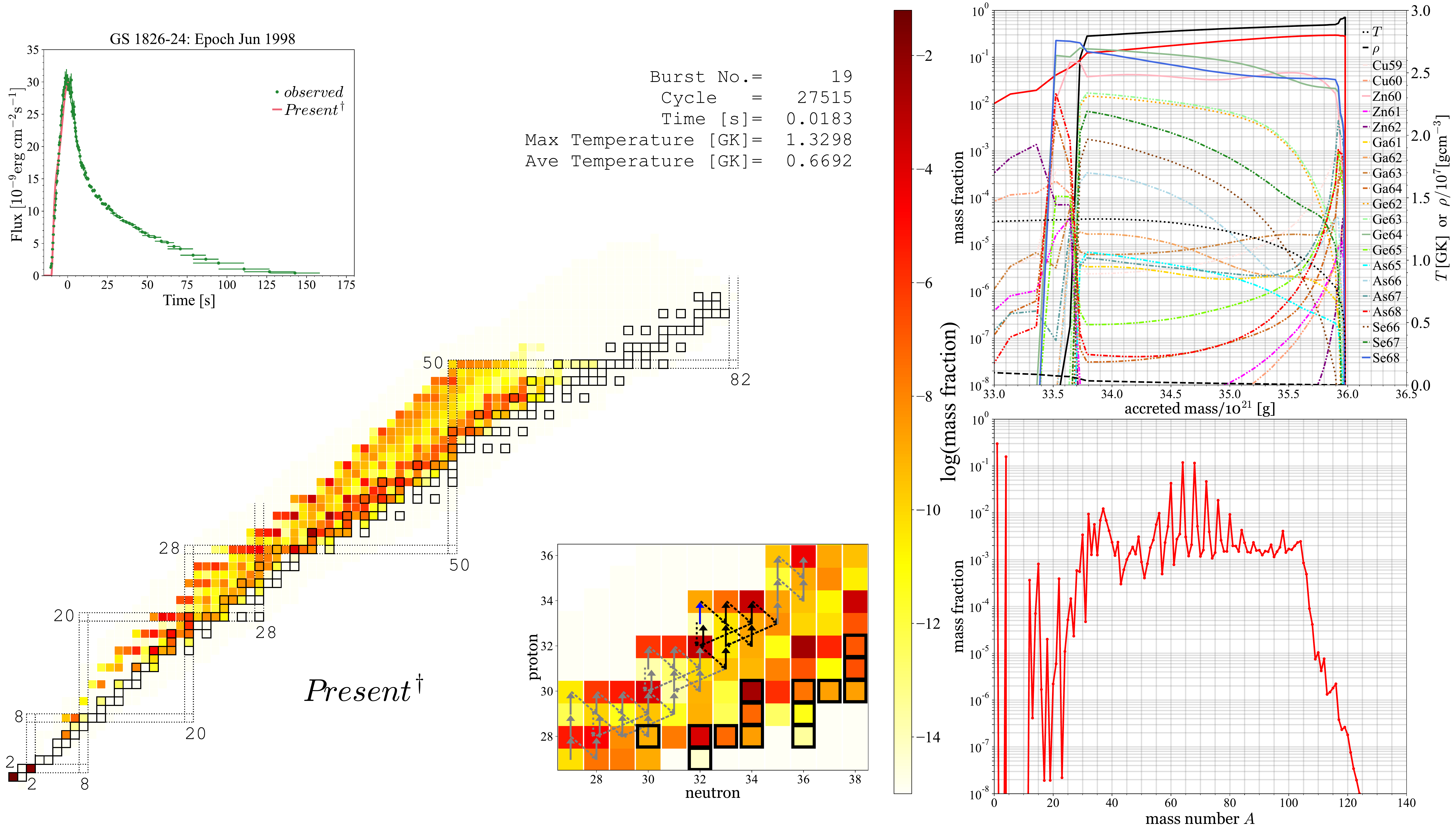}{18cm}{}}
\vspace{-10mm}
\caption{\label{fig:Peak2}{\footnotesize The nucleosynthesis and evolution of envelope corresponds to the moment at the immediate vicinity of the burst peak for \emph{Present}$^\ddag$ (\textsl{Top Panel}) and \emph{Present}$^\dag$ (\textsl{Bottom Panel}) scenarios. See Fig.~\ref{fig:Onset1} for further description.}}
\end{figure*}

\begin{figure*}
\gridline{\fig{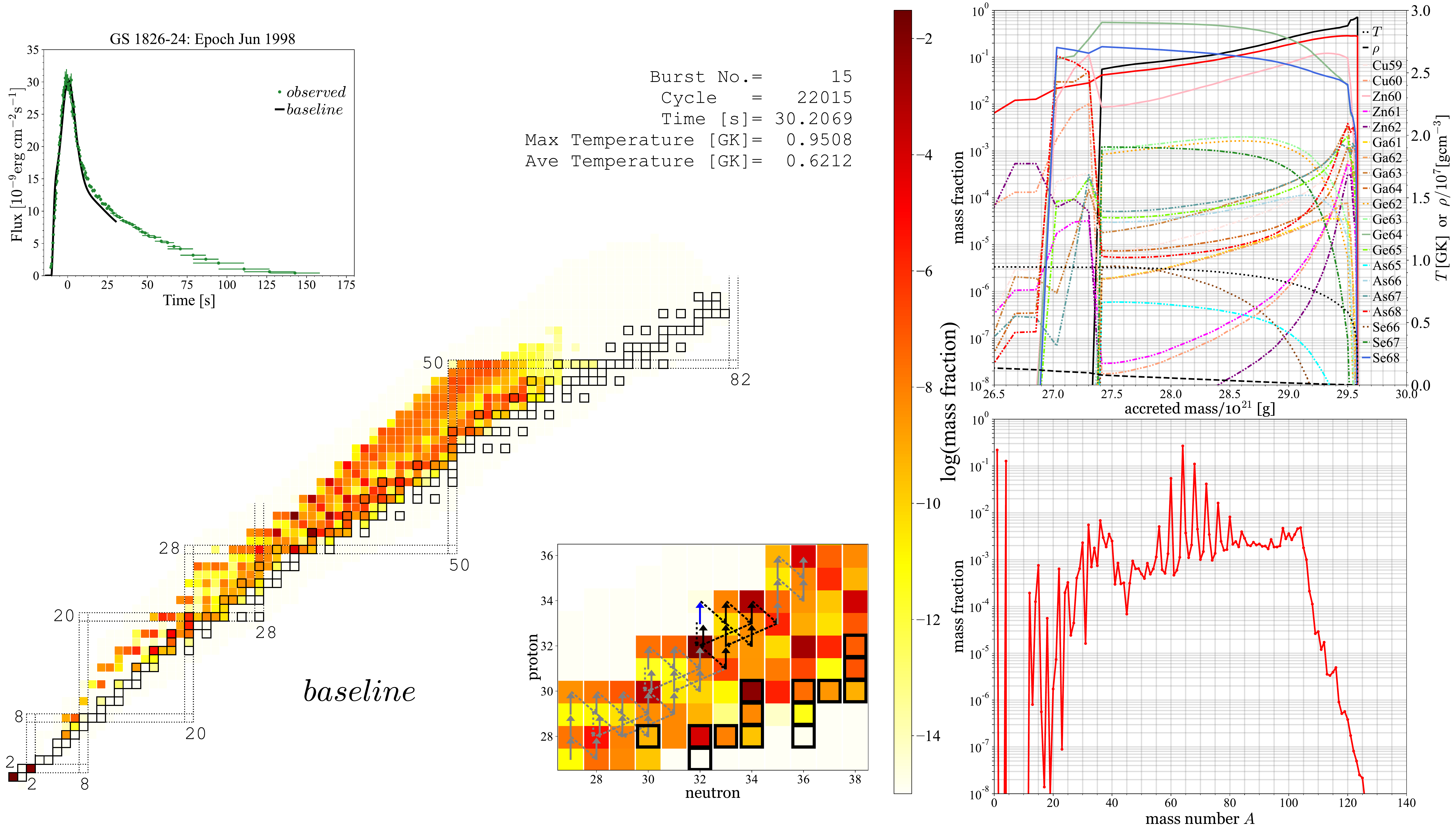}{18cm}{}}
\vspace{-10mm}
\gridline{\fig{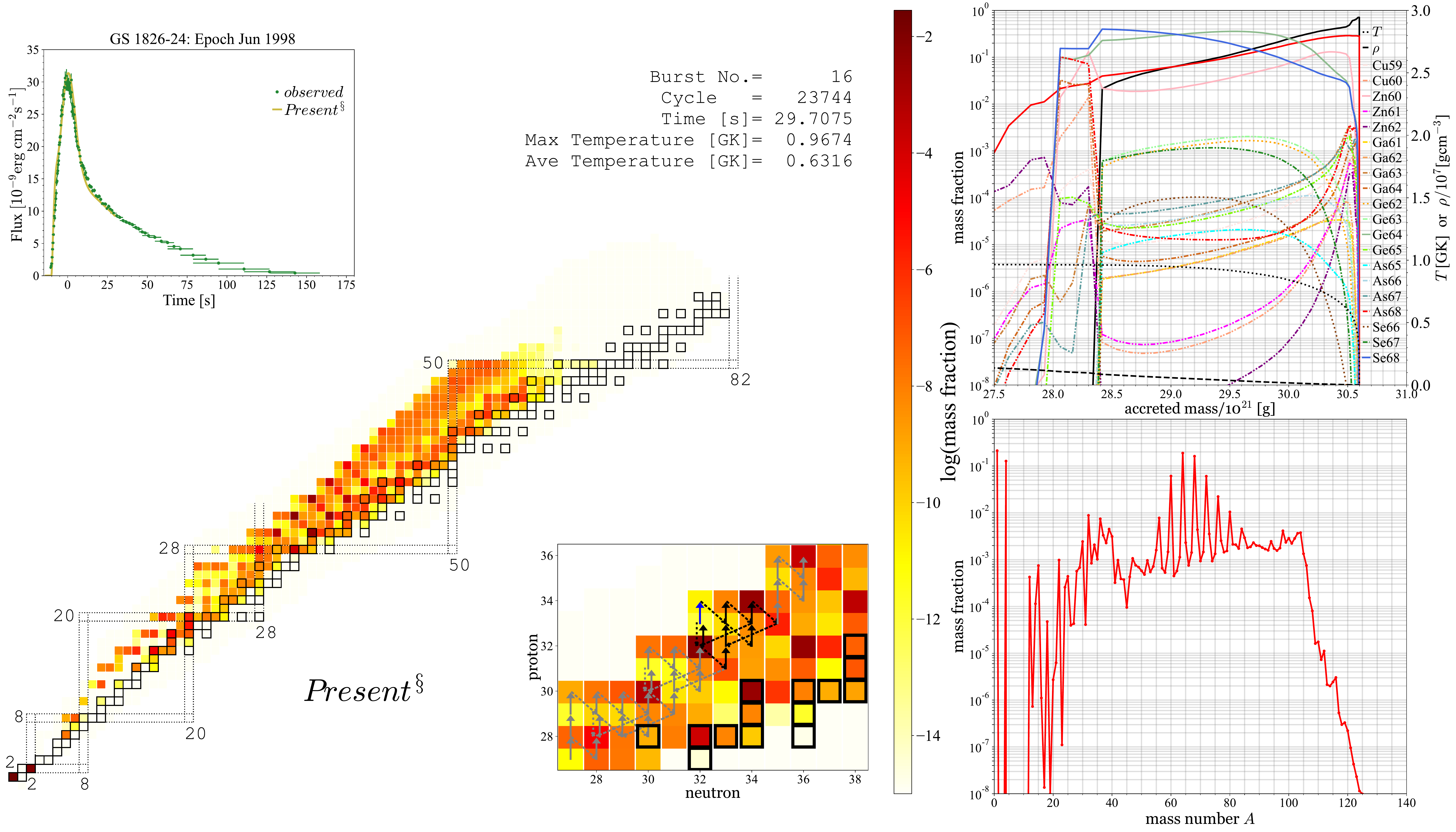}{18cm}{}}
\vspace{-10mm}
\caption{\label{fig:PostPeak1a}{\footnotesize The nucleosynthesis and evolution of envelope corresponds to the moment at around 30~s after the burst peak for \emph{baseline} (\textsl{Top Panel}) and \emph{Present}$^\S$ (\textsl{Bottom Panel}) scenarios. See Fig.~\ref{fig:Onset1} for further description.}}
\end{figure*}

\begin{figure*}
\gridline{\fig{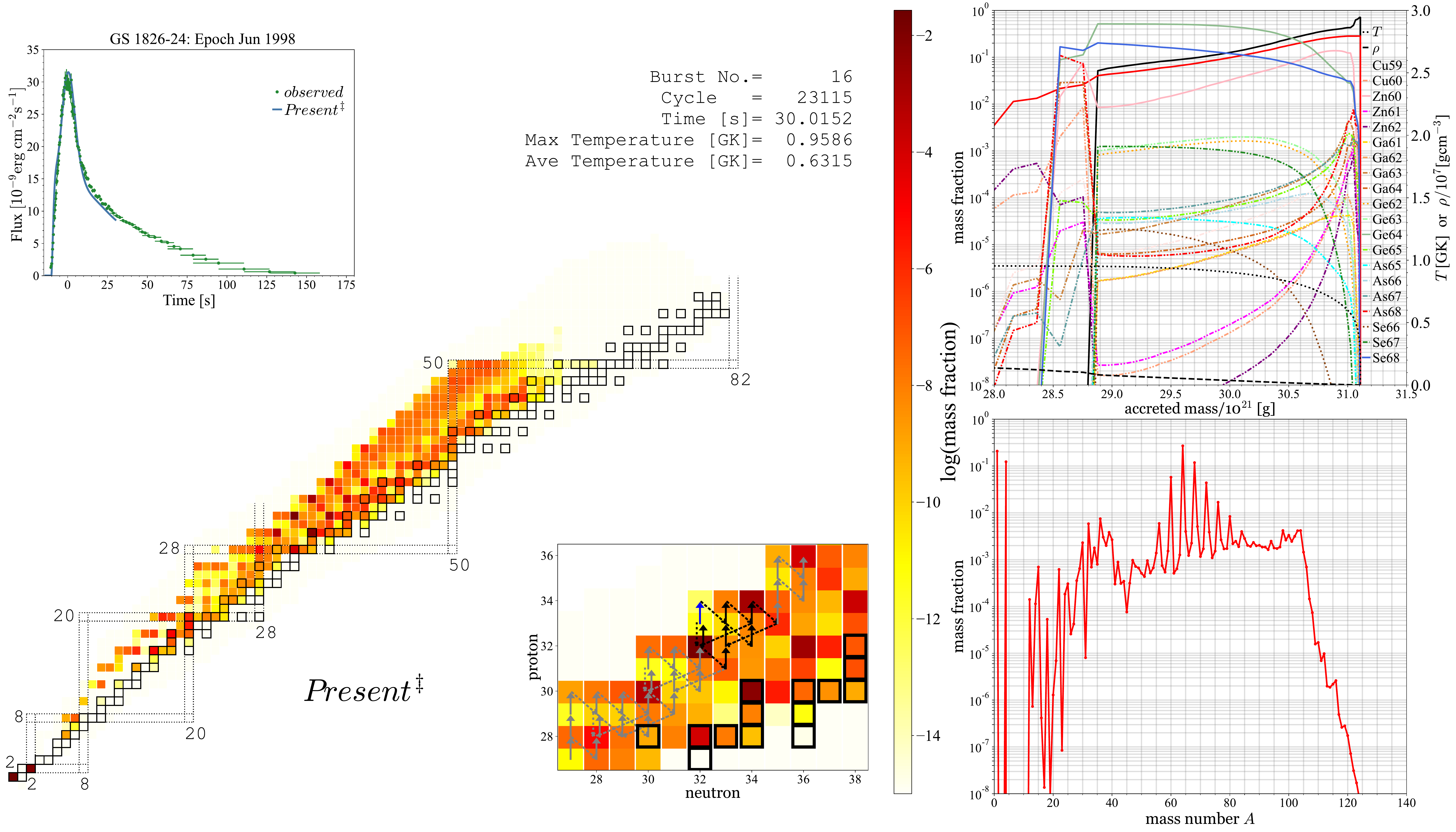}{18cm}{}}
\vspace{-10mm}
\gridline{\fig{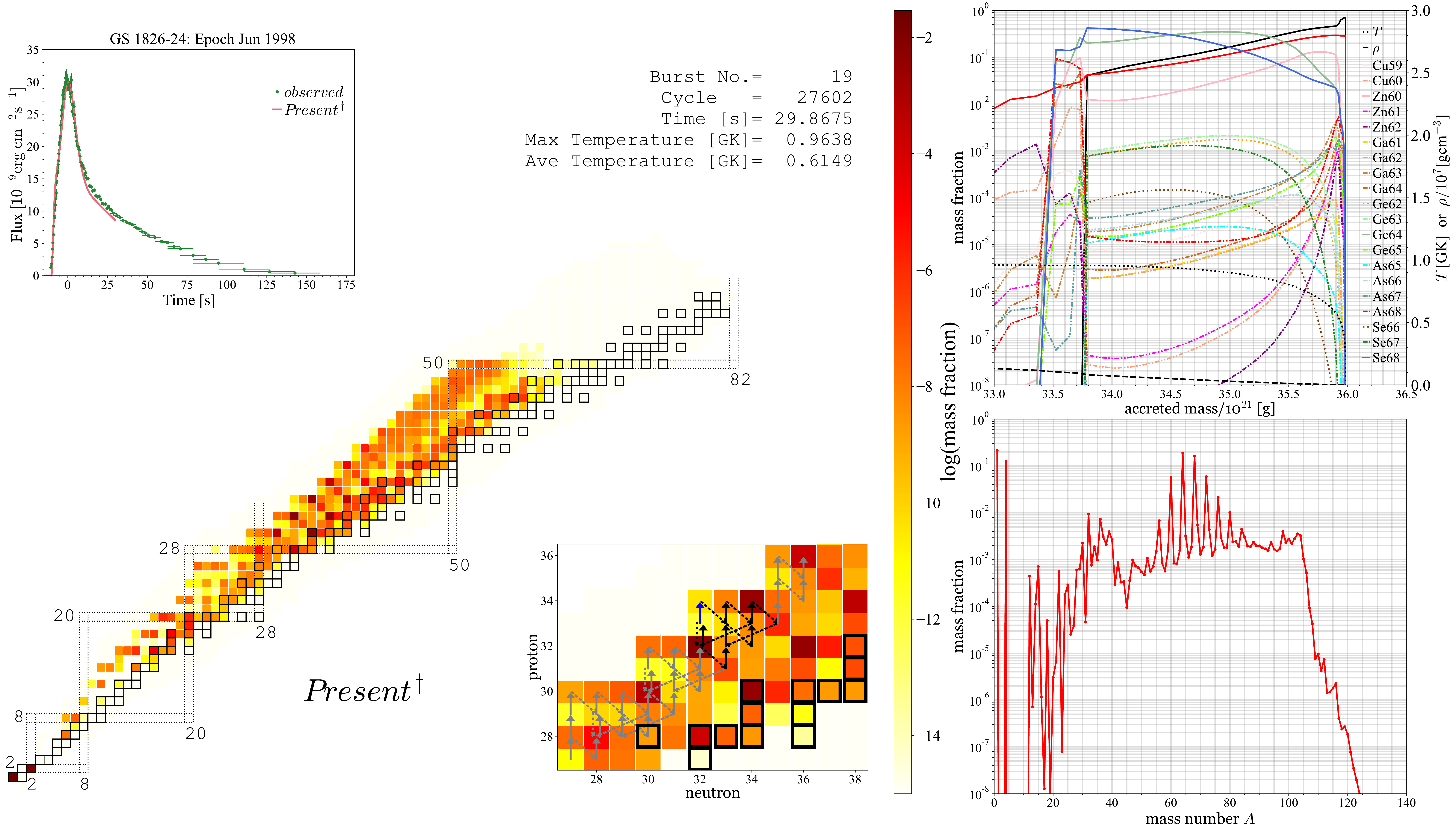}{18cm}{}}
\vspace{-10mm}
\caption{\label{fig:PostPeak1b}{\footnotesize The nucleosynthesis and evolution of envelope corresponds to the moment at around 30~s after the burst peak for \emph{Present}$^\ddag$ (\textsl{Top Panel}) and \emph{Present}$^\dag$ (\textsl{Bottom Panel}) scenarios. See Fig.~\ref{fig:Onset1} for further description.}}
\end{figure*}

\begin{figure*}
\gridline{\fig{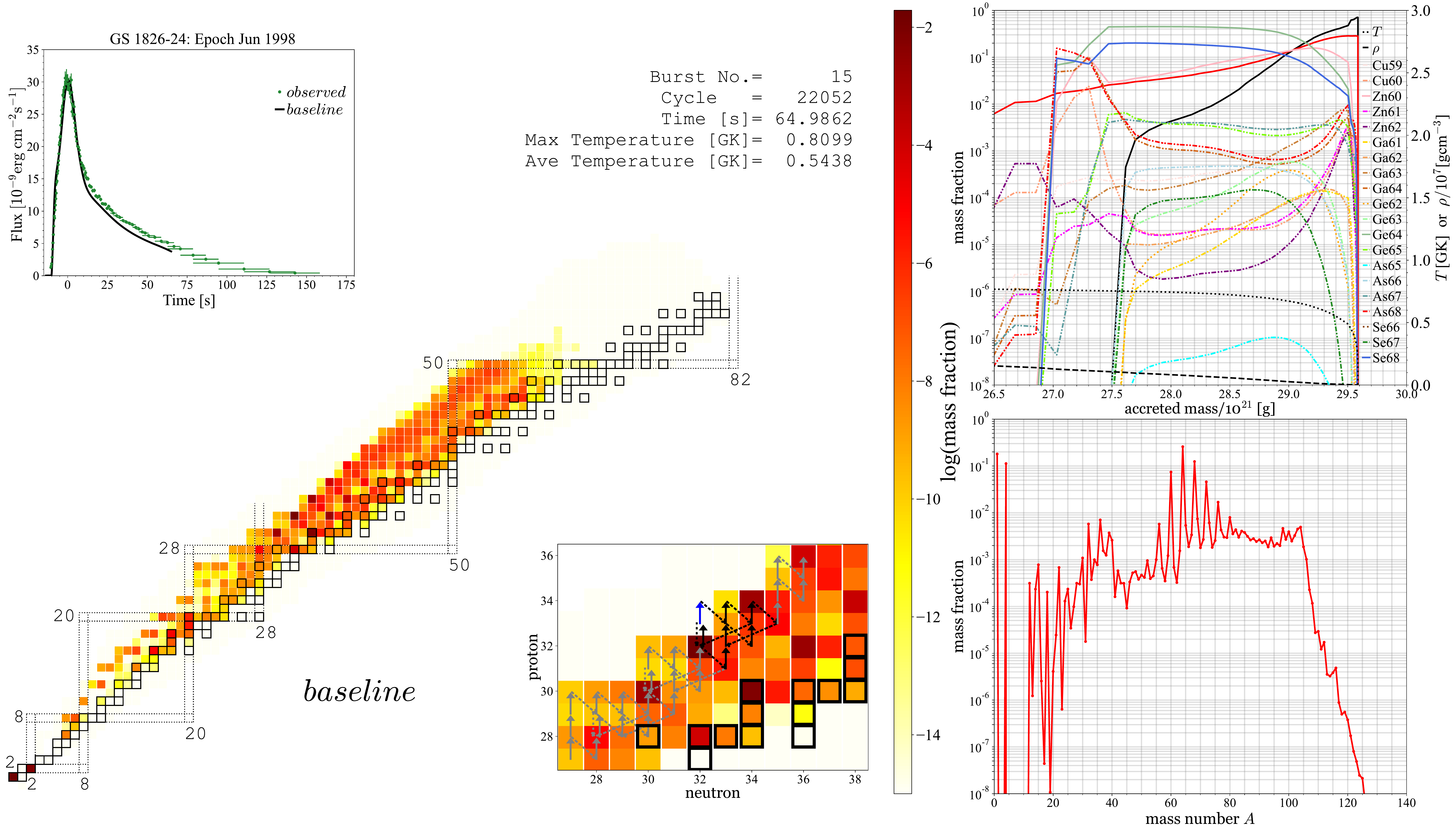}{18cm}{}}
\vspace{-10mm}
\gridline{\fig{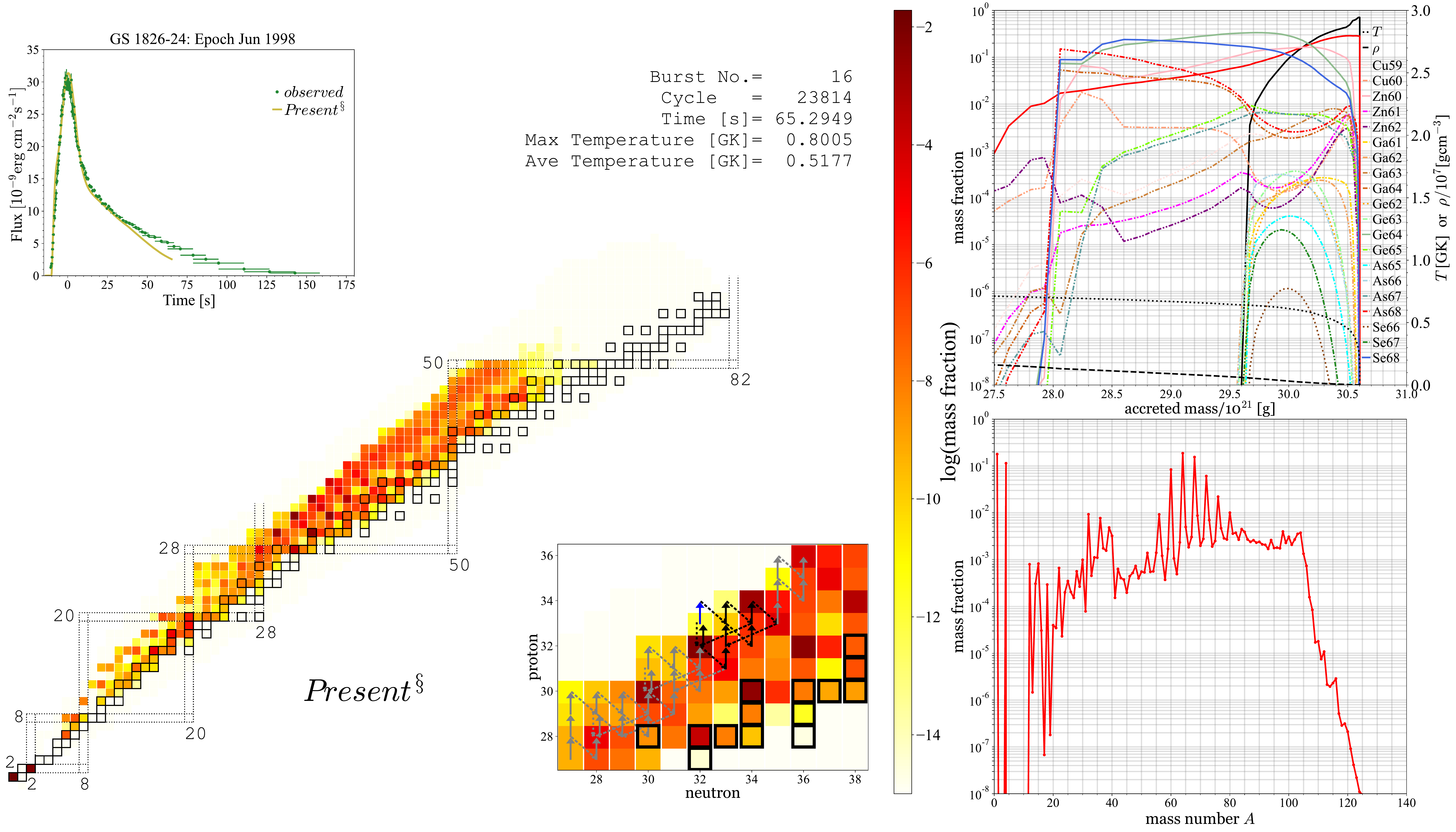}{18cm}{}}
\vspace{-10mm}
\caption{\label{fig:PostPeak2a}{\footnotesize The nucleosynthesis and evolution of envelope corresponds to the moment at around 65~s after the burst peak for \emph{baseline} (\textsl{Top Panel}) and \emph{Present}$^\S$ (\textsl{Bottom Panel}) scenarios. See Fig.~\ref{fig:Onset1} for further description.}}
\end{figure*}

\begin{figure*}
\gridline{\fig{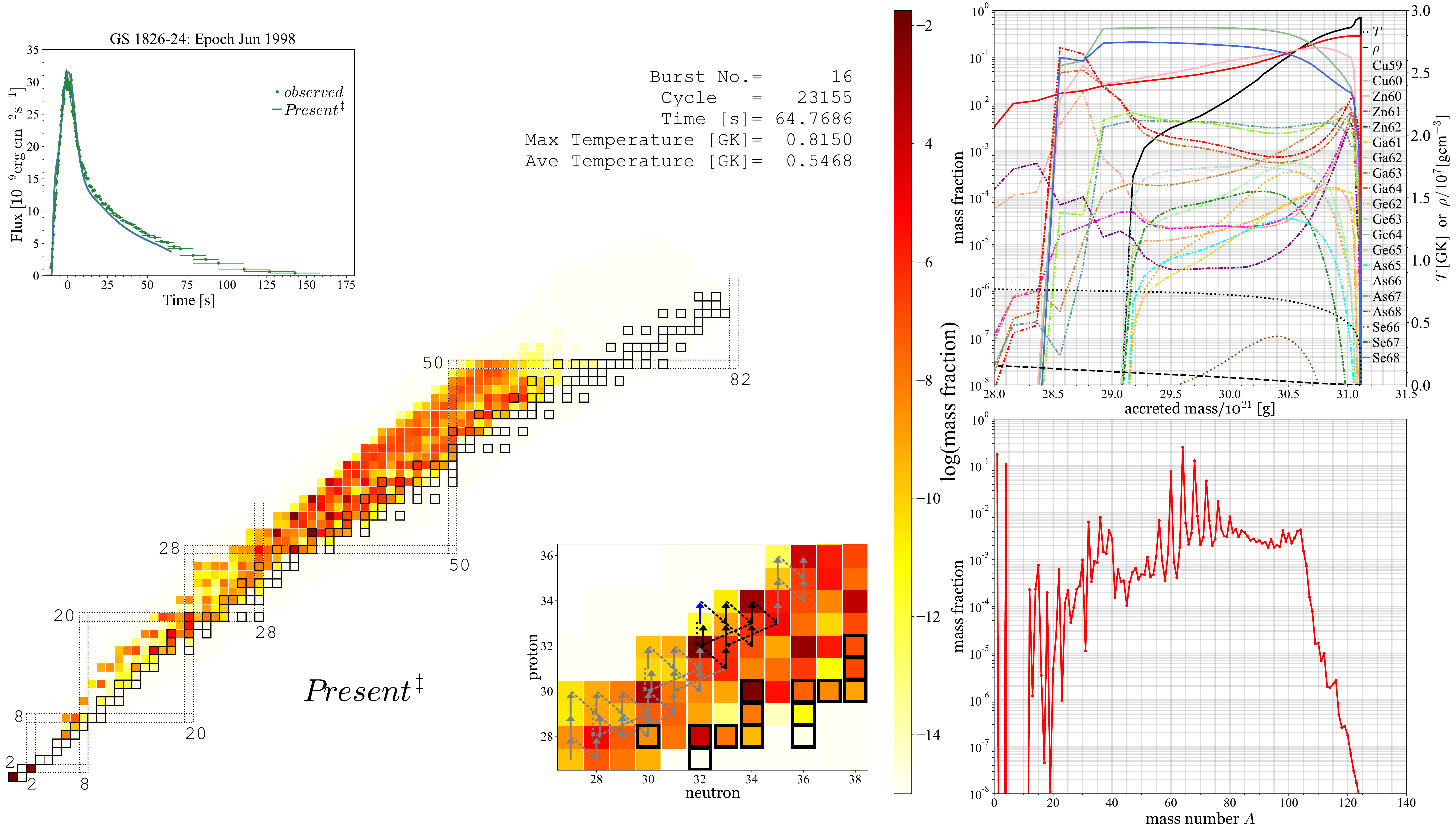}{18cm}{}}
\vspace{-10mm}
\gridline{\fig{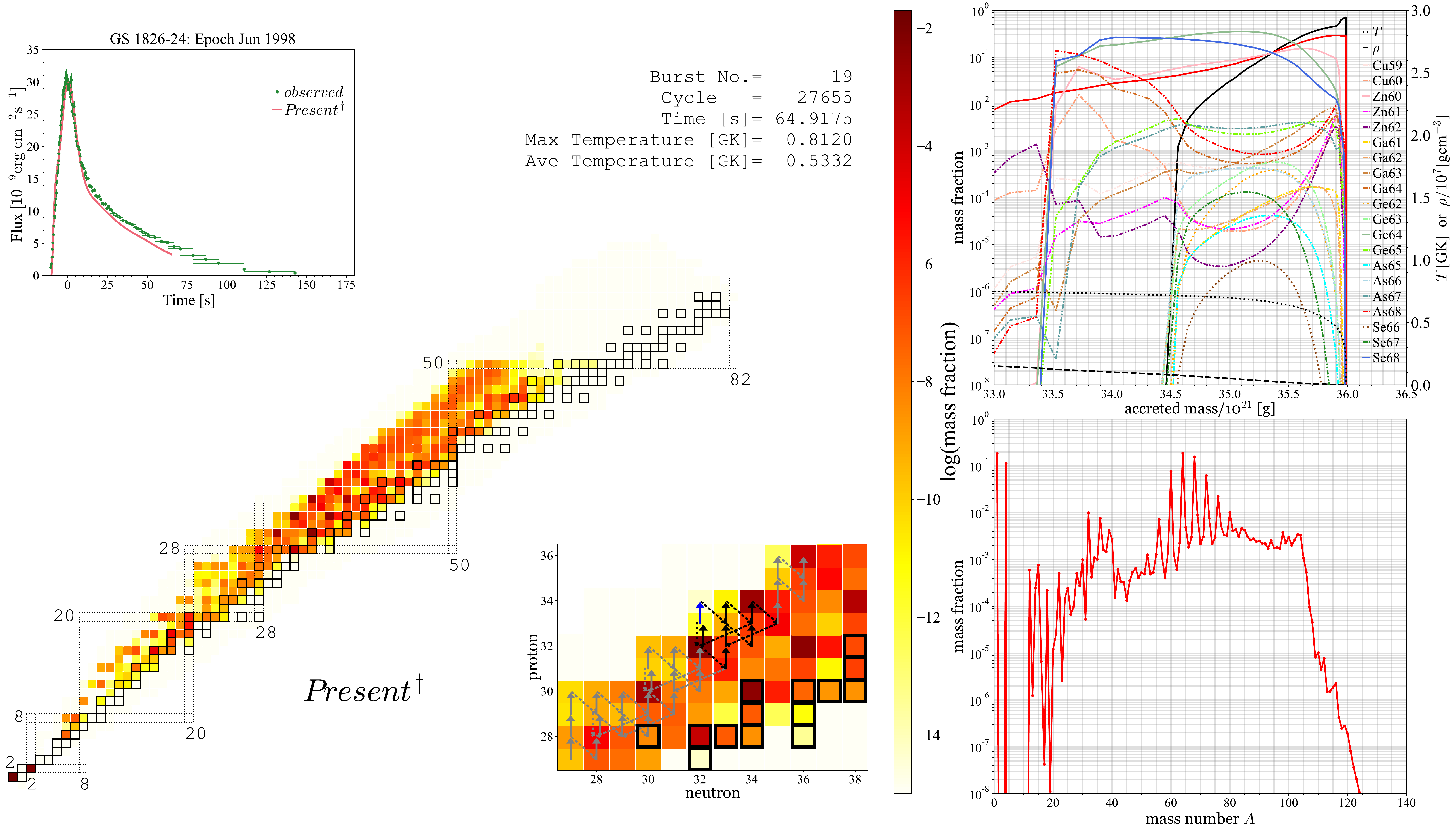}{18cm}{}}
\vspace{-10mm}
\caption{\label{fig:PostPeak2b}{\footnotesize The nucleosynthesis and evolution of envelope corresponds to the moment at around 65~s after the burst peak for \emph{Present}$^\ddag$ (\textsl{Top Panel}) and \emph{Present}$^\dag$ (\textsl{Bottom Panel}) scenarios. See Fig.~\ref{fig:Onset1} for further description.}}
\end{figure*}

We define the burst-peak time, $t=0$~s. The evolution time of light curve is respect to the burst-peak time. Figure~\ref{fig:Flux_GS1826} displays the best-fit modeled and observed XRB light curve profiles. The observed burst peak is located in the time regime $t=-2.5$~--~$2.5$~s (top left inset in Fig.~\ref{fig:Flux_GS1826}), and at the vicinity of the modeled light-curve peaks of \emph{baseline}, \emph{Present}$^\S$, \emph{Present}$^\ddag$, and \emph{Present}$^\dag$. The overall averaged flux deviations between the observed epoch and each of these theoretical models, \emph{baseline}, \emph{Present}$^\S$, \emph{Present}$^\ddag$, and \emph{Present}$^\dag$ in units of $10^{-9}\mathrm{erg~cm}^{-2}\mathrm{s}^{-1}$ are 
$1.175$, 
$1.159$, 
$1.149$, and
$1.252$,
respectively. 

All modeled light curves are less enhanced than the observed light curve at $t=8$~--~$125$~s due to the reduction in accretion rate. The updated $^{59}$Cu(p,$\gamma$)$^{60}$Zn and $^{61}$Ga(p,$\gamma$)$^{62}$Ge reaction rates can  induce enhancements at $t=8$~--~$30$~s and $t=35$~--~$125$~s, respectively \citep{Lam2021b}, see also the Supplemental Material of \citet{Hu2021}. The \emph{Present} $^{65\!\!}$As(p,$\gamma$)$^{66\!}$Se forward and reverse rates decrease the burst light curve at $t=40$~--~$100$~s (red line in Fig.~\ref{fig:Flux_GS1826}), whereas the lower limit of the \emph{Present} $^{65\!\!}$As(p,$\gamma$)$^{66\!}$Se forward and reverse rates produces a similar burst light curve profile as  \emph{baseline} (blue and black lines in Fig.~\ref{fig:Flux_GS1826}). Note that the lower limit of the \emph{Present} forward and reverse rates are based on a rather extreme $S_\mathrm{p}$($^{66\!}$Se) uncertainty. The burst tail at $t=50$~--~$80$~s generated from the $^{65\!\!}$As(p,$\gamma$)$^{66\!}$Se forward and reverse rates using $S_\mathrm{p}$($^{66\!}$Se) $=2.433$~MeV, $2.443$~MeV, and $2.507$~MeV could be just $\lesssim10^{-9}\,\mathrm{erg~cm}^{-2}\mathrm{s}^{-1}$ displaced from the burst tail of the \emph{Present}$^\dag$ scenario (red line in Fig.~\ref{fig:Flux_GS1826}) as long as the newly measured $S_\mathrm{p}$($^{66\!}$Se) is within a range of $\mytilde$50~keV close to the present $S_\mathrm{p}$($^{66\!}$Se) $=2.469$~MeV. The \emph{baseline} model that implements the NON-SMOKER $^{65\!\!}$As(p,$\gamma$)$^{66\!}$Se forward and reverse rates, can be a reference estimating the influence of the $S_\mathrm{p}$($^{66\!}$Se)$~=2.351$~MeV, 2.381~MeV, and $2.284$~MeV due to the rather close range of $S_\mathrm{p}$($^{66\!}$Se). Besides, the $^{65\!\!}$As(p,$\gamma$)$^{66\!}$Se forward and reverse rates generated from the $S_\mathrm{p}$($^{66\!}$Se) $=2.186$~MeV may further enhance the burst tail end to be $\sim\!\!1\times10^{-9}\mathrm{erg~cm}^{-2}\mathrm{s}^{-1}$ higher than the \emph{baseline} at $t=50$~--~$80$~s. The sensitivity study performed by \citet{Cyburt2016} on the influence of (p,$\gamma$) forward reaction rates does not exhibit that the $^{65\!\!}$As(p,$\gamma$)$^{66\!}$Se forward rate is influential, whereas the sensitivity study done by \citet{Schatz2017} indicate that the $^{65\!\!}$As(p,$\gamma$)$^{66\!}$Se reverse rate possibly impacts the burst tail. Our present study with the newly deduced $^{65\!\!}$As(p,$\gamma$)$^{66\!}$Se forward and reverse rates shows that the correlated forward and reverse rates characterizes the burst tail.

The updated $^{22}$Mg($\alpha$,p)$^{25}\!$Al reaction maintains its role in increasing the burst light curve at $t=16$~--~$60$~s (yellow line in Fig.~\ref{fig:Flux_GS1826}) even \emph{correlated influences} among dominant reactions are included in the \emph{Present}$^\S$ model. This finding agrees with the preliminary result of \citet{Lam2021b} shown in the Supplemental Material of \citet{Hu2021}.

From $t=130$~s onward, the \emph{baseline} and \emph{Present}$^{\S,\ddag,\dag}$ models successfully reproduce the tail end of the burst light curve of GS~1826$-$24. The modeled burst tail ends produced by \citet{Randhawa2020} are, however, over expanded which might be due to a somehow limited observed data that are selected for fitting the modeled burst light curves, see Fig.~4 in \citet{Randhawa2020} or in \citet{Hu2021}.


\subsection{Evolutions of accreted envelopes and \\ the respective nucleosyntheses} 
\label{sec:evo}

We perform a comprehensive study to understand the microphysics behind the differences among these modeled burst light curves with investigating the evolutions of accreted envelope regime where nuclei heavier than CNO isotopes are densely synthesized against the evolutions of the respective burst light curves of the 15$^\mathrm{th}$, 16$^\mathrm{th}$, 16$^\mathrm{th}$, and 19$^\mathrm{th}$ burst for the \emph{baseline}, \emph{Present}$^\S$, \emph{Present}$^\ddag$, and \emph{Present}$^\dag$ models, respectively. These selected bursts almost resemble the respective averaged light curve profile presented in Fig.~\ref{fig:Flux_GS1826}. The reference time of accreted envelope and nucleosynthesis in the following discussion is also relative to the burst-peak time, $t=0$~s.

\paragraph{The moment before and just before the onset.} 
The preceding burst leaves the accreted envelope with synthesized proton-rich nuclei, which go through $\beta^+$ decays, and enrich the region around long half-live stable nuclei, e.g., $^{32}$S, $^{36}$Ar, $^{40}$Ca, $^{60}$Ni, $^{64}$Zn, $^{68}$Ge, $^{72}$Se, $^{76}$Kr, and $^{80}$Sr, which are the remnants of waiting points. Meanwhile, the unburned hydrogen nuclei above the base of the accreted envelope keep the stable burning active until the freshly accreted stellar fuel stacks up, increasing the density of the accreted envelope due to the strong gravitational pull from the host neutron star of GS~1826$-$24 X-ray source for presetting the thermonuclear runaway conditions of the next XRB.

\begin{figure}[t]
\begin{center}
\includegraphics[width=8.5cm,angle=0]{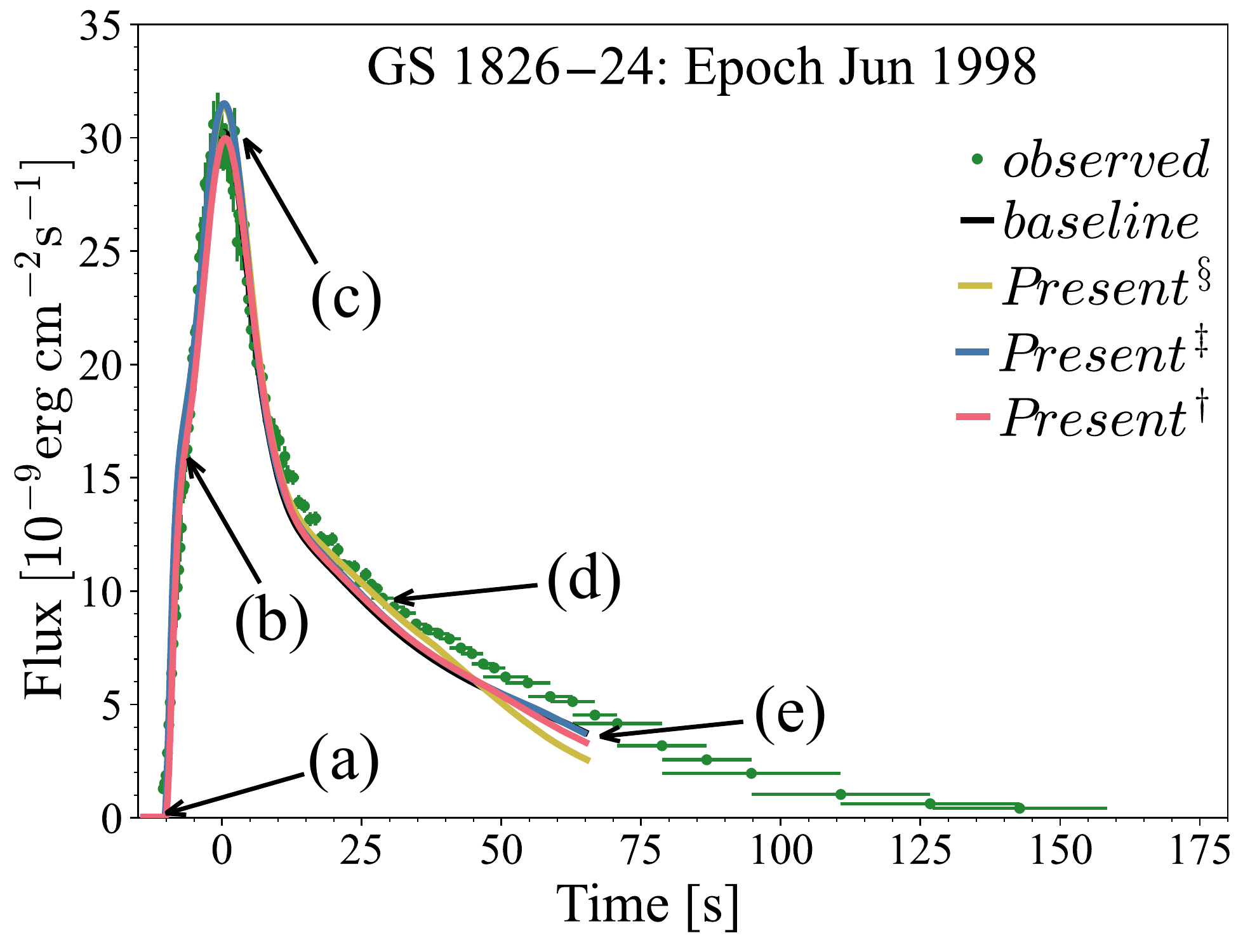}
\caption{{\footnotesize The evolution of a clocked burst for \emph{baseline} (15$^\mathrm{th}$ burst), \emph{Present}$^\S$ (16$^\mathrm{th}$ burst), \emph{Present}$^\ddag$ (16$^\mathrm{th}$ burst), and \emph{Present}$^\dag$ (19$^\mathrm{th}$ burst) with respect to the averaged-observed clocked burst of GS~1826$-$24 burster, at times (a) $t=-10.2$~s, (b) $t=-7.5$~s, (c) $t\approx 0$~s, (d) $t\approx 30$~s, and (e) $t\approx65$~s. The corresponding comparisons of averaged mass fractions for each mass number at the moments (a), (b), (c), (d), and (e) are presented in Fig.~\ref{fig:abundances}.}}
\label{fig:BurstEvo}
\end{center}
\end{figure}

At time $t=-10.2$~s, just before the onset of the succeeding XRB, the temperature of the envelope reaches a maximum value of $0.9$~GK for the \emph{baseline}, and \emph{Present}$^{\S,\ddag,\dag}$ scenarios, see Figs.~\ref{fig:Onset1}, \ref{fig:Onset2}, and \ref{fig:BurstEvo}(a). The $^{64}$Ge abundance is around one to five orders of magnitude higher than other surrounding isotopes, whereas the ratio of $^{68}$Se to $^{64}$Ge abundances is $\approx\!\!1.7$. Meanwhile, the evolution of $^{64}$Ge mass fraction in the mass coordinate of accreted envelope is qualitatively analogous to the $^{66\!}$Se mass fraction. The $^{64}$Ge abundance is comparable with the $^{60}$Zn and $^{66\!}$Se abundances (solid green, blue, and pink lines in Figs.~\ref{fig:Onset1}, ..., \ref{fig:PostPeak2b}). These four factors indicate that $^{64}$Ge is still a significant waiting point. The reaction flow passes through $^{68}$Se and advances to heavier proton-rich nuclei region meanwhile the degenerate envelope is on the brink of the onset of XRB.

For the \emph{baseline} scenario, the 2p-capture on $^{64}$Ge waiting point is not yet developed, whereas for the \emph{Present}$^\ddag$ scenario, the 2p-capture on $^{64}$Ge is weak. The reaction flows of these two scenarios follow the weak GeAs~II and sub-GeAs~II cycles and mainly break out at $^{69}$Se, see Fig.~\ref{fig:Cycles_GeAs} for the reaction paths in the GeAs cycles. For the \emph{Present}$^\S$ and \emph{Present}$^\dag$ scenarios, the 2p-capture on $^{64}$Ge has already established, and the break-out flow at $^{69}$Se is rather similar to the \emph{baseline} and \emph{Present}$^\ddag$ scenarios as the $^{68}$Se and $^{69}$Se abundances for these four scenarios are rather similar. The strong 2p-capture on $^{64}$Ge in the \emph{Present}$^\S$ and \emph{Present}$^\dag$ scenarios causes more than two orders of magnitude of $^{65\!\!}$As and $^{66\!}$Se cumulated in the GeAs cycles compared to the \emph{baseline} and \emph{Present}$^\ddag$ scenarios (dot-dashed cyan and dotted dark brown lines in top right insets of each panel in Figs.~\ref{fig:Onset1} and \ref{fig:Onset2}). Using either the upper (or lower) limits of the new $^{64}$Ge(p,$\gamma$)$^{65\!\!}$As reaction rate \citep{Lam2016} could mildly enhance (or reduce) the strength of 2p-capture on $^{64}$Ge of drawing the synthesized materials from $^{64}$Ge. Note that the GeAs cycles are still considered weak as compared to the NiCu and ZnGa cycles, nevertheless, the reaction flows in the GeAs cycles of the \emph{Present}$^\S$ and \emph{Present}$^\dag$ scenarios is stronger than the ones in \emph{baseline} and \emph{Present}$^\ddag$ scenarios. The high $^{66\!}$Se abundance in the \emph{Present}$^\S$ and \emph{Present}$^\dag$ scenarios are due to the implementation of the \emph{Present} $^{65\!\!}$As(p,$\gamma$)$^{66\!}$Se forward (and reverse) reaction rate, which is a factor of $\mytilde1.7$ higher than (a factor of $\mytilde4.5$ lower than) the NON-SMOKER $^{65\!\!}$As(p,$\gamma$)$^{66\!}$Se forward (reverse) rate at $T=0.9$~GK used in \emph{baseline}, and is about two orders ($\mytilde1.5$ order) of magnitude higher than the lower limit of \emph{Present} $^{65\!\!}$As(p,$\gamma$)$^{66\!}$Se forward (reverse) rate used in \emph{Present}$^\ddag$, see Fig.~\ref{fig:rp_65As_66Se}.

\paragraph{The moment at $t=-7.5$~s before the burst peak.} 
Overall the light curves of the \emph{baseline}, \emph{Present}$^\ddag$, \emph{Present}$^\S$, and \emph{Present}$^\dag$ scenarios rise to $15.4$, $16.3$, $16.2$, and $14.9$ in units of $10^{-9}\mathrm{erg~cm}^{-2}\mathrm{s}^{-1}$, respectively (top left insets of each panel in Figs.~\ref{fig:PrePeak1} and \ref{fig:PrePeak2}, and Fig.~\ref{fig:BurstEvo}(b)). The modeled light curve is quantitatively comparable with the observed bolometric flux. As the temperature of the envelope rises to around $1.1$~GK, the 2p-capture on $^{64}$Ge waiting point and the weak GeAs cycles starts establishing in the \emph{baseline} and \emph{Present}$^\ddag$ scenarios, which are about $2.5$~s later than the \emph{Present}$^\S$ and \emph{Present}$^\dag$ scenarios because both $^{65\!\!}$As(p,$\gamma$)$^{66\!}$Se rates used in the \emph{baseline} and \emph{Present}$^\ddag$ scenarios are up to (or more than) a factor of $2.5$ in $T=0.4$~--~$1.1$~GK lower than the one used in the \emph{Present}$^\S$ and \emph{Present}$^\dag$ scenarios (Fig.~\ref{fig:rp_65As_66Se}). Also, the reverse $^{65\!\!}$As(p,$\gamma$)$^{66\!}$Se rate used in \emph{baseline} is about a factor of $8.5$ higher than the \emph{Present} reverse rate used in \emph{Present}$^{\S,\ddag,\dag}$, causing a rather low cumulation of $^{66\!}$Se in \emph{baseline}.

The mass fractions of synthesized nuclei in the GeAs cycles evolve inversely with the rise of temperature of envelope along the mass zone, except $^{66\!}$Se (and $^{65\!\!}$As for the \emph{baseline} scenario). The evolution of productions of $^{65\!\!}$As, $^{66\!}$Se, $^{66}$As, $^{67}$Se, $^{67}$As, $^{68}$Se, and $^{68}$As up to this moment indicates that a strong reaction flow is formed along the 
$^{64}$Ge(p,$\gamma$)$^{65\!\!}$As (p,$\gamma$)$^{66\!}$Se($\beta^+\nu$)$^{66}$As(p,$\gamma$)$^{67}$Se($\beta^+\nu$)$^{67}$As(p,$\gamma$)$^{68}$Se($\beta^+\nu$) $^{68}$As(p,$\gamma$)$^{69}$Se path for these four scenarios. 
The grow of $^{65\!\!}$As mass fraction in the \emph{baseline} is due to the higher NON-SMOKER $^{65\!\!}$As(p,$\gamma$)$^{66\!}$Se reverse rate using the $S_\mathrm{p}$($^{66\!}$Se)~$=2.349$~MeV, indicating that some material is temporarily stored as $^{65\!\!}$As in the \emph{baseline} scenario. For the \emph{Present}$^{\S,\ddag,\dag}$ scenarios, more material is transmuted from $^{65\!\!}$As and temporarily stored as $^{66\!}$Se, which then decays via ($\beta^+\nu$) channel and quickly transmutes to $^{69}$Se and surge through the heavier proton-rich region. 





\paragraph{The moment at the vicinity of the burst peak.} 
The accreted envelopes of \emph{baseline}, \emph{Present}$^\ddag$, \emph{Present}$^\S$, and \emph{Present}$^\dag$ reach the maximum temperature at times $-0.9$~s ($1.33$~GK), $-1.1$~s ($1.34$~GK), $-1.5$~s ($1.34$~GK), and $-0.8$~s ($1.34$~GK), respectively. The rise of temperature in the envelope before the burst peak for all scenarios due to the mutual feedback between the nuclear energy generation and hydrodynamics during the onset induces the release of material stored in dominant cycles of nuclei lighter than $^{64}$Ge. Due to the lower $^{65\!\!}$As(p,$\gamma$)$^{66\!}$Se reaction rates used in \emph{baseline} and \emph{Present}$^\ddag$, less material is drawn to $^{68}$Se causing the production of $^{64}$Ge surpasses $^{68}$Se in these two scenarios. In fact, both $^{64}$Ge and $^{68}$Se have already been produced and cumulated when $t\approx-10.6$~s or $\approx0.5$~s before the onset.

The location of the observed burst peak could be at $\mathord\pm0.5$~s away from the modeled burst peaks of these four scenarios (top left insets of each panel in Figs.~\ref{fig:Peak1} and \ref{fig:Peak2}, and Fig.~\ref{fig:BurstEvo}(c)). Note that the advantage of impartially fitting the whole observed burst light curve helps us to avoid unexpectedly shifting the modeled burst peak further away from the thought location of observed burst peak. Such misalignment with the observed burst peak occurs in the modeled burst peaks produced by \citet{Randhawa2020} models, see Fig.~4 in \citet{Randhawa2020} or in \citet{Hu2021}.


We find that the \emph{Present}$^\S$ model uses the latest $^{22}$Mg($\alpha$,p)$^{25}\!$Al reaction rate \citep{Hu2021} which extends the dominance of $^{22}$Mg(p,$\gamma$)$^{23}\!$Al reaction up to $1.67$~GK. This causes the reaction flow in the \emph{Present}$^\S$ scenario mainly follows the $^{22}$Mg(p,$\gamma$)$^{23}\!$Al(p,$\gamma$)$^{24}\!$Si path at the $^{22}$Mg branch point that is faster for the reaction flow to synthesize more proton-rich nuclei nearer to proton dripline than the $^{22}$Mg($\alpha$,p)$^{25}\!$Al(p,$\gamma$)$^{26}\!$Si path and gives rise to more hydrogen burning at later time.

\paragraph{The moment at $t\approx30$~s after the burst peak.} 
The consequence of more hydrogen burning caused by the latest $^{22}$Mg($\alpha$,p)$^{25}\!$Al reaction rate, which is almost one order of magnitude lower than the NON-SMOKER $^{22}$Mg($\alpha$,p)$^{25}\!$Al reaction rate, manifests on the enhancement of burst light curve at time regime $t=16$~--~60~s for the \emph{Present}$^\S$ scenario (top left insets of each panel in Figs.~\ref{fig:PostPeak1a} and \ref{fig:PostPeak1b}, and Fig.~\ref{fig:BurstEvo}(d)). Such enhancement is consistent with that was found by \citet{Hu2021}, and exhibits the role of important reactions found by \citet{Cyburt2016} improving the modeled burst light curve.

\paragraph{The moment at $t\approx65$~s after the burst peak.} 
The modeled burst light curves diverge at $t\approx 45$~s and reach a distinctive difference at $t\approx65$~s (top left insets of each panel in Figs.~\ref{fig:PostPeak2a} and \ref{fig:PostPeak2b}, and Fig.~\ref{fig:BurstEvo}(e)). Materials that have been released since around the burst-peak period from cycles in $sd$-shell, e.g., the NeNa, SiP, SCl, and ArK cycles, in bottom $p\!f$-shell, e.g., the CaSc cycle, have passed through the NiCu, ZnGa, and weak GeAs cycles, enrich the region beyond Ge and Se. The 2p-capture on $^{64}$Ge for \emph{baseline} and \emph{Present}$^\ddag$ only lasts until $t=21.4$~s and $t=35.8$~s, respectively, whereas for \emph{Present}$^\S$ and \emph{Present}$^\dag$, this 2p-capture lasts until $t=49.1$~s and $t=58.6$~s, respectively. The longer time the 2p-capture on $^{64}$Ge extends in the XRBs, the more material is transferred via the $^{64}$Ge(p,$\gamma$)$^{65\!\!}$As(p,$\gamma$)$^{66\!}$Se($\beta^+\nu$)$^{66}$As(p,$\gamma$)$^{67}$Se($\beta^+\nu$)$^{67}$As(p,$\alpha$) $^{68}$Se($\beta^+\nu$)$^{68}$As(p,$\alpha$)$^{69}$Se path to surge through the region above Se with intensive (p,$\gamma$)-($\beta^+\nu$) reaction sequences depleting accreted hydrogen appreciably (solid black lines in the top right inset of each panel in Figs.~\ref{fig:PostPeak2a} and \ref{fig:PostPeak2b}). The H exhaustion in the envelope quenches the (p,$\gamma$) reactions, reduces the syntheses of proton-rich nuclei, and the produced proton-rich nuclei are left to sequential $(\beta^+\nu)$ decays increasing the production of daughter nuclei, e.g., $^{62}$Zn, $^{66}$Ge, $^{60}$Ni, and $^{64}$Zn, which are the daughter nuclei of the ZnGa and GeAs cycles, and $^{60}$Zn and $^{64}$Ge waiting points.

\subsection{Burst ashes}

\begin{figure*}
\gridline{
\fig{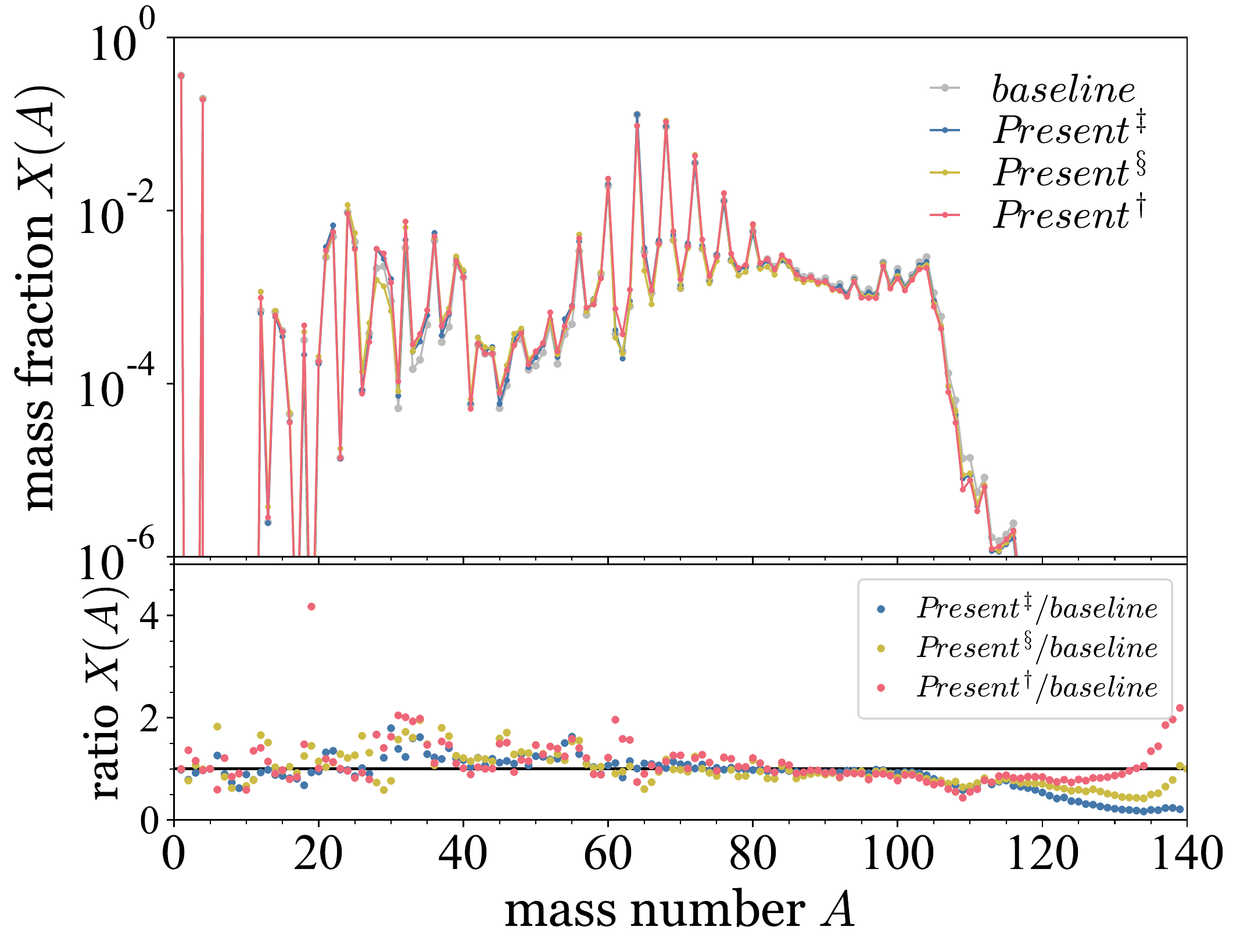}{8.5cm}{(a) The moment at $t=-10.2$~s just before the onset.} 
\fig{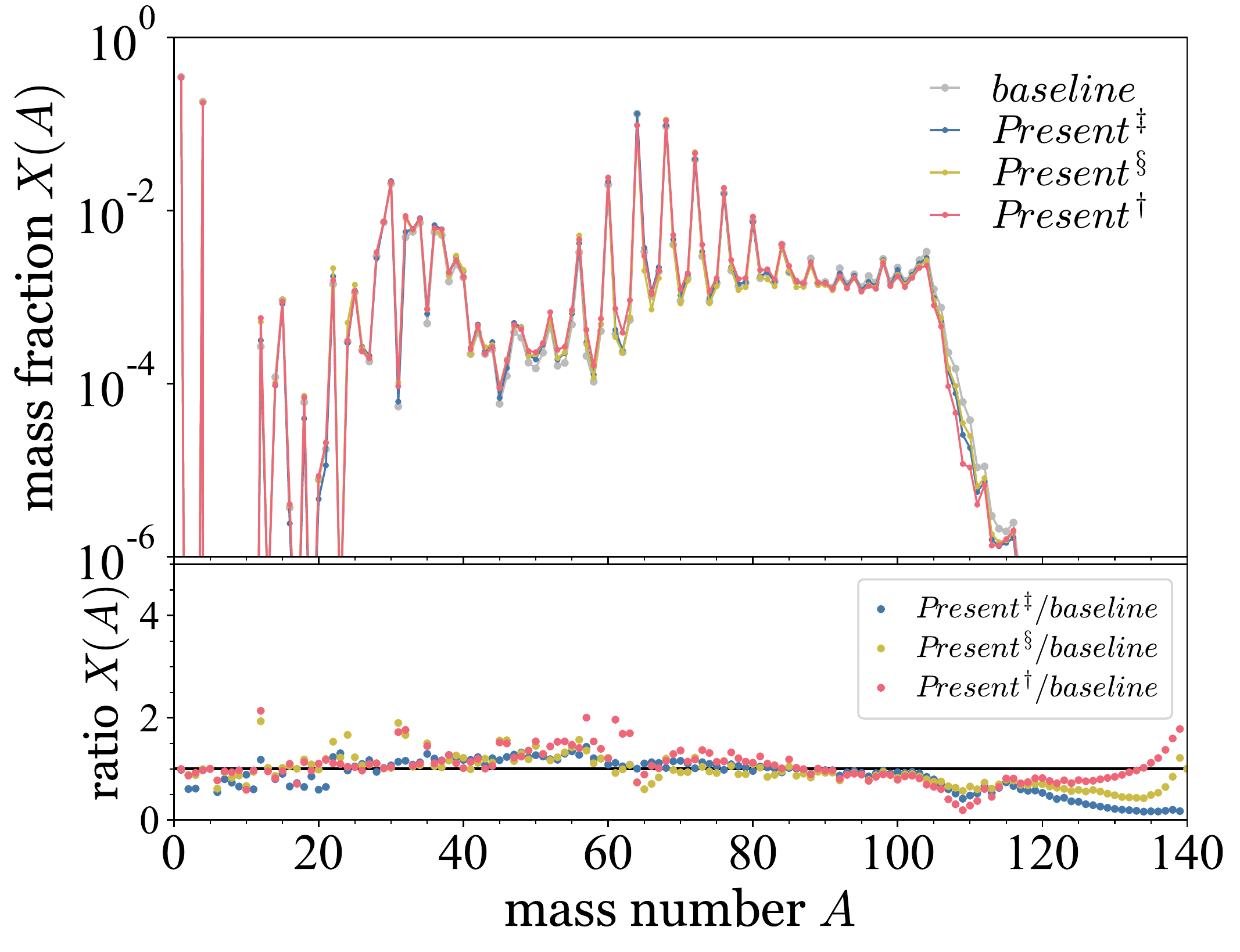}{8.5cm}{(b) The moment at $t=-7.5$~s before the burst peak.}
}
\gridline{
\fig{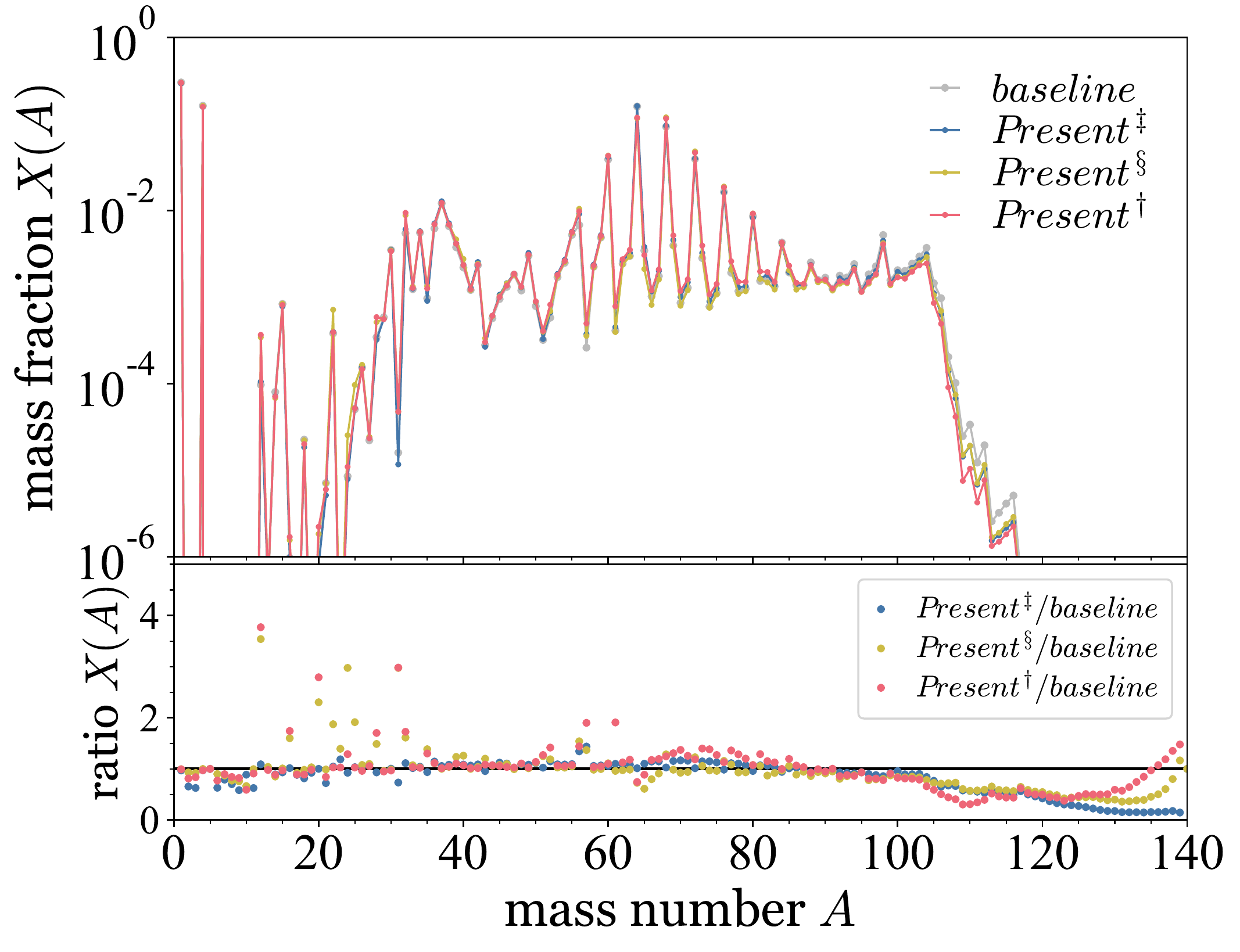}{8.5cm}{(c) The moment at $t \approx 0$~s, the vicinity of the burst peak.} 
\fig{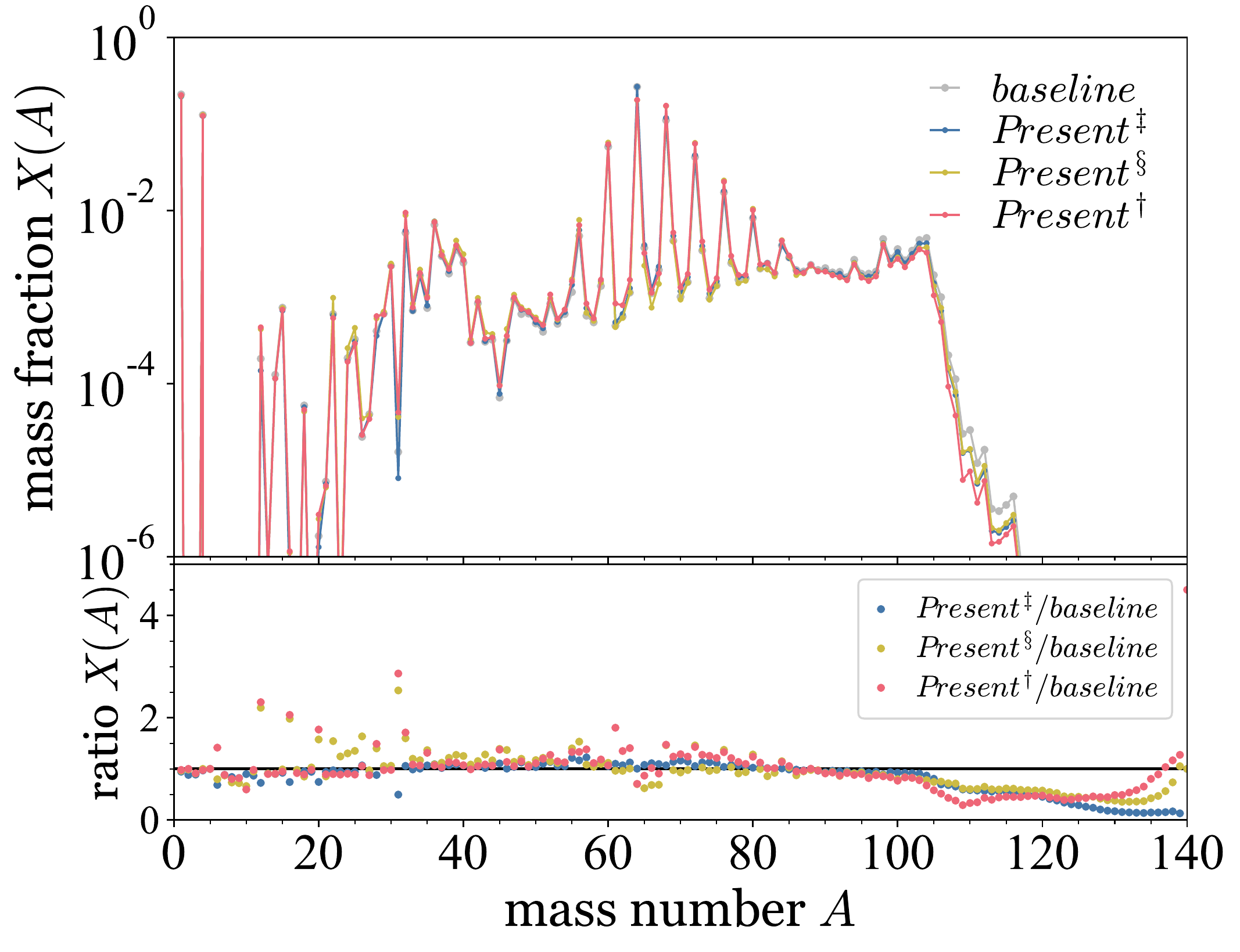}{8.5cm}{(d) The moment at $t \approx 30$~s after the burst peak.}
} 
\gridline{
\fig{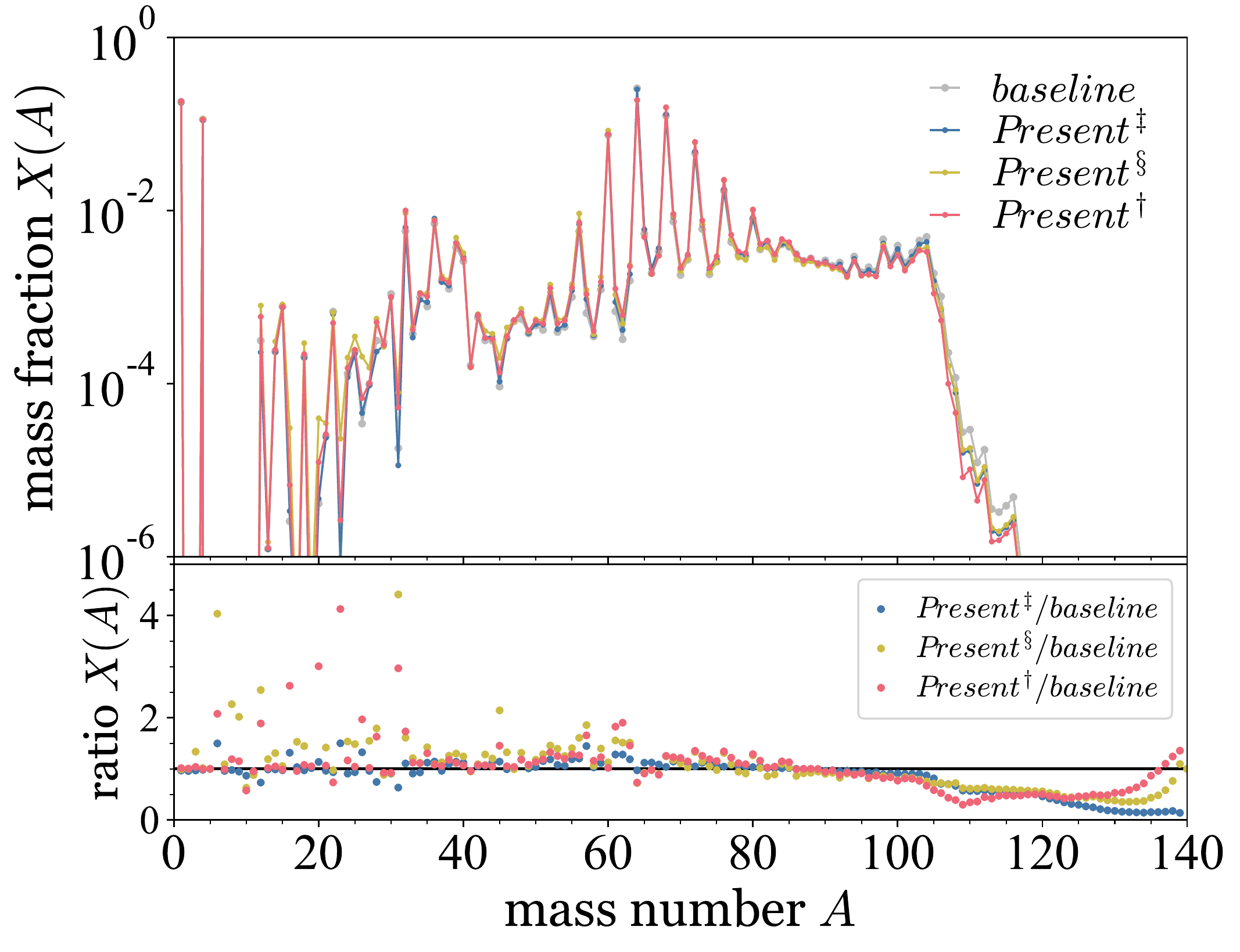}{8.5cm}{(e) The moment at $t \approx 65$~s after the burst peak.} 
\fig{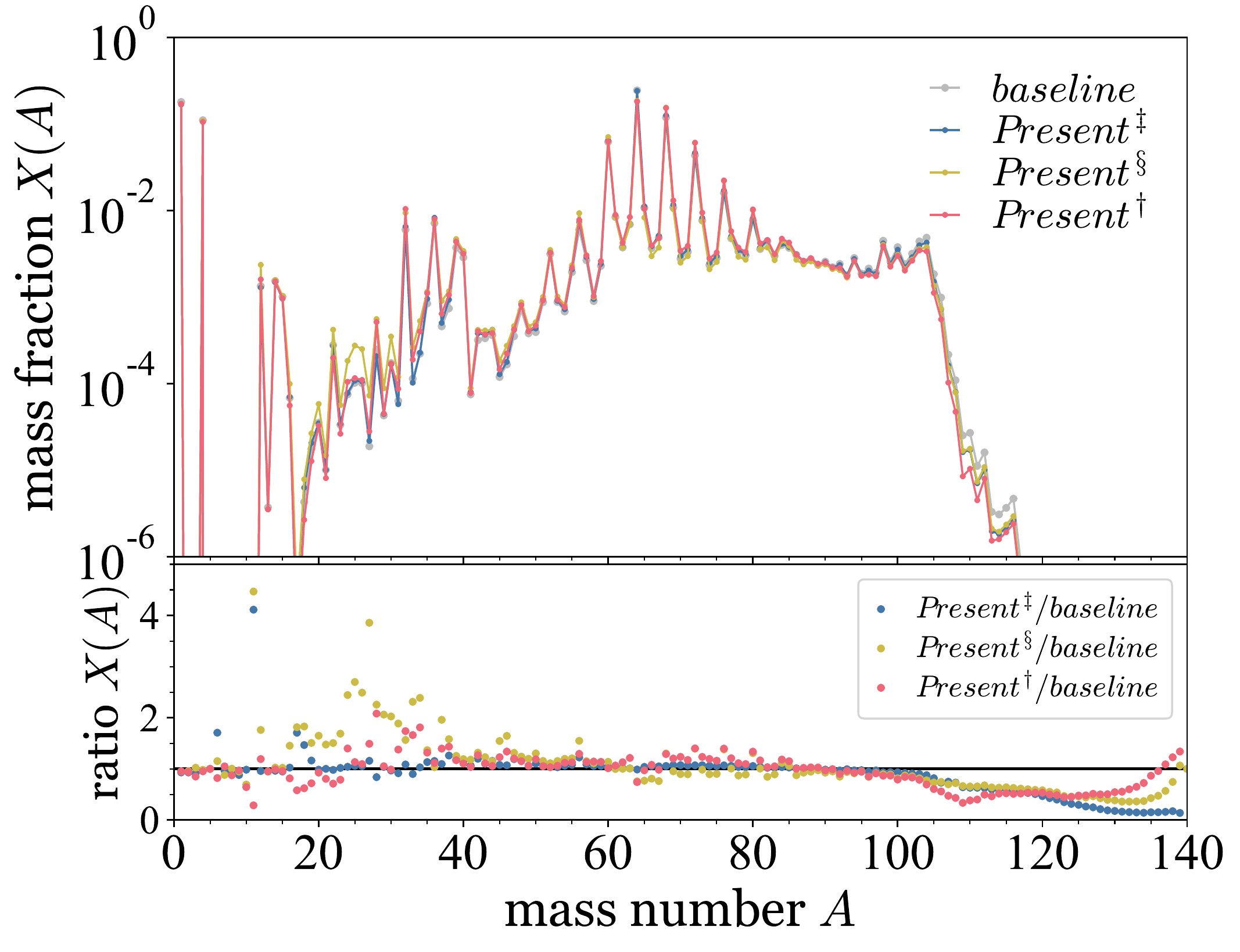}{8.5cm}{(f) The moment at $t \approx 150$~s after the burst peak.}
} 
\caption{\label{fig:abundances}{\footnotesize The averaged mass fractions for each mass number at times (a) $t=-10.2$~s, (b) $t=-7.5$~s, (c) $t\approx 0$~s, (d) $t\approx 30$~s, (e) $t\approx65$~s, and (f) $t\approx150$~s (\textsl{Top Panel} of each sub-figure). The comparison of averaged mass fractions between \emph{baseline} and \emph{Present}$^{\ddag,\S,\dag}$ is presented in the \textsl{Bottom Panel} of each sub-figure.}}
\end{figure*}


The \emph{baseline} and \emph{Present}$^{\S,\ddag,\dag}$ models that we construct reproduce the burst tail end of GS~1826$-$24 clocked burst from $t=130$~s onward with excellent agreement with the averaged-observed \emph{Epoch Jun 1998}. Nevertheless, the burst tail ends produced by \citet{Randhawa2020} both baseline and updated model are over expanded and misaligned with observation (\emph{Epoch Sept 2000}). Such deviation indicates that their modeled burst does not recess in accord with the observation and H-burning may somehow be still active in the envelope. The deviation could also be due to the limited data specifically chosen for fitting the modeled burst light curve or due to the astrophysical settings of both models, see Fig.~4 in \citet{Randhawa2020} or in \citet{Hu2021}. The one-zone model constructed by \citet{Schatz2017} successfully estimate a gross feature of the burst light curve influenced by a scaled $^{65\!\!}$As(p,$\gamma$)$^{66\!}$Se reaction rate (with $S_\mathrm{p}$($^{66\!}$Se) of $100$~keV uncertainty larger than the one used in NON-SMOKER $^{65\!\!}$As(p,$\gamma$)$^{66\!}$Se rate), however, the H-burning diminishes earlier than their baseline model due to the implemented extreme parameters \citep{Schatz2021}, causing a more rapid decrease of the burst tail end.

The compositions of burst ashes generated from the \emph{baseline} and \emph{Present}$^{\S,\ddag,\dag}$ models are presented in Fig.~\ref{fig:abundances}(f). The temperature of accreted envelope decreases from $0.8$~GK to $0.64$~GK starting from $t=65$~s to $150$~s. The weak feature of the latest $^{22}$Mg($\alpha$,p)$^{25}\!$Al reaction rate, which is more than two order of magnitudes lower than the $^{22}$Mg(p,$\gamma$)$^{23}$Al reaction at $T=0.8$~--~$0.64$~GK, allows more nuclei to be produced via the $^{22}$Mg(p,$\gamma$)$^{23}$Al(p,$\gamma$)$^{24}$Si reaction flow. The material in the reaction flow is then stored in the dominant cycles in $sd$-shell nuclei, e.g., the SiP, SCl, and ArK cycles. As H is then almost depleted, decreasing nuclear energy generation from (p,$\gamma$) reactions and causing the drop in temperature, the reaction flow is less capable to break out from cycles in $p\!f$-shell nuclei. The synthesized materials are kept in these cycles until the end of the burst. Therefore, the implementation of the latest $^{22}$Mg($\alpha$,p)$^{25}\!$Al in the \emph{Present}$^\S$ model increases the production of hot CNO cycle and $sd$-shell nuclei up to a factor of $4.5$ (for $^{12}$C which could be the main fuel for Type-I X-ray superburst; \citealt{Cumming2006}). Meanwhile, compared to the \emph{baseline} model, the abundance of $^{12}$C isotope is increased about a factor of $4.2$ based on the \emph{Present}$^\ddag$ model. These nuclei mainly are the remnants (daughter nuclei) left over from the proton-rich nuclei in the dominant cycles of $p$- and $sd$-shell nuclei. Such enrichment of $sd$-shell nuclei enhances the light nuclei abundances that eventually sink to the neutron-star surface.

We notice that a periodic increment exhibits in the remnant of $A=64$~--~$88$ nuclei up to a factor of $1.4$, with leading increment of waiting points remnants, $A=64$, $68$, $72$, $76$, $80$, $84$, and $88$. This indicates that a set of even weaker cycles resembling the weak GeAs II and sub-II cycles exists at waiting points heavier than $^{68}$Se. Nonetheless, the 2p-capture on waiting point feature is extremely weak for waiting points heavier than $^{64}$Ge, e.g., $^{68}$Se, $^{72}$Kr, $^{76}$Sr, $^{80}$Zr. The periodic increment of \emph{Present}$^\S$ is weaker than \emph{Present}$^\ddag$ and \emph{Present}$^\dag$ because materials are still stored in the dominant cycles of $sd$-shell nuclei due to the weak $^{22}$Mg($\alpha$,p)$^{25}\!$Al reaction. Moreover, the abundances of nuclei $A>100$ decreases for the \emph{Present}$^{\S,\ddag,\dag}$ models as more materials are kept in the cycles of $sd$- and $p\!f$-shell nuclei.



\subsection{Neutron-star mass-radius relation}

\begin{figure}[t]
\begin{center}
\includegraphics[width=\columnwidth,angle=0]{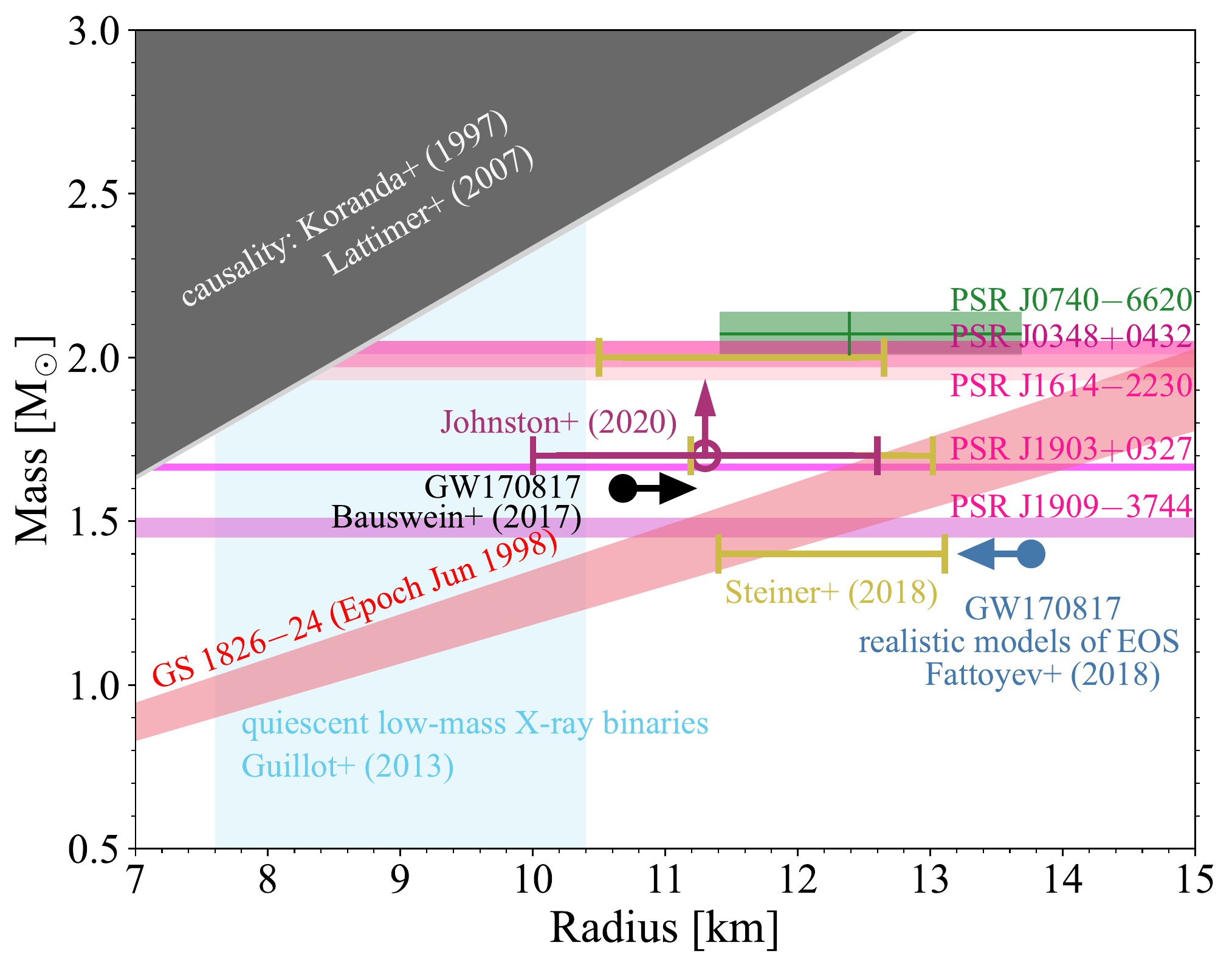}
\caption{\label{fig:compactness}{\footnotesize The $\mathrm{M}_{\mathrm{NS}}$-$\mathrm{R}_{\mathrm{NS}}$ constraints based on observed pulsars with uncertainties $\delta{\mathrm{M}}_\mathrm{NS}<0.05\,\mathrm{M}_\odot$: PSR~J0348$+$0432~\citep{Antoniadis2013}, PSR~J1614$-$2230~\citep{Demorest2010}, PSR~J1903$+$0327 and PSR~J1909$-$3744~\citep{Arzoumanian2018}; Bayesian estimation of $\mathrm{M}_{\mathrm{NS}}$-$\mathrm{R}_{\mathrm{NS}}$ of PSR~J0740$+$6620~\citep{Riley2021}; $(1+z)$ deduced from the \emph{Present}$^\S$ best-fitted modeled light curve for the GS~1826$-$24 clocked bursts and from \citet{Johnston2020} based on three epochs; quiescent low-mass X-ray binaries~\citep{Guillot2013}, causality~\cite{Koranda1997,Lattimer2007}, and conditional constraints from the GW170817 neutron star merger~\citep{Bauswein2017} and realistic models of equation of state (EOS; \citealt{Fattoyev2018,Steiner2018}).}}
\label{fig:NScompactness}
\end{center}
\end{figure}

Using the best-fit modeled burst light curve produced from the \emph{baseline} and \emph{Present}$^{\S,\ddag,\dag}$ models, we estimate $(1+z)=1.26^{+0.04}_{-0.05}$, of which the uncertainty is based on the averaged deviation (2~\%) of comparing the modeled light curves with the averaged-observed light curve of \emph{Epoch Jun 1998}. The range of host neutron-star mass-radius relation, $\mathrm{M}_{\mathrm{NS}}$-$\mathrm{R}_{\mathrm{NS}}$, of the GS~1826$-$24 X-ray source is then estimated using our best-fit $(1+z)$ and Eq.~(1) of \citet{Johnston2020}. The deduced range of $\mathrm{M}_{\mathrm{NS}}$-$\mathrm{R}_{\mathrm{NS}}$ for GS~1826$-$24 is compared with recently assessed $\mathrm{M}_{\mathrm{NS}}$-$\mathrm{R}_{\mathrm{NS}}$ constraints, i.e., the $\mathrm{M}_{\mathrm{NS}}$ of PSR~J0348$+$0432 and PSR~J1614$-$2230 deduced by \citet{Antoniadis2013} and \citet{Demorest2010}, respectively, based on Shapiro time delay (overlapping pink strips in Fig.~\ref{fig:compactness}), the $\mathrm{M}_{\mathrm{NS}}$-$\mathrm{R}_{\mathrm{NS}}$ of PSR~J0740$+$6620 estimated recently by \citet{Riley2021} using Bayesian analysis (green zone in Fig.~\ref{fig:compactness}), the $\mathrm{M}_{\mathrm{NS}}$ of PSR~J1903$+$0327 and PSR~J1909$-$3744 compiled by \citeauthor{Arzoumanian2018} (\citeyear{Arzoumanian2018}; distinctive light pink strips in Fig.~\ref{fig:compactness}), the $\mathrm{R}_\mathrm{NS}$ range of neutron stars with $\mathrm{M}_\mathrm{NS}=1.4$, $1.7$, and $2\mathrm{M}_\odot$ proposed by \citeauthor{Steiner2018} (\citeyear{Steiner2018}; three yellow lines in Fig.~\ref{fig:compactness}), the $\mathrm{M}_{\mathrm{NS}}$-$\mathrm{R}_{\mathrm{NS}}$ of GS~1826$-$24 proposed by \citet{Johnston2020} using Markov chain Monte-Carlo method to estimate the properties of three epochs (dark purple line in Fig.~\ref{fig:compactness}), lower and upper limits of $\mathrm{R}_{\mathrm{NS}}$ given by \citeauthor{Bauswein2017} (\citeyear{Bauswein2017}; black dot in Fig.~\ref{fig:compactness}) and by \citeauthor{Fattoyev2018} (\citeyear{Fattoyev2018}; blue dot in Fig.~\ref{fig:compactness}) based on the study of GW170817 neutron-star merger.

The presently estimated range of $\mathrm{M}_{\mathrm{NS}}$-$\mathrm{R}_{\mathrm{NS}}$ for GS~1826$-$24 (light red zone in Fig.~\ref{fig:compactness}) overlaps with the constraints suggested by \citet{Johnston2020} and \citet{Steiner2018} (Fig.~\ref{fig:compactness}), presuming that its $\mathrm{M}_{\mathrm{NS}}$-$\mathrm{R}_{\mathrm{NS}}$ is likely in the range of $\mathrm{M}_\mathrm{NS}\gtrsim 1.7\mathrm{M}_\odot$ and $\mathrm{R}_\mathrm{NS} \sim\!\!12.4$ -- 13.5~km. This suggests that the radius of PSR~J1903$+$0327 could be close to the range of $12.4\lesssim\mathrm{R}_\mathrm{NS}/\mathrm{km}\lesssim13.5$, and neutron stars with $\mathrm{M}_\mathrm{NS}\approx 1.7\mathrm{M}_\odot$ could be less compact than that were estimated by \citet{Guillot2013}. 

We emphasize that as the present neutron-star mass-radius constraint is based on $(1+z)=1.26^{+0.04}_{-0.05}$ deduced from XRB models with averaged deviations between modeled and observed burst light curves up to $1.252\times10^{-9}\mathrm{erg~cm}^{-2}\mathrm{s}^{-1}$, a $(1+z)$ factor deduced from a well match modeled and observed burst light curve with much lower deviation is highly desired to provide a more reasonable constraint.


\section{Summary and conclusion}
\label{sec:summary}

We use the self-consistent RHB theory with DD-ME2 interaction to calculate the MDE for isospin $I=1/2, 1, 3/2$, and 2 $p\!f$-shell mirror nuclei of $A=41$~--~$75$. Then, a set of $S_\mathrm{p}$($^{66\!}$Se) values are obtained from folding the experimental $^{66}$Ge and $^{65}$Ge nuclear masses and theoretical MDEs, or the experimental $^{66}$Ge and $^{65\!\!}$As nuclear masses and theoretical MDEs. The $S_\mathrm{p}$($^{66\!}$Se$)=2.469\mathord\pm0.054$~MeV is selected to be the reference. The 54-keV uncertainty is deduced from the rms deviation of comparing both theoretical and updated experimental MDEs (AME2020). Then, using the full $p\!f$-model space shell-model calculations with the GXPF1a Hamiltonian, we choose $S_\mathrm{p}$($^{66\!}$Se$)=2.469$~MeV as a reference, and conservatively assume the uncertainty to be 100~keV to estimate the extent of the new $^{65\!\!}$As(p,$\gamma$)$^{66\!}$Se forward and reverse reaction rates to cover other estimated $S_\mathrm{p}$($^{66\!}$Se), i.e., $S_\mathrm{p}$($^{66\!}$Se) $=2.433$~MeV, $2.443$~MeV, and $2.507$~MeV, whereas the forward and reverse reaction rates based on $S_\mathrm{p}$($^{66\!}$Se) $=2.186$~MeV, $2.351$~MeV, and $2.284$~MeV are calculated as well. We comprehensively study the influence of the new rate on the burst light curve of GS~1826$-$24 clocked burster, nucleosynthesis in and evolution of the accreted envelope, and burst-ash abundances at the burst tail end. Future precisely measured $S_\mathrm{p}$($^{66\!}$Se) with uncertainty lower than or $\approx50$~keV can confirm our present findings and predictions.

We find that the duration of 2p-capture on $^{64}$Ge and weak GeAs cycles are affected by the new $^{65\!\!}$As(p,$\gamma$)$^{66\!}$Se forward and reverse reaction rates. The longer time the 2p-capture on $^{64}$Ge maintains in a clocked XRB of the GS~1826$-$24 X-ray source, the more material is transferred via the $^{64}$Ge(p,$\gamma$)$^{65\!\!}$As(p,$\gamma$)$^{66\!}$Se($\beta^+\nu$)$^{66}$As(p,$\gamma$)$^{67}$Se($\beta^+\nu$)$^{67}$As(p,$\alpha$) $^{68}$Se($\beta^+\nu$)$^{68}$As(p,$\alpha$)$^{69}$Se path to reach the region heavier than Se of which intensive (p,$\gamma$)-($\beta^+\nu$) reaction sequences ascertainably burn accreted hydrogen, release nuclear energy, and thus increase the burst light curve. Meanwhile, the status of $^{64}$Ge as an important and historic waiting point is affirmed by analogizing the evolution of $^{64}$Ge production with the synthesis of $^{66\!}$Se along the mass coordinate of accreted envelope, and the comparable abundances of $^{64}$Ge, $^{60}$Zn, and $^{66\!}$Se in the accreted envelope.

We also include the new $^{22}$Mg($\alpha$,p)$^{25}\!$Al reaction rate in our study, and find that its influence on the clocked XRB at time regime $t=16$~s~--~$60$~s and burst-ash compositions is stronger than other considered reactions, i.e., the new $^{65\!\!}$As(p,$\gamma$)$^{66\!}$Se forward and reverse, $^{56}$Ni(p,$\gamma$)$^{57}$Cu(p,$\gamma$)$^{58}$Zn, $^{55}$Ni(p,$\gamma$)$^{56}$Cu, and $^{64}$Ge(p,$\gamma$)$^{65\!\!}$As reactions. The affected burst-ash compositions consist of nuclei in hot CNO cycle, $sd$ and $p\!f$ shells, up to $A=140$, except $A=87,\dots,96$. Both \emph{Present}$^{\S}$ and \emph{Present}$^{\dag}$ models show that the abundances of nuclei $A=64$, $68$, $72$, $76$, and $80$ are affected up to a factor of $1.4$. The inclusion of the updated $^{22}$Mg($\alpha$,p)$^{25}\!$Al reaction rate in \emph{Present}$^{\S}$ influences the production of $^{12}$C up to a factor of $4.5$ that could be the main fuel for superburst. A set of noticeably periodic increment of burst-ash abundances exists in the region heavier than $^{64}$Ge with waiting points leading the increment, indicating that the resemblance of 2p-capture and weak GeAs cycles also coexists in the region during the thermonuclear runaway.

We remark that the impartial fit on the whole timespan of burst light curve permits us to produce a set of modeled burst light curves with excellent agreement with the observed \emph{Epoch Jun 1998} at burst peak and tail end, and the distinguished feature of the light curve is reproduced. The averaged deviation of between modeled and observed burst light is only up to $1.252\times10^{-9}\mathrm{erg~cm}^{-2}\mathrm{s}^{-1}$. The best-fit modeled bursts, which diminish accordingly with observation and with considerably low discrepancy, presumably provides a set of more convincing burst-ash abundances that are not observed. This permits us to understand the nucleosyntheses that happen during the thermonuclear runaway in the accreted envelope. The modeled burst light curve produced by \citet{Randhawa2020} is, however, rather extensive and the H-burning does not recede accordingly with observation, and also the modeled burst peak is misaligned with observation. The one-zone model constructed by \citet{Schatz2017} successfully obtains the influence of $^{65\!\!}$As(p,$\gamma$)$^{66\!}$Se reaction rate (with larger $S_\mathrm{p}$($^{66\!}$Se) compared to the NON-SMOKER $^{65\!\!}$As(p,$\gamma$)$^{66\!}$Se rate) on a gross feature of burst light curve, however, the H-burning recedes earlier than their baseline model, accelerating the recession of burst tail end.

In this work, we presume the radius of the host neutron star of GS~1826$-$24 to be in the range of $\sim\!\!12.4$ -- 13.5~km as long as its mass $\gtrsim\!1.7\mathrm{M}_\odot$. Besides, the radius of PSR~J1903$+$0327 could be likely close to the range of $12.4\lesssim\mathrm{R}_\mathrm{NS}/\mathrm{km}\lesssim13.5$, and future observations could shed light on the estimation of neutron-star radii coupled with $\mathrm{M}_\mathrm{NS}\approx 1.7\mathrm{M}_\odot$. This presumption indicates that the neutron-star compactness estimated by \citet{Guillot2013} is somehow more compact.


We are deeply grateful to H. Schatz for reading our manuscript and for providing thoughtful and constructive suggestions to improve the manuscript, and to B. A. Brown for providing the previous and updated $S_\mathrm{p}$($^{66\!}$Se) and for reading and suggesting constructive thoughts to the part related to the $S_\mathrm{p}$($^{66\!}$Se). We thank W. J. Huang for providing the information of AME. We also thank the anonymous referee for the helpful comments and remarks on this manuscript. We are very thankful to N. Shimizu for suggestions in tuning the \textsc{KShell} code at the PHYS\_T3 (Institute of Physics) and QDR4 clusters (Academia Sinica Grid-computing Centre) of Academia Sinica, Taiwan, to D. Kahl for checking and implementing the newly updated $^{56}$Ni(p,$\gamma$)$^{57}$Cu reaction rate, to M. Smith for using the Computational Infrastructure for Nuclear Astrophysics, and to J. J. He for fruitful discussion.
This work was financially supported by 
the Strategic Priority Research Program of Chinese Academy of Sciences (CAS, Grant Nos. XDB34000000 and XDB34020204) 
and National Natural Science Foundation of China (No. 11775277). 
We are appreciative of the computing resource provided by the Institute of Physics (PHYS\_T3 cluster) and the Academia Sinica Grid-computing Center (ASGC) Distributed Cloud resources (QDR4 cluster) of Academia Sinica, Taiwan. 
Part of the numerical calculations were performed at the Gansu Advanced Computing Center. 
YHL gratefully acknowledges the financial supports from the Chinese Academy of Sciences President's International Fellowship Initiative (No. 2019FYM0002) and appreciates the laptop (Dell M4800) partially sponsored by Pin-Kok Lam and Fong-Har Tang during the pandemic of Covid-19.
A.H. is supported by the Australian Research Council Centre of Excellence for Gravitational Wave Discovery (OzGrav, No. CE170100004) and for All Sky Astrophysics in 3 Dimensions (ASTRO 3D, No. CE170100013), and US National Science Foundation under Grant No. PHY-1430152 (JINA Center for the Evolution of the Elements).





\end{document}